%% file: 0.main.tex
\documentclass[manuscript,screen]{acmart}
\renewcommand\footnotetextcopyrightpermission[1]{} 

\usepackage{xcolor}
\usepackage{enumitem}
\usepackage{dirtytalk}
\usepackage[super]{nth}
\usepackage{nicefrac}
\usepackage{bbm}
\usepackage{lscape}

\usepackage{tikz,colortbl}
\usetikzlibrary{calc}
\usepackage{zref-savepos}
\newcounter{NoTableEntry}
\renewcommand*{\theNoTableEntry}{NTE-\the\value{NoTableEntry}}

\newcommand*{\notableentry}{%
  \multicolumn{1}{@{}c@{}|}{%
    \stepcounter{NoTableEntry}%
    \vadjust pre{\zsavepos{\theNoTableEntry t}}
    \vadjust{\zsavepos{\theNoTableEntry b}}
    \zsavepos{\theNoTableEntry l}
    \hspace{0pt plus 1filll}%
    \zsavepos{\theNoTableEntry r}
    \tikz[overlay]{%
      \draw[black] 
        let
          \n{llx}={\zposx{\theNoTableEntry l}sp-\zposx{\theNoTableEntry r}sp},
          \n{urx}={0},
          \n{lly}={\zposy{\theNoTableEntry b}sp-\zposy{\theNoTableEntry r}sp},
          \n{ury}={\zposy{\theNoTableEntry t}sp-\zposy{\theNoTableEntry r}sp}
        in
        (\n{llx}, \n{lly}) -- (\n{urx}, \n{ury}) 
      ;
    }%
  }%
}

\AtBeginDocument{%
  \providecommand\BibTeX{{%
    \normalfont B\kern-0.5em{\scshape i\kern-0.25em b}\kern-0.8em\TeX}}}

\setcopyright{acmcopyright}
\acmDOI{10.1145/1122445.1122456}

\acmConference[Woodstock '18]{Woodstock '18: ACM Symposium on Neural Gaze Detection}{June 03--05, 2018}{Woodstock, NY}

\begin{document}

\title{A Network Science perspective of Graph Convolutional Networks: A survey}

\author{Mingshan Jia}
\email{mingshan.jia@uts.edu.au}
\orcid{1234-5678-9012}

\author{Bogdan Gabrys}
\email{bogdan.gabrys@uts.edu.au}
\author{Katarzyna Musial}
\email{katarzyna.musial-gabrys@uts.edu.au}
\affiliation{%
  \institution{University of Technology Sydney}
  \streetaddress{15 Broadway}
  \city{Sydney}
  \state{NSW}
  \country{Australia}
  \postcode{2007}
}



\renewcommand{\shortauthors}{Jia et al.}

\begin{abstract}

The mining and exploitation of graph structural information have been the focal points in the study of complex networks. Traditional structural measures in Network Science focus on the analysis and modelling of complex networks from the perspective of network structure, such as the centrality measures, the clustering coefficient, and motifs and graphlets, and they have become basic tools for studying and understanding graphs. In comparison, graph neural networks, especially graph convolutional networks (GCNs), are particularly effective at integrating node features into graph structures via neighbourhood aggregation and message passing, and have been shown to significantly improve the performances in a variety of learning tasks. These two classes of methods are, however, typically treated separately with limited references to each other. In this work, aiming to establish relationships between them, we provide a network science perspective of GCNs. Our novel taxonomy classifies GCNs from three structural information angles, i.e., the layer-wise message aggregation scope, the message content, and the overall learning scope. Moreover, as a prerequisite for reviewing GCNs via a network science perspective, we also summarise traditional structural measures and propose a new taxonomy for them. Finally and most importantly, we draw connections between traditional structural approaches and graph convolutional networks, and discuss potential directions for future research.
\end{abstract}

\ccsdesc[500]{Networks~Topology analysis and generation}
\ccsdesc[500]{Networks~Network structure}
\ccsdesc[500]{Computing methodologies~Machine learning}
\keywords{Graph Convolutional Networks, Network Science, graph structural measures}

\maketitle
\pagestyle{plain}

\input{1.intro}
\input{2.preliminaries}
\input{3.graph_structrual_measures}
\input{4.local_structure_in_GNN}

\input{5.conclusion}


\begin{acks}
The authors thank Yu-Xuan Qiu, Joakim Skarding and Xiaohan Zhang for their helpful comments and discussions. This work was supported by the Australian Research Council, Grant No. DP190101087: \say{Dynamics and Control of Complex Social Networks}.
\end{acks}

\bibliographystyle{ACM-Reference-Format}
\bibliography{references}









\end{document}

%% file: 1.intro.tex
\section{Introduction}


Networks or graphs are a general language for modelling and analysing complex systems that are abstracted as entities and their connections \cite{barabasi2016network, newman2018networks}. In the representation of networks, domain data is no longer only being a set of isolated data points but also contains important information about the relationships among them. The entities are related to each other according to the structure of the network, and modelling these relational structures allows us to build more accurate models of the domain data. Various types of real-world data can naturally be modelled as networks, such as social networks representing social actors and their relationships \cite{musial2013social}, molecular graphs representing chemical atoms and their bonds \cite{huber2007graphs}, transportation networks representing infrastructures and traffic flow \cite{bell1997transportation}, control flow graphs representing code blocks and their executions \cite{moller2012static}, etc. 

Although networks are very powerful at modelling relational data, processing them is significantly more difficult, mainly due to their intricate topological structures. Compared to other common data formats such as images or text, network data does not have a starting or an ending point that can be defined in Euclidean space, nor the essential notion of spacial locality and proximity. Therefore, understanding and exploiting graph structure has always been a core theme in analysing complex networks. Traditional network science approaches are mostly structure-related heuristics, such as various types of node centralities \cite{lu2016vital} for node-level analysis, common neighbours similarity and its variants \cite{martinez2016survey} for link-level analysis, and motifs \cite{milo2002network} and significance profile \cite{milo2004superfamilies} for graph-level analysis. These methods, along with others, have become the standard tools for analysing graphs and have been used in all kinds of networks. Certainly, these approaches have their limitations. First is their applicability --- each is effective for examining specific properties but falls short of capturing other structural aspects. Another drawback is that most heuristic approaches focus on graph structures while overlooking the rich information that could be contained in nodes or on edges \cite{liu2020introduction}.

Another mainstream class of methods is grounded in deep learning on graphs, especially the recently emerging and quickly gaining in popularity graph convolutional networks (GCNs) \cite{wu2020comprehensive}. GCNs are generalised from the notion of Convolutional Neural Networks (CNNs) \cite{albawi2017understanding}, redefining them for non-Euclidean graph data. GCNs ingeniously combine graph structure and node/edge features via neighbourhood message aggregation and a structure-based propagation scheme. Being a rapidly evolving area of research, a large number of graph convolutional network approaches have emerged in recent years, aiming to improve its expressivity, scalability, or targeting specific tasks or types of networks \cite{abadal2021computing}. However, there are still many challenges and opportunities in this field. Some of the key open problems include developing more powerful and efficient GCN architectures, extending these models to handle temporal, multi-layered, or other more complex graph data, and improving the interpretability and transparency of GCN models.

\begin{figure}[t]
\centerline{\includegraphics[scale = 1.05]{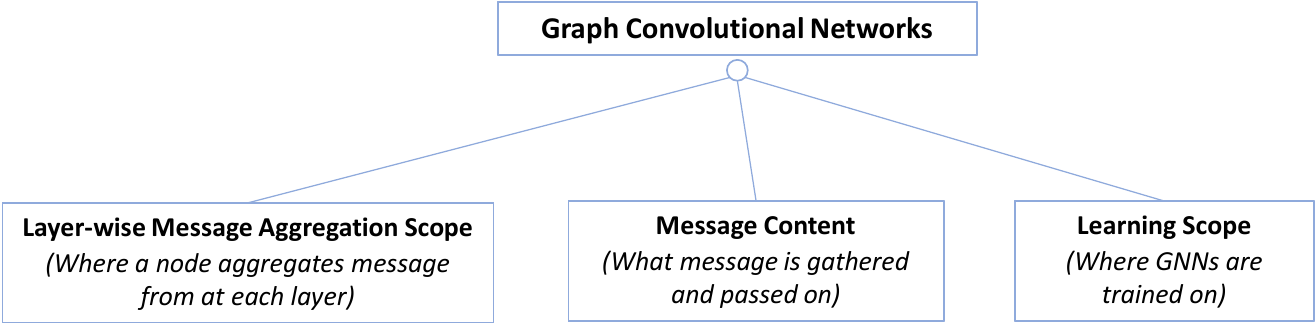}}
\caption{Taxonomy of graph convolutional networks from structural perspectives.}
\label{fig:gcn_taxonomy}
\vspace{-0mm}
\end{figure}

Traditional structural measures of Network Science are direct and efficient tools for analysing and understanding complex networks, while graph convolutional networks are deep learning models designed specifically for graph data in order to address various learning problems. As discussed, the two classes of methods have their own strengths and weaknesses. Surprisingly, they are very often treated separately with relatively limited references to each other. Network science researchers may be sceptical about the lack of explainability of deep learning approaches, while deep learning researchers tend to overlook the advance in traditional non-learning approaches. We believe, however, that with the established foundations of traditional structural measures in Network Science, and GCNs emerging as a new powerful class of methods, there would be great benefits to be realised from a closer integration and awareness of the two communities. On the one hand, GCNs gracefully incorporate node features, which are largely overlooked in traditional structural measures, into the structure of graphs, and achieve state-of-the-art performances in various tasks. On the other hand, traditional network science notions, being the foundations of understanding and characterising complex networks, are also indispensable in studying GCNs. Different types of structural measures are being exploited in the recent advance of GCNs as well \cite{bouritsas2020improving, jin2020gralsp, li2020distance, xu2021automorphic}. Therefore, in this work, we aim to link the two classes of methods together by comprehensively reviewing each of them, proposing new taxonomies and discussing their connections. 

Along with the phenomenal development of GCNs, many survey articles appeared to summarise and review them. Some have a broad scope that covers graph representation learning \cite{hamilton2017representation} or graph deep learning \cite{bronstein2017geometric, zhang2020deep,wu2020comprehensive} in general. Some others are focused on specific aspects, such as the design pipeline or the composition modules of graph neural networks \cite{zhou2020graph}, the dynamic mechanisms \cite{skarding2021foundations}, or the learning on limited labelled samples \cite{xie2022self}. However, there still lacks an examination that focuses on how graph structure information (which is the main focus of traditional network science approaches), is exploited in graph convolutional networks. Thus, in this work, we propose new taxonomies of GCNs from the perspective of graph local structure, and at the same time, review the latest works that improve graph neural networks through exploiting local structural information. Specifically, we propose to summarise graph convolutional networks from three structural angles, i.e., the scope of layer-wise message aggregation, the content of the message being passed on, and the overall scope of learning on graphs (Figure~\ref{fig:gcn_taxonomy}). 

\begin{figure}[t]
\centerline{\includegraphics[scale = 1.05]{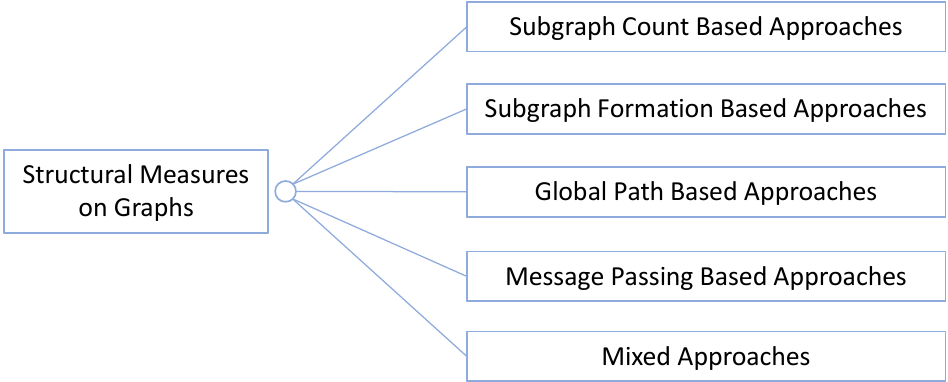}}
\caption{Structural measures on graphs.}
\label{fig:taxonomies_overall}
\vspace{-0mm}
\end{figure}

Moreover, a systematic understanding of traditional graph structural approaches is the prerequisite for thoroughly reviewing GCNs via a network science perspective. Therefore, before jumping into the sphere of graph neural networks, we first summarise and classify non-learning graph structural measures. The study of graph structures is so ubiquitous that they often appear in different terms, such as the big family of centrality measures \cite{lu2016vital, rodrigues2019network, das2018study}, the popular notion of motifs \cite{milo2002network} and graphlets \cite{milenkovic2008uncovering} and the set of subgraph formation measurements such as the clustering coefficient \cite{watts1998collective}, the closure coefficient \cite{yin2019local}, the square clustering coefficient \cite{lind2005cycles}, etc. Existing surveys on structure measurements only cover one or two sets of those notions, and fail to unite them from an overarching perspective or to draw connections and comparisons between them. Therefore, in this work with a focus on graph structure, we also propose a new taxonomy that brings all these concepts together. Specifically, we 
group existing graph structural measures into five categories: subgraph count based measures, subgraph formation based measures, global path based measures, message passing based measures, and hybrid measures ( Figure~\ref{fig:taxonomies_overall}). 
More importantly, through summarising both the traditional structural measures and the graph convolutional network approaches, we could draw connections between the two, strengthen the understanding and analysis of GCNs and lead to insightful discussions about potential research avenues. 

To summarise, the main contributions of this survey are as follows:
\begin{itemize}
    \item We propose a new taxonomy that brings together various types of traditional structure-based approaches. We make a clearer distinction between the concepts of local and global, and we first introduce and summarise the category of subgraph formation based approaches.
    \item We propose a novel taxonomy of graph convolutional networks, with a focus on the exploitation of graph structural information. The taxonomy categorises GCNs from three structural information angles, i.e., the layer-wise message aggregation scope, the message content, and the overall learning scope. We review and summarise the latest GCN approaches with a structural focus, and provide a thorough analysis of the time and space complexities.
    \item We draw connections between the graph convolutional networks and the traditional structure-based approaches, and discuss three potential future research avenues in the joint area.  
\end{itemize} 

The rest of this survey is organised as follows: In Section~\ref{sec:background}, we introduce and compare two pairs of concepts, i.e., local and global, and motifs and graphlets. In Section~\ref{sec:traditional}, we present the five categories of graph structural measures and discuss four open problems. In Section~\ref{sec: GNN}, we introduce the novel taxonomy of graph convolutional networks, and discuss their time and space complexities. In Section~\ref{sec:discussion}, we discuss the connections between the traditional structural measures and the graph convolutional networks, and present some potential research directions. Finally, we conclude the article in Section~\ref{sec:conclusion}. 

%% file: 2.preliminaries.tex
\section{Preliminaries and Background} \label{sec:background}

This section introduces preliminary concepts that are helpful for understanding the proposed taxonomies. 

\subsection{Local vs. Global}
 When discussing graph structural measures, we need first to distinguish what is local and what is global. Previous works \cite{donges2009complex,jackson2017economic, martinez2016survey, ma2012biological} either only focus on where the measures are defined by dividing them into two or three categories: (i) the "local", "micro" or "individual" level; (ii) the "global", "macro" or "aggregate" level; and (iii) sometimes at the "mesoscopic", "quasi-local" or "subnetwork" level; or they are defined solely based on the scope of information involved in their computation. This, however, leads to some confusion. For example, the betweenness centrality is defined for nodes (at the node-level) but requires global information to compute. Should it be termed a local measure or a global measure? Similarly, the average clustering coefficient is defined at the network-level, but only needs local information at each node --- calculating the local clustering coefficient at each node, then averaging over all nodes.  

Therefore, we propose the following terms to distinguish both at what level the measures are defined and the scope of information that is needed to calculate them: \\
\begin{itemize}
    \item \textit{Local-level measure} is a measurement defined on a node-level or link-level (the link here also includes the non-existing or potential link which is often used in a link prediction task). Thus, it can be further divided into a \textit{node-level measure} and a \textit{link-level measure}.
    
    \item \textit{Network-level measure} is a measurement defined for the entire network. 
    
    \item \textit{Local structural measure} is a measurement whose computation only involves the nearby neighbourhood of a node, i.e., within a range of k-hop away from a node. In most cases, k is less than or equal to $4$. Many traditional measures only care about the immediate neighbourhood around a node, and we name them as \textit{Strict-local structural measures}.
    
    \item \textit{Global structural measure} is a measurement that involves the global information in computation. This type of measurement often involves the computation of paths between nodes in the network.
\end{itemize}

Now, when we revisit the betweenness centrality, it is both a local-level and a global structural measure. The average clustering coefficient, on the other hand, is both a network-level measure and a local structural measure. Notice that the average clustering coefficient involves the extra step of averaging over all nodes. Indeed, it is $n$ times the complexity of computing the local clustering coefficient at a single node. However, any local-level measure can easily have an extended definition at the network-level through averaging over all nodes or edges. Moreover, in the practice of network analysis, local-level measures are often calculated for the entire network, looping over all nodes or all edges. Therefore, when defining local or global structural measures, we choose to exclude this aggregation or averaging step. 

To summarise, we use the terms \say{local-level} and \say{global-level} to distinguish where the measure is defined, and we use the terms \say{local structural} and \say{global structural} to distinguish the scope of information involved in the computation, before the aggregation/averaging step.  


\subsection{Motifs vs. Graphlets} \label{sec:motif_vs_graphlet}
Next, we distinguish three similar concepts that are later used in our taxonomies, i.e., subgraphs, motifs and graphlets. A subgraph, as the name implies, is a smaller graph whose node set and edge set are subsets of those of the original graph. We then recap the notions of motifs \cite{milo2002network} and graphlets \cite{milenkovic2008uncovering} according to the papers that proposed them. 

\begin{table}[t]
\centering
\caption{Some 3-node and 4-node motifs in directed networks\cite{milo2002network}. Motifs containing bidirectional edges are not included. }
\label{tab:motifs}
\begin{tabular}{|c|c|c|}
\hline
Motif & Designation              & Type of network                                                                               \\ \hline
\includegraphics[width=2.8cm, height=1.5cm, keepaspectratio]{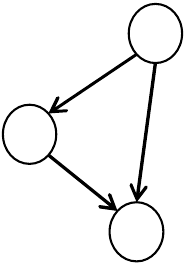}    & 3-node feed-forward loop & \begin{tabular}[c]{@{}c@{}}Gene regulation network \\ Neural network\\ Electronic circuits (forward logic chips)\end{tabular}    \\ \hline
\includegraphics[width=2.8cm, height=1.5cm, keepaspectratio]{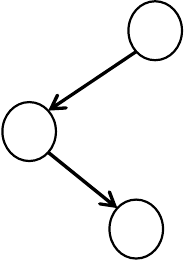}   & 3-chain                  & Food webs                                                                                    \\ \hline
\includegraphics[width=2.8cm, height=1.5cm, keepaspectratio]{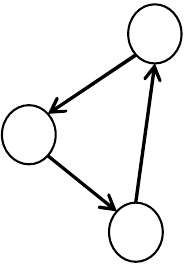}      & 3-node feedback loop     & \begin{tabular}[c]{@{}c@{}}Gene regulation network \\ Neural network\\ Electronic circuits (forward logic chips)\end{tabular}   \\ \hline
\includegraphics[width=2.8cm, height=1.5cm, keepaspectratio]{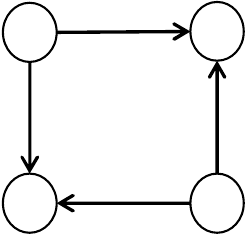}      & Bi-fan                   & \begin{tabular}[c]{@{}c@{}}Gene regulation network\\ Neural network\\ Electronic circuits (forward logic chips) \\Electronic circuits II\end{tabular} \\ \hline
\includegraphics[width=2.8cm, height=1.5cm, keepaspectratio]{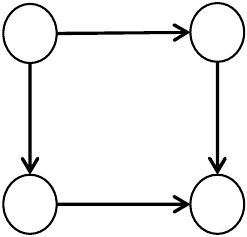}      & Bi-parallel              & \begin{tabular}[c]{@{}c@{}}Neural network\\ Food webs\\ Electronic circuits (forward logic chips)\end{tabular}          \\ \hline
\includegraphics[width=2.8cm, height=1.5cm, keepaspectratio]{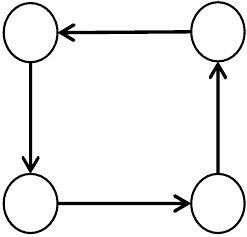}      & 4-node feedback loop     & Electronic circuits II                                                                       \\ \hline
\end{tabular}
\end{table}

Network motifs \cite{milo2002network} are subgraphs that recur much
more frequently in the real network than in an ensemble of randomised networks.
They are defined at the network-level, in order to uncover the basic building blocks of directed networks across domains. Subgraphs having a $p$-value less than $0.01$ are deemed as motifs, where $p$ is the probability of the subgraph appearing more times in randomised networks than in the real network. The statistical significance of a motif can also be captured by the Z-score, which is calculated as follows:

$$Z_{i}=\left(N_{i}^{\mathrm{real}}-\bar{N}_{i}^{\mathrm{rand}}\right) / \operatorname{std}\left(N_{i}^{\text {rand }}\right),$$ 

where $N_{i}^{\mathrm{real}}$ is the number of subgraphs of type $i$ in the real network, and $N_{i}^{\text {rand }}$ is the number of subgraphs of type $i$ in a randomised network. A natural downside of this approach, however, is that it needs to generate a large number of random networks (e.g. 100s or 1000s) using a certain configuration model. The original work only focuses on 3-node and 4-node directed subgraphs, finding that particular subgraphs such as 3-node feed-forward loop, 3-node feedback loop, bi-fan, bi-parallel, and 4-node feedback loop are significant building blocks in several different types of directed networks (Table~\ref{tab:motifs}).

\begin{figure}[t]
\centerline{\includegraphics[scale = 0.8]{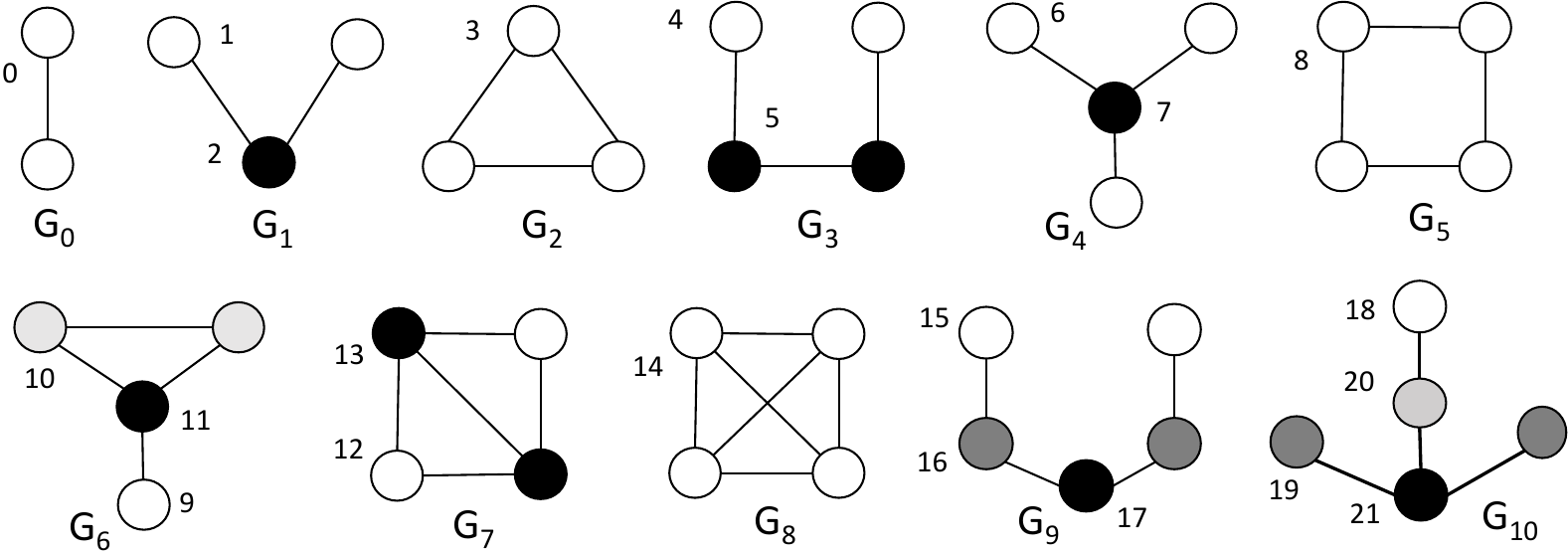}}
\caption{Graphlets and their orbits \cite{milenkovic2008uncovering}}
\label{fig:graphlets}
\vspace{0mm}
\end{figure}

Graphlets \cite{milenkovic2008uncovering}, are nonisomorphic induced subgraphs around a focal node. In the original work, it is defined for undirected networks. A key difference between motifs and graphlets is that graphlets are defined at node-level. The term automorphism orbits, or orbits for short, are used to distinguish different positions of the focal node in a subgraph. 
Therefore, when subgraph size is limited to a range of 2 to 5 nodes, there are 73 different orbits on 30 different subgraphs. We recap graphlets with the orbits in Figure~\ref{fig:graphlets} (in order to save some space, the majority of 5-node graphlets are omitted). It is worth mentioning that the idea of counting induced subgraphs is also extended to the link-level, leading to the notion of edge orbits \cite{hovcevar2016computation}. Taking graphlet $G_1$ in Figure~\ref{fig:graphlets} for example, there exist two (node) orbits denoted '1' and '2', respectively. In contrast, there is only one edge orbit in it since the two edges are structurally equivalent.

\begin{figure}[h]
\centerline{\includegraphics[scale = 0.55]{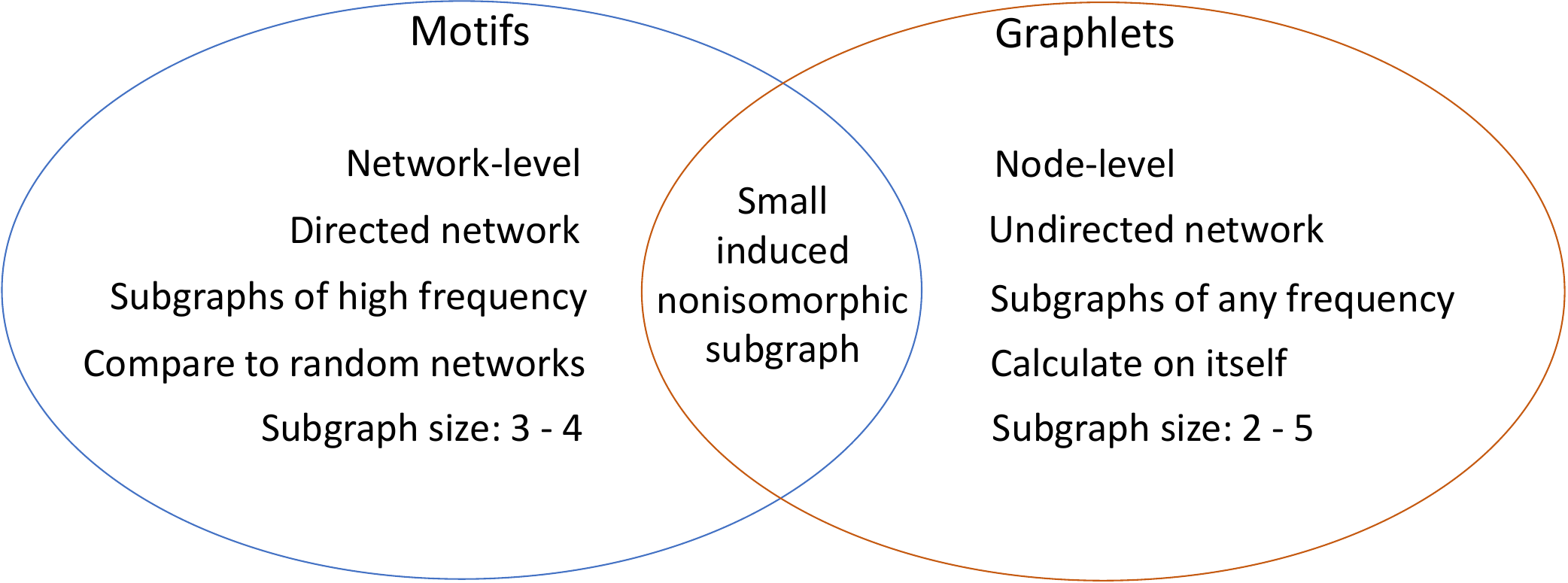}}
\caption{Motifs vs. Graphlets}
\label{fig:motif_graphlet}
\vspace{0mm}
\end{figure}

To summarise, motifs and graphlets are both small induced subgraphs, but they are different in the following aspects (Figure~\ref{fig:motif_graphlet}):
motifs are defined at the network-level while graphlets are defined at the node-level; motifs are proposed for directed networks while graphlets are for undirected networks; motifs are discovered from comparing real networks to randomised networks with the same degree sequence while graphlets are calculated on the network itself; lastly, motifs contain 3 - 4 nodes while graphlets have 2 -5 nodes. Notice that most of the analyses stop at 4 or 5 nodes because a subgraph containing more than 5 nodes would become too complicated for us to enumerate and interpret all possible subgraphs or orbits. For example, a 6-node induced subgraph leads up to 112 different types of subgraphs and 407 different orbits. Taking link directions into consideration, there are $1,530,843$ subgraphs and $9,031,113$ orbits \cite{ribeiro2021survey}.

%% file: 3.graph_structrual_measures.tex
\section{Graph structural measures} \label{sec:traditional}
In order to set up the context of reviewing graph convolutional networks from a Network Science perspective, we first summarise traditional graph structural measures and propose a novel taxonomy for them, which will later be used in our categorisation and analysis of GCNs. Specifically, We divide existing structural measures into five categories (see Figure~\ref{fig:taxonomies_overall}): 

\begin{itemize}
    \item \textit{Subgraph Count Based Approaches}. These measures are defined based on the number of a particular subgraph or subgraphs. 
    
    \item \textit{Subgraph Formation Based Approaches}. In this category, the measures are defined by the ratio of the numbers of two subgraphs: one contains fewer edges (or nodes) and is viewed as the formation base of another. 
    
    \item \textit{Global Path Based Approaches}. As the name implies, these measures are based on unbounded paths. They involve the calculation of shortest paths or all paths originating from a node to any node in the entire graph.   
    
    \item \textit{Message Passing Based Approaches}. Unlike previous categories, message passing-based approaches utilise graph structural information in an implicit manner: every node is initialised with an importance score. Then iteratively, each node updates its score through aggregating the scores of its neighbours. Graph Neural Network approaches (see more in Section~\ref{sec: GNN}) can be viewed as transforming this traditional message passing approach into a learnable process. 
    
    \item \textit{Hybrid Approaches}. These measures are simply some combinations of the previous four categories. 
\end{itemize}

We now explain the logic behind our taxonomy. The first two categories both cover a local area of the whole network (within a certain distance from the focal node, or containing a limited number of nodes and edges). The first category --- subgraph count based approaches --- is built from counting the number of particular local structures. For example, the number of neighbours, local paths or subgraphs. The second category --- subgraph formation based approaches --- is uniquely defined based on the ratio of two subgraphs and thus bears the meaning of measuring the formation of certain local structures. To have both of them in the taxonomy instead of combining them into one category is to stress their differences. 

Then, the third category expands its scope to the entire network. We name it global path based approaches instead of just global approaches. This is because all global approaches involve either the calculation of shortest paths or all paths
originating from a node to any other node in the entire graph. Notice here that a path is also a particular type of subgraph. However, a local path or bounded path, such as a 2-path or 3-path, belongs to the category of subgraph count based measures, whereas a global path or unbounded path is in this category. We choose to differentiate the third category from the previous two categories from the perspective of the covered scope.  

Next, the fourth category --- message passing based approaches --- is based on the idea of propagating information along the edges. It is a different form to the abovementioned three categories because it does not calculate any type of subgraphs or global paths. Instead, the structure is utilised in an implicit way. Every node is initialised with an importance score. Then iteratively, each node updates its score through aggregating the scores of its neighbours. Although these four categories are largely different from each other, there are many approaches that combine them together, which are naturally put into the fifth category --- mixed approaches.  


\subsection{Subgraph Count Based Approaches} \label{sec:subgraph-count}
\begin{figure}[h]
\centerline{\includegraphics[scale = 1]{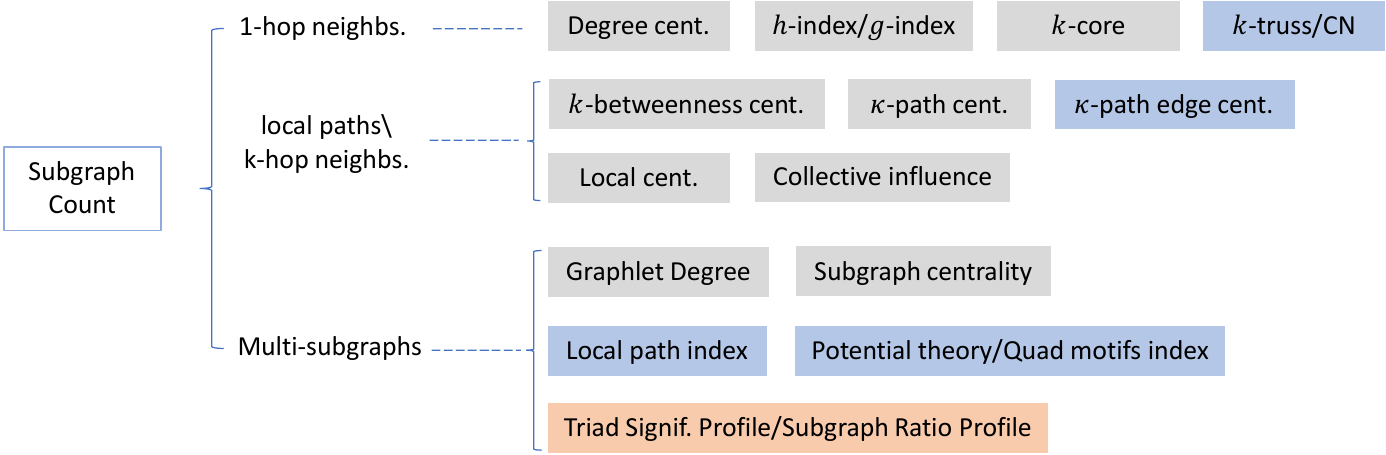}}
\caption{Subgraph count based measures.}
\label{fig:sc}
\vspace{0mm}
\end{figure}

Subgraph count based measures are based on the number of a particular subgraph or subgraphs. We further divide them into three subclasses, i.e., measures defined on 1-hop neighbours, measures defined on k-hop neighbours/local paths, and measures defined on multi-subgraphs. Figure~\ref{fig:sc} gives the detailed categorisation. The colour of the block differentiates where the approach is defined: grey is on the node-level, blue is on the link-level, and orange is on the network-level. 

\subsubsection{1-hop neighbours} As the name implies, the calculations within this category only require the immediate neighbourhood around a node or a link. \label{sec:1-hop}
\begin{itemize}[leftmargin=*]
    \item[--] \textbf{Degree centrality}. Through calculating the number of nodes directly connected to a node, the degree centrality is an easy and straightforward way to assess the importance or influence of the node\cite{freeman1978centrality}. In order to render it within the range of (0,1], it is often normalised by the size of the network minus one. Mathematically, the normalised degree centrality of node $i$ is defined as: 
    \begin{equation}
        \Theta_D(i) = \frac{d_i}{n-1}.
    \end{equation}
    Despite being so simple, the degree centrality has been widely applied in various domains. For example, in customer networks, the degree centrality is used to find opinion leaders \cite{risselada2016indicators}, and in biomedical semantic networks, it is effective in selecting crucial information for summarising a disease treatment \cite{zhang2011degree}. Some interesting extensions of the degree centrality include the in-degree/out-degree centrality in directed networks, the strength centrality and weighted strength centrality in weighted networks \cite{candeloro2016new} and the cross-layer degree centrality in multi-layered networks \cite{brodka2011degree}.
    \item[~]
    
    \item[--] \textbf{$h$-index/$g$-index}. $h$-index is proposed to evaluate the impact of an individual's research output: A researcher has an index of $h$ if $h$ of his or her papers have at least $h$ citations \cite{hirsch2005index}. It is then used as a centrality measure in networks, and named as lobby index or $l$-index \cite{korn2009lobby}. The $l$-index of a node is the largest integer $k$ such that the node has at least $k$ neighbours with a degree of at least $k$. Egghe argued that the influence of highly cited papers is underplayed in the $h$-index, and proposed a $g$-index to overcome this disadvantage \cite{egghe2006theory}. After ranking a researcher's papers according to their citations, the $g$-index is defined as the highest rank $g$ such that the top $g$ papers together have at least $g^2$ citations. From its definition, the $g$-index is always greater than or equal to the $h$-index. To address the same issue, an $e$-index is proposed to complement the $h$-index for excess citations\cite{zhang2009index}. Recently, a local $h$-index centrality is proposed to identify influential spreaders by simultaneously considering the $h$-index values of the node and its neighbours \cite{liu2018leveraging}: $\Theta_{LH}(i) = h(i) + \sum_{j \in Ni}h(j)$.
    \item[~]
    
    \item[--] \textbf{$k$-core} \cite{kitsak2010identification}. Instead of only calculating the number of 1-hop neighbours at one node (as in the degree centrality) or at both the node and its neighbours (as in the $h$-index), a $k$-core or coreness takes into account the number of neighbours at every node. Specifically, the $k$-core is defined as a subgraph in which all nodes of a degree smaller than k have been removed and the remaining nodes have a degree of at least $k$. A node located in a higher $k$-core is deemed more important than a node in a lower $k$-core. The $k$-core is calculated through the $k$-shell decomposition \cite{carmi2007model} which incrementally (from 1 to $k$) removes nodes with degree less than $k$ (which in turn results in lowering the degree of remaining nodes) until no more nodes need to be removed. Given that the degree centrality, the $h$-index and the coreness are all based on the number of 1-hop neighbours, Lü et al. further revealed their relationships through proposing the high-order $h$-indices \cite{lu2016h}. Bae et al. further propose a neighbourhood coreness that considers both the degree of a node and the coreness of its neighbours \cite{bae2014identifying}:
    \begin{equation}
        \Theta_{NC}(i) = \sum_{j \in N(i)}ks(j).
    \end{equation}
    The assumption is that a node having more connections to the neighbours located in the core of the network is more influential.
    \item[~]

    \item[--] \textbf{$k$-truss/Common neighbours}. A $k$-truss is a subgraph where every edge is contained in at least $k-2$ triangles\cite{cohen2008trusses, wang2012truss}. It is found through counting the number of common neighbours of a pair of nodes that forms an edge, i.e., the number of triangles that the edge participates in. The $k$-truss is also a $(k+1)$-core. Counting common neighbours around a pair of nodes that have not formed an edge (a non-edge) is also a basic approach in a link prediction task \cite{liben2007link}. There is a big family of similar approaches based on the number of neighbours around non-edges, such as the Adamic-Adar index, the resource allocation index, the preferential attachment index, among others \cite{martinez2016survey}. Notice that both $k$-truss and Common Neighbours-like approaches are defined at the link-level. The block colour is therefore blue in Figure~\ref{fig:sc}.         
    
\end{itemize}

\subsubsection{local paths/k-hop neighbours} \label{sec: local paths}
The group of methods in this category requires the calculation of local paths or k-hop neighbours.

\begin{itemize}[leftmargin=*]
    \item[--] \textbf{$k$-betweenness centrality} \cite{borgatti2006graph}. The $k$-betweenness centrality or bounded-distance betweenness centrality is a variation of the well-known betweenness centrality that limits the length of shortest paths to a predefined value $k$. Specifically, the $k$-betweenness centrality of any node $i$ is calculated by: 
    \begin{equation}
        \Theta_{B_k}(i)=\sum_{s, t \in V} \frac{\sigma_k(s, t \mid i)}{\sigma_k(s, t)},
    \end{equation}
    where $\sigma_k(s, t)$ is the number of shortest paths of length at most $k$ between a node pair $s$ and $t$, and $\sigma_k(s, t \mid i)$ is the number of those paths that pass through node $i$. The reason for proposing a bounded-distance betweenness centrality is that in some networks, long paths are rarely used for the propagation of influence.
    \item[~]

    \item[--] \textbf{$\kappa$-path centrality} \cite{alahakoon2011k}. Instead of limiting the length of shortest paths between node pairs, the $\kappa$-path centrality assumes that message traversals are along random simple paths of length at most $k$, and proposes to calculate the sum of the probability that a message originating from any possible node goes through the focal node. The $\kappa$-path centrality of node $i$ is defined as:
    \begin{equation}
        \Theta_{P_k}(i)=\sum_{s \in V} \frac{\sigma_{k}(s \mid i)}{\sigma_k(s)},
    \end{equation}
    where $s$ are all the possible source nodes, $\sigma_{k}(s \mid i)$ is the number of $k$-paths originating from $s$ and passing through $i$, and $\sigma_k(s)$ is the overall number of $k$-paths originating from $s$. In order to calculate it more efficiently in large networks, a randomised approximation algorithm called RA-$\kappa$path is also proposed. \cite{alahakoon2011k} 
    \item[~]
    
    \item[--] \textbf{$\kappa$-path edge centrality} \cite{de2012novel}. Moving the $\kappa$-path centrality definition to link-level, we then have the $\kappa$-path edge centrality. The $k$-path edge centrality of any given edge $e$ is defined as the sum of the frequency with which a message originated from any possible node traverses $e$, assuming that the message traversals are along random simple paths of length at most $k$:
    \begin{equation} \label{eq:k-path}
        \Theta_{P_k}(e)=\sum_{s \in V} \frac{\sigma_{k}(s \mid e)}{\sigma_k(s)}.
    \end{equation}
    Quite similar to Equation~\ref{eq:k-path}, only here $\sigma_{s}^{\kappa}(e)$ is the number of $\kappa$-paths originating from $s$ that go over the edge $e$. The original $\kappa$-path edge centrality is very expensive to compute in large networks with a big $k$, therefore two randomised approximations have been further proposed, i.e., ERW-$\kappa$path and WERW-$\kappa$path \cite{de2012novel}.
    \item[~]
    
    \item[--] \textbf{Local centrality} \cite{chen2012identifying}. Local centrality, sometimes summarised as LocalRank \cite{lu2016vital} utilises the information within a node's 4-hop neighbourhood. Concretely, the local centrality of node $i$ is defined as: 
    \begin{align}
        \Theta_{LR}(i)=\sum_{j \in N(i)} Q(j), && Q(j)=\sum_{k \in N(j)} R(k),
    \end{align}
    where $N(i)$ and $N(j)$ are the set of 1-hop neighbours of node $i$ and $j$, and $R(k)$ is the number of both 1-hop and 2-hop neighbours of node $k$. It is said to perform better than betweenness centrality and almost as well as closeness centrality to identify influential nodes under the setting of a SIR model, with only a time complexity of $O(n\langle k\rangle^{2})$.
    \item[~]
    
    \item[--] \textbf{Collective influence} \cite{morone2015influence}. Collective influence (CI) is another interesting method that takes higher-order neighbourhoods into consideration. The idea is to find those nodes that will cause the biggest drop in the \say{energy function} when removed. Specifically, the level $k$ collective influence of a node $i$ is defined as: 
    \begin{equation}
        \Theta_{{CI}_k}(i)=(d_{i}-1) \sum_{j \in N_k(i)}(d_{j}-1),
    \end{equation}
    where $N_k(i)$ is $k$-hop neighbours of a node $i$. After applying the collective influence score, the paper finds that a large number of previously neglected weakly connected nodes (nodes of lower degree) emerge among the optimal influencers \cite{morone2015influence}.
\end{itemize}

\subsubsection{Multi-subgraphs} Methods of this category involve the count of multiple different subgraphs. They can be at the node level, the link level or the network level.

\begin{itemize}[leftmargin=*]
    \item[--] \textbf{Graphlet degree} \cite{milenkovic2008uncovering}. As discussed in Section~\ref{sec:motif_vs_graphlet}, graphlets are nonisomorphic induced subgraphs around a node. Graphlet degree is a 73-dimensional vector formed by all different orbits in the subgraphs of size 2-5 nodes. The paper discovers that in PPI networks, nodes grouped together under this measure belong to the same protein complexes, perform the same biological functions and have the same tissue expressions. Some interesting extensions of graphlets include the dynamic graphlets for temporal networks\cite{hulovatyy2015exploring}, the directed graphlets for directed networks\cite{aparicio2016extending}, the coloured graphlets for heterogeneous networks\cite{gu2018homogeneous}, and the typed-edge graphlets for edge-labelled networks \cite{jia2022encoding}.
    \item[~]
    
    \item[--] \textbf{Subgraph centrality} \cite{estrada2005subgraph}. Subgraph centrality focuses on subgraphs captured by closed walks of different lengths around a given node. For example, when the walk length is $4$, three types of subgraphs are covered, which are $2$-cliques, $2$-paths, and $4$-cycles. The number of closed walks of length $k$ around node $i$ can be calculated from the $i$\textsuperscript{th} diagonal entry of the $k$\textsuperscript{th} power of the adjacency matrix. When the walk becomes unbounded, the subgraph centrality of node $i$ is calculated by:
    \begin{equation}
        \Theta_{S}(i)=\sum_{k=0}^{\infty} \frac{\mu_{k}(i)}{k !}, 
    \end{equation}
    where $\mu_{k}(i)=\left(\mathbf{A}^{k}\right)_{i i}$. 
    It is shown to be more discriminative than many popular centrality measures such as the degree centrality, the betweenness and the eigenvector centrality.
    \item[~]
    
    \item[--] \textbf{Local path index} \cite{lu2009similarity}.
    Extended from common neighbours, the local path index counts both the number of $2$-paths and $3$-paths between a pair of nodes. The approach is proposed for link prediction, and therefore focuses on non-connected node pairs. Concretely, the local path index of a node pair $i$ and $j$ is defined as:
    \begin{equation}
        \Theta_{LP}(i,j)=A_{ij}^{2}+\epsilon A_{ij}^{3},
    \end{equation}
    where $\epsilon$ is a weigh parameter for $3$-paths. The paper finds out that the local path index remarkably outperforms common neighbours and can reach a competitive accuracy to the Katz index where all paths are considered. Some other works compare 3-paths approaches against 2-paths approaches in link prediction and find out that $3$-path approaches perform better in PPI networks and food webs \cite{muscoloni2018local, kovacs2019network, zhou2021experimental}.
    \item[~]
    
    \item[--] \textbf{Potential theory/Quad motifs index}. The potential theory aims to predict links in directed networks. By counting the numbers of $4$ different directed 2-paths and $8$ different directed 3-paths around a pair of nodes, the paper finds out that a link has a higher probability of appearing if it could generate more bi-fan subgraphs \cite{zhang2013potential}. Very similar to the idea of potential theory, the quad motifs index proposes to count particularly three types of directed open-quadriad (3-paths) subgraphs: two of them are the bases for bi-parallel subgraphs and the other one is for bi-fan \cite{hu2019quad}. Specifically, the quad motifs index of a pair of nodes $i$ and $j$ is defined as: 
    \begin{equation}
        \Theta_{QM}(i, j)=\alpha \times s_{F}(i, j)+\frac{(1-\alpha)}{2}\left(s_{P1}(i, j) + s_{P2}(i, j)\right),
    \end{equation}
    where $s_F(i, j)$ is the contribution from the bi-fan base while $s_{P1}(i, j)$ and $s_{P2}(i, j)$ are the contributions from two bi-parallel bases. Together with the local path index, it is interesting to see that $3$-path subgraphs are of particular importance in link prediction. 
    \item[~]
    
    \item[--] \textbf{Triad significance profile/Subgraph ratio profile} \cite{milo2004superfamilies}. Extended from network motifs \cite{milo2002network}, the triad significance profile (TSP) is constructed from normalised $Z$ scores of $13$ different directed 3-node subgraphs.
    \begin{align}
        TSP = \{SP_1, SP_2, ..., SP_{13}\}, &&
        \mathrm{SP}_{i}= \nicefrac{Z_{i}}{\left(\Sigma Z_{i}^{2}\right)^{1 / 2}}.
    \end{align}
    $Z_i$ is in turn calculated from comparing with an ensemble of randomised networks with the same degree sequence, i.e., $Z_{i}=\left(N_{i}^{\mathrm{real}}-\bar{N}_{i}^{\mathrm{rand}}\right) / \operatorname{std}\left(N_{i}^{\text {rand }}\right)$. Subgraph ratio profile (SRP), on the other hand, is built from $6$ undirected $4$-node subgraphs ($G_3$ to $G_8$ in Figure~\ref{fig:graphlets}) :
    \begin{align}
        SRP = \{SRP_1, SRP_2, ..., SRP_{6}\}, && \mathrm{SRP}_{i}= \nicefrac{\Delta_{i}}{\left(\Sigma \Delta_{i}{ }^{2}\right)^{1 / 2}}.
    \end{align}
    Unlike TSP, SRP uses the abundance of each subgraph relative to random networks, i.e., $\Delta_{i}=\frac{N \mathrm{real}_{i}-\left\langle N \mathrm{rand}_{i}\right\rangle}{N \mathrm{real}_{i}+\left\langle N \operatorname{rand}_{i}\right\rangle+\varepsilon}$. Previously seemingly unrelated networks are found to belong to several superfamilies with very similar significance profiles. Notice also that these two approaches are defined on network-level, not on node or link-level as we have seen often.  
\end{itemize}

\subsection{Subgraph Formation Based Approaches}
\begin{figure}[h]
\centerline{\includegraphics[scale = 1]{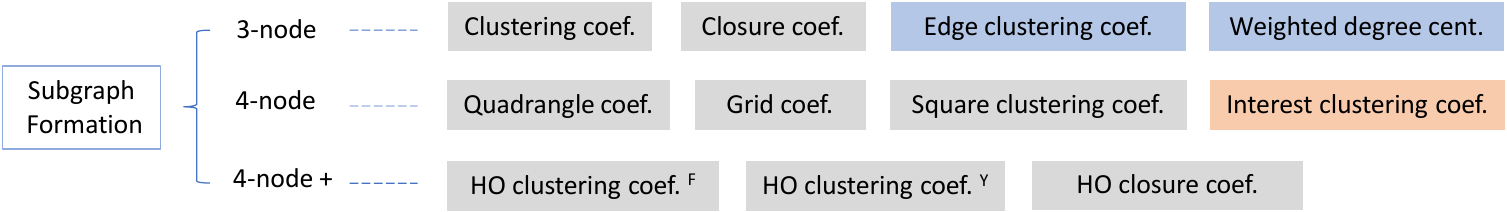}}
\caption{Subgraph formation based measures.}
\label{fig:SF}
\vspace{0mm}
\end{figure}

Subgraph formation based measures view a subgraph being built from another less complex subgraph, i.e., with one link, multiple links, or one node less. We further divide them 
into three categories according to the size of the subgraph, 3-node, 4-node and 4-node plus (Figure~\ref{fig:SF}). Most of these approaches are defined at node-level, except that the edge clustering coefficient is at link-level and the interest clustering coefficient is at network-level. 

\subsubsection{3-node subgraph} The 3-node subgraph is the simplest yet most important category in the taxonomy. 

\begin{itemize}[leftmargin=*]
    \item[--] \textbf{Clustering coefficient} \cite{watts1998collective}. The clustering coefficient is the first and most influential measure in this category. It measures the extent to which the neighbours of a node connect to each other. From a structural formation perspective, it measures the formation of triangles upon open-triads (also called wedges). Specifically, the clustering coefficient of node $i$ is defined as the ratio between the number of triangles containing node $i$ (denoted $T(i)$) and the number of open triads (denoted $OT(i)$): 
    \begin{equation} \label{eqn:clustering}
    \mathcal{C}_C(i) =\frac{T(i)}{O T(i)}=\frac{\frac{1}{2} \sum_{j \in N(i)}|N(i) \cap N(j)|}{\frac{1}{2} d_{i}\left(d_{i}-1\right)}.
    \end{equation}
    Due to its significance and simplicity in definition, the clustering coefficient has been widely applied in studying complex networks \cite{pavlopoulos2011using, brust2012clustering, said2018cc} and extended to directed networks \cite{fagiolo2007clustering, ahnert2008clustering}, weighted networks \cite{barrat2004architecture, onnela2005intensity, zhang2005general} and signed networks \cite{kunegis2009slashdot, costantini2014generalization}. 
    \item[~]
    
    \item[--] \textbf{Closure coefficient} \cite{yin2019local}.
    The closure coefficient measures the formation of triangles from a new perspective, i.e., with the focal node located at the end of an open-triad. Different from the ordinary centre node perspective in clustering coefficient (orbit 2 of $G_1$ in Figure~\ref{fig:graphlets}, denoted as $G_1^{(2)}$), the focal node in closure coefficient serves as the end node of an open triad (orbit type $G_1^{(1)}$). The closure coefficient of node $i$ is calculated as the fraction of open triads ($OT_E(i)$), where $i$ serves as the end node, that actually forms triangles:
    \begin{equation} \label{eqn_lcc}
    \mathcal{C}_E(i)=\frac{2 \cdot T(i)}{OT_E(i)}=\frac{\sum_{j \in N(i)}|N(i) \cap N(j)|}{\sum_{j \in N(i)}(d_j-1)}.
    \end{equation}
    Despite this subtle difference in definition, the closure coefficient has very different properties compared to the clustering coefficient. It has been extended to directed networks \cite{yin2020measuring,jia2020closure} and weighted networks \cite{jia2021directed}.  
    \item[~]

    \item[--] \textbf{Edge clustering coefficient} \cite{wang2011identification}. Defined on link-level, the edge clustering coefficient (ECC) evaluates to what extent nodes cluster around the focal edge. From a structure formation view, it measures the formation of triangles upon this link. The edge clustering coefficient of an edge $e_{ij}$ is defined as:
    \begin{equation}
        \mathcal{C}_e(i, j)=\frac{T(i,j)}{\min \left(d_{i}-1, d_{j}-1\right)},
    \end{equation}
    where $T(i, j)$ is the number of triangles that $e_{ij}$ participates, and $\min \left(d_{i}-1, d_{j}-1\right)$ is the number of triangles that edge could possibly form. Based on ECC, a node centrality measure is then defined as the sum of the edge clustering coefficients of all edges connected to it, i.e., $\mathcal{C}_N(i)=\sum_{j \in N_{i}} \mathcal{C}_e(i, j)$. This measure has been proven to be more efficient for identifying essential proteins than many other centrality measures. 
    \item[~]
    
    \item[--] \textbf{Weighted degree centrality} \cite{tang2013predicting}. Weighted degree centrality (WDC) is also proposed to identify essential proteins. Although this name seems to suggest a close relation to the degree centrality, it is in fact an extension of the edge clustering coefficient. This approach is different in that it takes into account not only the PPI graph data but also the gene expression data. Specifically, a weight of an interaction is calculated as:
    \begin{equation}
        \mathcal{C}_w(i, j) = \mathcal{C}_e(i, j) + r(i', j'),
    \end{equation}
    where $\mathcal{C}_e(i, j)$ is the edge clustering coefficient from the graph data, and $r(i', j')$ is the Pearson correlation coefficient calculated from the gene expression data. Similarly, the weighted degree centrality of a node is then defined as: $\Theta_W(i) = \sum_{j \in N_{i}}\mathcal{C}_w(i, j)$.
    This approach essentially integrates node features when analysing networks.    
\end{itemize}

\subsubsection{4-node subgraph} 4-node subgraphs are much more complicated than the 3-node subgraphs. There are in total 6 different subgraphs and 11 different orbits in 4-node subgraphs (Figure~\ref{fig:graphlets}). 

\begin{itemize}[leftmargin=*]
    \item[--] \textbf{Quadrangle coefficients} \cite{jia2021measuring}. Many real networks (such as PPI networks, neural networks and food webs) are naturally rich in quadrangles. Quadrangle coefficients, or i-quad coefficient and o-quad coefficient, are thus proposed to measure the formation of quadrangles upon open-quadriads (3-paths). As there are two orbits in an open-quadriad ($G_3^{(5)}$ and $G_3^{(4)}$), i-quad coefficient has the focal node at $G_3^{(5)}$ while o-quad coefficient has the focal node at $G_3^{(4)}$. Specifically, the quadrangle coefficients of node $i$ are defined as:
    \begin{align}
        \mathcal{C}_I(i) = \frac{2 Q(i)}{OQI(i)}, &&
        \mathcal{C}_O(i) = \frac{2 Q(i)}{OQO(i)},
    \end{align}
    where $Q(i)$ is the number of quadrangles; $OQI(i)$ and $OQI(i)$ are number of open-quadriads with $i$ as the inner node and outer node respectively. They are found to be more efficient than 3-node measures in classifying networks and predicting links.
    \item[~]
    
    \item[--] \textbf{Grid coefficients} \cite{caldarelli2004structure}. Grid coefficients, including the primary grid coefficient and the secondary grid coefficient, also aim to measure the formation of 4-cycles.  The primary grid coefficient measures the formation of \say{primary quadrilaterals} upon a node and three of its 1-hop neighbours, which is essentially the formation of chordal cycles ($G_7$) from tailed-triangles (orbit $G_6^{(11)}$). Concretely, the primary grid coefficient of node $i$ is defined as:
    \begin{equation}
        \mathcal{C}_{G_{p}}(i)= \frac{Q_{p}(i)}{d_{i} (d_{i} - 1)(d_i - 2) / 2}, 
    \end{equation}
    where $G_{p}(i)$ is the number of chordal-cycles containing $i$ and the denominator is the number of possible chordal-cycles built from a node and its three neighbours. The secondary coefficient measures the formation of \say{secondary quadrilaterals} from a node, two of its 1-hop neighbours and one of its 2-hop neighbours:
    \begin{equation}
        \mathcal{C}_{G_{s}}(i)= \frac{Q_{s}(i)}{d_{i,2nd} d_{i} (d_i - 1) / 2}.
    \end{equation}
    Notice, however, in this definition the 2-hop neighbour could be at orbit $G_3^{(4)}$ or at orbit $G_{10}^{(20)}$. The latter essentially involves 5 nodes in total. 
    
    \item[~]

    \item[--] \textbf{Square clustering coefficient}. As triangles (3-cycles) are absent in bipartite networks, the square clustering coefficient is proposed to measure the formation of 4-cycles in the context of bipartite networks \cite{lind2005cycles}. What is unusual about this approach is that it views 4-cycles being built from node overlapping instead of node connection. Specifically, the square coefficient of node $i$, with a pair of its neighbours $m$ and $n$, is calculated as:
    \begin{equation}
        \mathcal{C}_S(i | m, n) = \frac{Q_{imn}}{(d_m - \eta_{imn})(d_n - \eta_{imn}) + Q_{imn}},
    \end{equation}
    where $Q_{imn}$ is the number of 4-cycles containing nodes $i$, $m$, $n$; and $\eta_{imn} = 1 + q_{imn}$ if $m$ and $n$ are not connected (or $\eta_{imn} = 2 + q_{imn}$ if $m$ and $n$ are connected). Zhang et al. \cite{zhang2008clustering} later proposed a modified version of square clustering coefficient: $\mathcal{{C}}_{S_Z}(i | m, n) = \frac{Q_{imn}}{(d_m - \eta_{imn}) + (d_n - \eta_{imn}) + Q_{imn}}$. With this minor change at the denominator, 4-cycles are now built from connecting nodes. It is mainly applied in community detection.
    \item[~]
    
    \item[--] \textbf{Interest clustering coefficient} \cite{trolliet2020interest}. An interest clustering coefficient is introduced to measure the \say{clustering of interest links} in directed social networks. It argues that the best way of defining a relationship between two individuals is through common interests, i.e., two individuals having links towards a common neighbour will have a higher chance to follow other common neighbours. From a structural view, the interest clustering coefficient essentially measures the formation of bi-fan subgraphs (Table~\ref{tab:motifs}) upon open bi-fans: 
    \begin{equation}
        \mathcal{C}_I = \frac{4 \cdot \# \; \mbox{bifan}}{\# \; \mbox{open-bifan}}.
    \end{equation}
    Note that this metric is defined at network-level. The paper finds out that the interest clustering coefficient of Twitter is higher than the traditional directed clustering coefficient, and further proves its usage in a link recommendation task.
\end{itemize}

\begin{table}
\caption{Metrics for 3-node and 4-node subgraph formation.}
\label{tab:measures}
\begin{tabular}{|l|c|c|c|} 
\hline
\textbf{\begin{tabular}[c]{@{}l@{}}3-node/4-node \\ subgraph formation\end{tabular}} & \textbf{Undirected networks}            & \textbf{Directed networks}   & \textbf{Weighted networks}                                                     \\ \hline
\includegraphics[width=4cm, height=3cm, keepaspectratio]{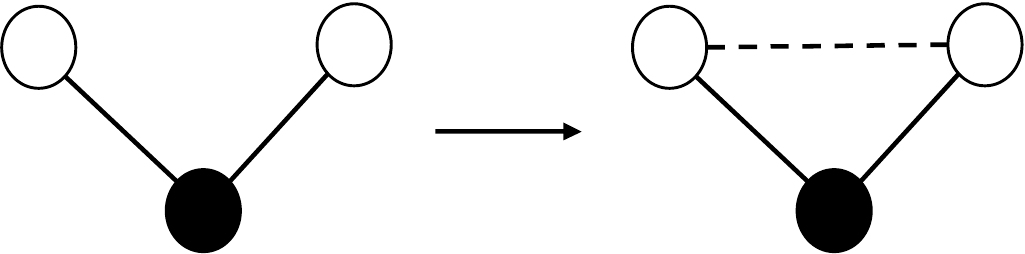}  & clustering coef.\cite{watts1998collective}   & directed clustering coef.\cite{fagiolo2007clustering, ahnert2008clustering}  & \begin{tabular}[c]{@{}l@{}}wgted. clustering coef.\cite{barrat2004architecture, onnela2005intensity, zhang2005general}\\ wgted. signed clustering coef.\cite{kunegis2009slashdot, costantini2014generalization}\\ wgted. directed clustering coef.\cite{fagiolo2007clustering}\end{tabular} \\ \hline
\includegraphics[width=4cm, height=3cm, keepaspectratio]{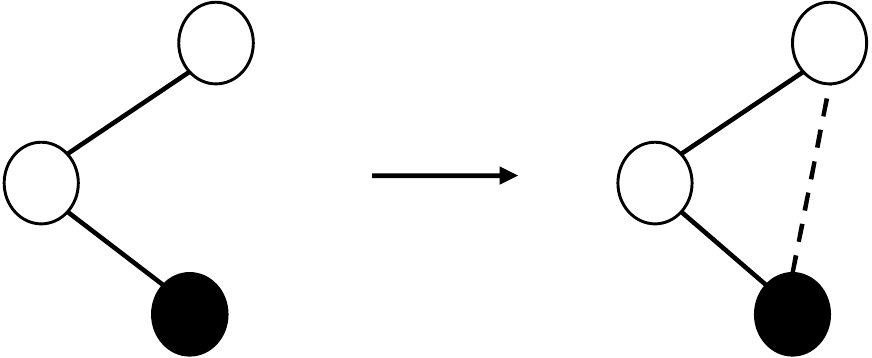}   & closure coef.\cite{yin2019local}  & directed closure coef. \cite{yin2020measuring,jia2020closure}  & weighted closure coef. \cite{jia2021directed} \\ \hline
\includegraphics[width=4cm, height=3cm, keepaspectratio]{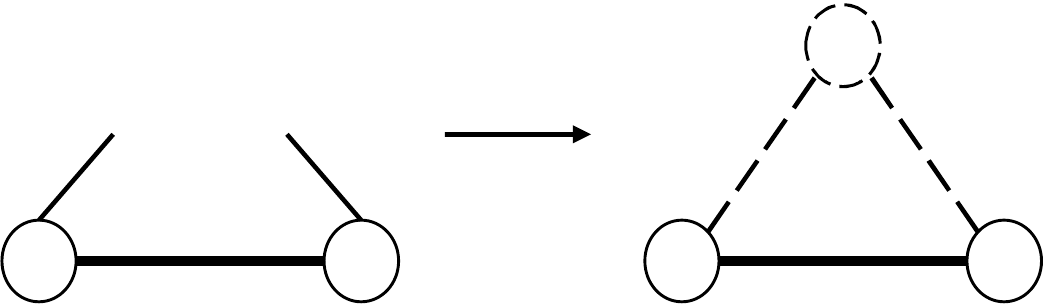} & edge clustering coef.\cite{wang2011identification}  & \notableentry  & \notableentry  \\ \hline
\includegraphics[width=4cm, height=3cm, keepaspectratio]{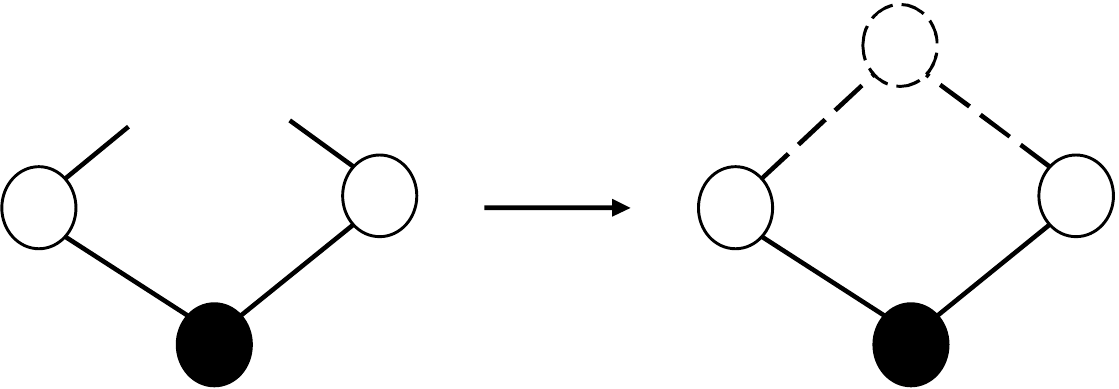} & \begin{tabular}[c]{@{}l@{}} higher-order clustering\\ coef. (Fronczak)\cite{fronczak2002higher}\\  higher-order clustering\\ coef. (Abdo)\cite{abdo2006clustering}\end{tabular} & None    & None  \\ \hline
\includegraphics[width=4cm, height=3cm, keepaspectratio]{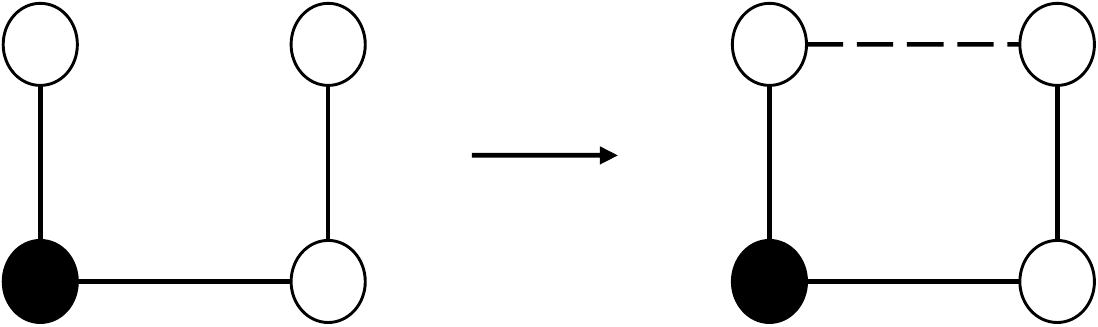} & \begin{tabular}[c]{@{}l@{}} square clustering coef. \\ (Lind \cite{lind2005cycles}, Zhang \cite{zhang2008clustering}) \\ i-quad coef. \cite{jia2021measuring} \\ primary grid coef. \cite{caldarelli2004structure}  \end{tabular} & None   & None   \\ \hline
\includegraphics[width=4cm, height=3cm, keepaspectratio]{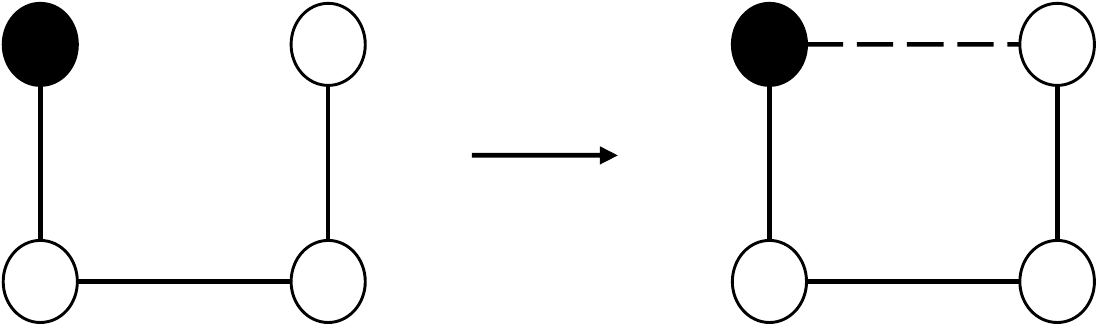}  & o-quad coef. \cite{jia2021measuring}  & None  & weighted o-quad coef. \cite{jia2021measuring}  \\ \hline
\includegraphics[width=4cm, height=3cm, keepaspectratio]{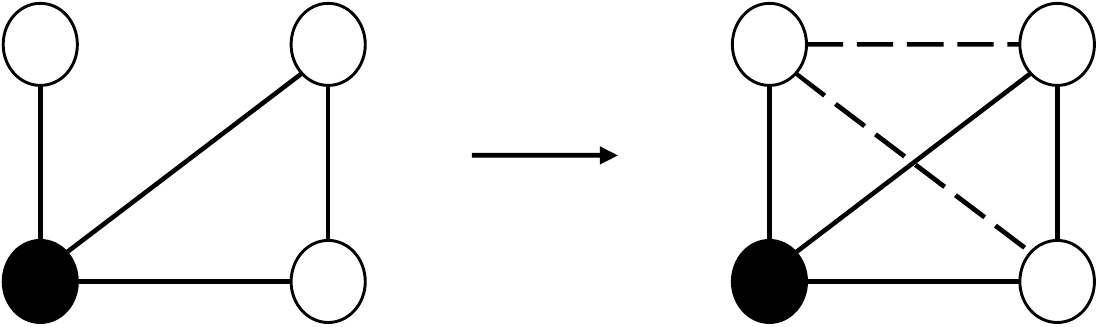}  & \begin{tabular}[c]{@{}l@{}}higher-order clustering\\  coef. (Yin)\cite{yin2018higher}\end{tabular}   & None  & None \\ \hline
\includegraphics[width=4cm, height=3cm, keepaspectratio]{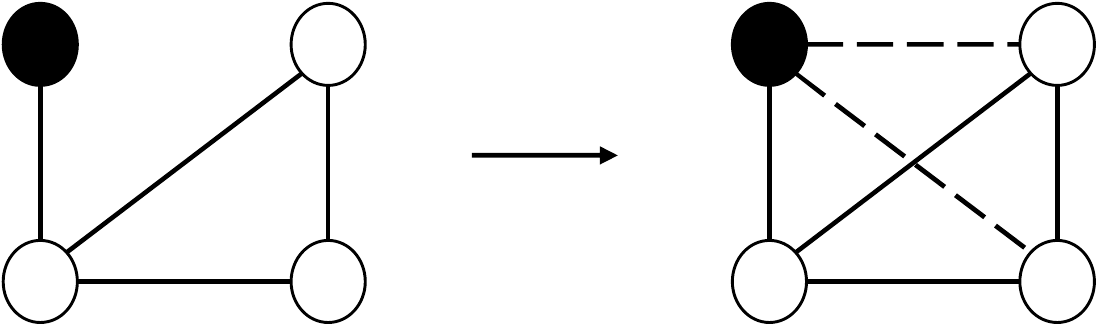}  & \begin{tabular}[c]{@{}l@{}}higher-order closure\\  coef. (Yin)\cite{yin2019local}\end{tabular}   & None  & None \\ \hline
\end{tabular}
\end{table}

\subsubsection{Beyond 4-node subgraph} Some approaches are introduced with a variable subgraph size. In actual usage, however, due to high complexity, they seldom go beyond the size of 6 nodes. 

\begin{itemize}[leftmargin=*]
    \item[--] \textbf{Higher-order clustering coefficients\textsuperscript{F}}  \cite{fronczak2002higher}. Fronczak et al. propose the higher clustering coefficients to evaluate the probabilities that the shortest paths between any two neighbours of node $i$ equals $k$, when all paths passing through node $i$ are neglected. Particularly, a clustering coefficient of order $k$ for node $i$ is calculated as:
    \begin{equation}
        \mathcal{C}_{H_F}(i \mid k) = \frac{2E(i \mid k)}{d_i(d_i - 1)},
    \end{equation}
    where $E(i \mid k)$ denotes the number of shortest paths of length $k$ between $i$'s neighbours. When $k$ equals 1, it degrades to the standard clustering coefficient, and when $k$ equals 2, it measures the formation of 4-cycles. Note that each pair of neighbours could have multiple shortest paths of the same length, and only one of them should be counted so that the value of higher-order clustering coefficients is bounded by 1.
    \item[~]
    
    \item[--] \textbf{Higher-order clustering coefficient\textsuperscript{Y}} \cite{yin2018higher}. The higher-order clustering coefficient proposed by Yin et al. is another generalisation of the traditional clustering coefficient. It aims to measure the formation of higher-order cliques. Specifically, a $k$\textsuperscript{th}-order clustering coefficient of node $i$ is defined as the probability that a $k$-clique plus an edge incident to $i$ (termed as $k$-wedge) forms a $(k+1)$-clique:
    \begin{equation}
        \mathcal{C}_{H_Y}(i \mid k) = \frac{k \cdot |C_{k+1}(i)|}{|W_k(i)|} = \frac{k \cdot |C_{k+1}(i)|}{(d_i - k + 1)|C_k(i)|}, 
    \end{equation}
    where $C_{k+1}(i)$ is the set of $(k+1)$-cliques containing node $i$, and $W_k(i)$ is the set of $k$-wedges with $i$ as the centre node. The properties of higher-order clustering coefficient in random graph and the small-world model have also been thoroughly investigated \cite{yin2018higher}. 
    \item[~]

    \item[--] \textbf{Higher-order closure coefficient} \cite{yin2019local}. Higher-order closure coefficient measures the formation of higher-order cliques from a different perspective, i.e., the focal node being the end-node of a $k$-wedge (instead of the centre-node). The $k$\textsuperscript{th}-order closure coefficient of node $i$ is thus defined as the fraction of end-node based $k$-wedges that are closed (a closed $k$-wedge is a $(k+1)$-clique):
    \begin{equation}
        \mathcal{C}_{H_E}(i \mid k) = \frac{k \cdot |C_{k+1}(i)|}{|W_k'(i)|} = \frac{k \cdot |C_{k+1}(i)|}{\sum_{j \in N(i)}\left[C_k(j)-(k-1) C_k(i)\right]}, 
    \end{equation}
    where $C_{k+1}(i)$ is the set of $(k+1)$-cliques containing node $i$, and $W_k(i)'$ is the set of $k$-wedges with $i$ as the end-node. Higher-order closure coefficient is proven to be useful in finding seeds for personalised PageRank community detection.
\end{itemize}
An illustrative summary for most abovementioned approaches is given in Table~\ref{tab:measures}.

\subsection{Global Path Based Approaches} \label{sec:glocal-path} 
\begin{figure}[h]
\centerline{\includegraphics[scale = 1]{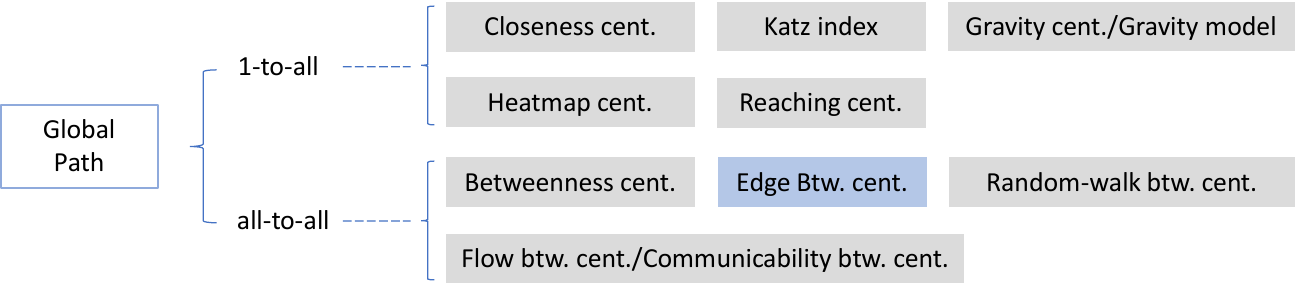}}
\caption{Global path based measures.}
\label{fig:GP}
\vspace{0mm}
\end{figure}

Global path based approaches require structural information across the whole network in the form of unbounded paths between nodes. One set of methods is based on the paths from one node to all other nodes, such as the well known closeness centrality and Katz index; another set of methods is based on paths between all node pairs, represented by the betweenness centrality (Figure~\ref{fig:GP}).

\subsubsection{One-to-all} The approaches of this type involve the paths from one node to all other nodes. They are also referred to as radial measures. 

\begin{itemize}[leftmargin=*]
    \item[--] \textbf{Closeness centrality} \cite{freeman1978centrality}. Being one of the most classic centrality measures, closeness centrality is defined as the reciprocal of the average shortest path distance from a focal node $i$ to all other nodes:
    \begin{equation}
        \Theta_C(i) = \frac{|V| - 1}{\sum_{j \in V, j \neq i}d(i, j)}.
    \end{equation}
    Obviously, the original definition is not suitable for graphs with more than one connected component. To address this problem, a modified version of the closeness centrality is defined as \cite{wasserman1994social}:
    \begin{equation}
        \Theta_{C'}(i) = \frac{n-1}{|V| - 1}\frac{n - 1}{\sum_{j = 1}^{n-1}d(i, j)},
    \end{equation}
    where $n$ is the number of nodes in one connected component.
    Due to its intuitiveness in definition, the closeness centrality keeps being applied and extended in various fields. Some recent works include the neighbourhood closeness centrality in predicting essential proteins \cite{li2018united}, and the backward/forward closeness in studying global value chains \cite{guan2020closeness}.   
    \item[~]
    
    \item[--] \textbf{Katz index} \cite{katz1953new}. Unlike the closeness centrality that focuses on shortest paths, Katz centrality of a node considers all paths reaching other nodes with longer paths contributing less. Concretely, the Katz centrality of a node $i$ is calculated as:
    \begin{equation}
        \Theta_K(i) = \sum_j \sum_{k=1}^{\infty}\beta^k\mathbf{A}_{ij}^k,
    \end{equation}
    where $k$ is a path length and $\beta$ is an attenuation parameter in a range $(0,\frac{1}{\lambda})$, $\lambda$ being the largest eigenvalue of $\mathbf{A}$. Further, the overall matrix $\mathbf{M} =\sum_{k=1}^{\infty} (\beta \cdot \mathbf{A})^{k}$ is an weighted ensemble of all paths.  Thus, $\mathbf{M}_{ij}$ represents the weighted sum of paths from $i$ to $j$ in all possible hops. Note that this definition is naturally suitable in directed networks and a recent work proposes to generate node embedding of a directed graph by performing a singular value decomposition on the Katz index matrix \cite{ou2016asymmetric}.   
    \item[~]
    
    \item[--] \textbf{Gravity model \cite{li2019identifying} /Gravity centrality} \cite{ma2016identifying} . 
    Inspired by Newton's gravity law formula, a gravity model is proposed by viewing the degree of a node as its mass and the shortest path length between two nodes as their distance:
    \begin{equation}
        \Theta_G(i) = \sum_{j \in V, j \neq i} \frac{d_i \cdot d_j}{d(i, j)^2}.
    \end{equation}
    In order to make it easier to compute in large networks, a modified version limits the radius from the entire network to a certain length. Adopting a similar idea, the gravity centrality is introduced by regarding the coreness of a node as its mass, and the shortest path length between nodes as their distance:
    \begin{equation}
        \Theta_G'(i) = \sum_{j \in N_k(i)} \frac{ks(i) \cdot ks(j)}{d(i, j)^2},
    \end{equation}
    where $N_k(i)$ is the neighbourhood of node $i$ within $k$-hops, and $ks(i)$ is the coreness of node $i$. The two approaches are shown to be effective in identifying influential spreaders through analyses of the SIR model on real networks. 
    
    \item[~] 
    
    \item[--] \textbf{Heatmap centrality} \cite{duron2020heatmap}. Heatmap centrality measures the influence of a node by comparing the farness of the node with the average farness of its neighbours. Farness, the reciprocal of closeness, is defined as the sum of the lengths of shortest paths from a node to all other nodes, i.e., $f(i) = \sum_{j \in V, j \neq i}d(i,j)$. Therefore, the heatmap centrality of node $i$ is quantified as:
    \begin{equation}
        \Theta_{HM}(i) = f(i) - \frac{\sum_{j \in N(i)}f(j)}{|N(i)|}.
    \end{equation}
    The intuition of this metric is that if a node has a smaller farness than its neighbours, the probability of information passing through it is higher. Note that according to heatmap centrality, a top-ranked node of influence should have the most negative value. Although the definition of heatmap centrality is more related to the closeness centrality, it is revealed that it is highly correlated with the betweenness centrality.
    \item[~]
    
    \item[--] \textbf{Reaching centrality} \cite{mones2012hierarchy}. Reaching centrality aims to rank the influence of a node in directed networks. Intuitively, the reaching centrality of node $i$ is quantified as the proportion of nodes that can be reached by the node via outgoing edges, i.e., the number of nodes with a directed distance from $i$, divided by $|V| - 1$. Further, a global reaching centrality is then defined as:
    \begin{equation}
        GRC = \frac{\sum_{i \in V} [\Theta_R^{max} - \Theta_R(i)]}{|V| - 1},
    \end{equation}
    where $\Theta_R^{max}$ is the largest reaching centrality of all nodes. The meaning of $GRC$ is the difference between the maximum reaching centrality and the average reaching centrality. Global reaching centrality is used as a hierarchy measure for directed networks and is shown to be capable of capturing the degree of hierarchy in both synthetic and real networks. 
\end{itemize}

\subsubsection{All-to-all} The approaches here involve the count of paths between all node pairs, and among them the ones that pass through a focal node or edge. They are also referred to as medial measures. 
\begin{itemize}[leftmargin=*]
    \item[--] \textbf{Betweenness centrality} \cite{freeman1977set}. Betweenness centrality, or more precisely, the shortest-path betweenness centrality is one of the best-known centrality measures. The betweenness centrality of node $i$ is quantified as the sum of the fraction of all-pairs shortest paths going through $i$:
    \begin{equation}
        \Theta_{B}(i)=\sum_{s, t \in V} \frac{\sigma(s, t \mid i)}{\sigma(s, t)},
    \end{equation}
    where $\sigma(s, t \mid i)$ is the number of shortest paths between node pair $s$ and $t$ that pass through node $i$, and $\sigma(s, t)$ is the number of all shortest paths between $s$ and $t$. It is often normalised by $\frac{(|V|-1)(|V|-2)}{2}$, in order to be compared in different networks. The betweenness centrality has also been generalised to directed networks\cite{white1994betweenness} and weighted networks \cite{opsahl2010node}.  
    \item[~]
    
    \item[--] \textbf{Edge betweenness centrality} \cite{girvan2002community}. With a small modification on the original betweenness centrality, Girvan and Newman propose an edge betweenness centrality in order to detect a community structure in complex networks. The edge betweenness centrality of an edge $e$ is quantified as the sum of the fraction of all-pairs shortest paths passing through $e$:
    \begin{equation}
        \Theta_{B}(e)=\sum_{s, t \in V} \frac{\sigma(s, t \mid e)}{\sigma(s, t)},
    \end{equation}
    According to the definition, edges which lie between communities will have large edge betweenness. Therefore, the underlying communities of the network would be uncovered by removing edges of high edge betweenness centrality. It has been widely applied in a community detection task, and some recent applications include the study of anti-vaccination sentiment on Facebook \cite{hoffman2019s} and the analysis of microbial diversity in  marine sediment \cite{hoshino2020global}.  
    \item[~]
    
    \item[--] \textbf{Flow betweenness centrality \cite{freeman1991centrality}/ Communicability betweenness centrality} \cite{estrada2009communicability}. A major limitation of the betweenness centrality is that it exclusively focuses on the shortest paths. In real situations, however, information often takes a more circuitous path randomly or intentionally \cite{stephenson1989rethinking}. The flow betweenness addresses this issue by considering all paths between nodes. Specifically, the flow betweenness centrality of a node $i$ is defined as:
    \begin{equation}
        \Theta_F(i) = \sum_{s,t \in V}\frac{\phi (s, t \mid i)}{\phi (s, t)},
    \end{equation}
    where $\phi (s, t \mid i)$ is the maximum flow between $s$ and $t$ that passes through $i$, and $\phi (s, t)$ is the total flow between $s$ and $t$. The maximum flow is in turn calculated by the minimum cut capacity \cite{ford2015flows}. Having established the notion of \say{capacity } on links, the flow betweenness centrality is naturally suitable for weighted networks. Instead of treating each path equally, the communicability betweenness centrality proposes to reduce the weight for longer paths:
    \begin{equation}
        \frac{2}{(n-1)(n-2)}\sum_{s,t \in V}\frac{\sum_{k=0}^{\infty}\frac{1}{k!}\mu^k(s,t \mid i)}{\sum_{k=0}^{\infty}\frac{1}{k!}\mu^k(s,t)},
    \end{equation}
    where $\mu^k(s, t \mid i)$ is the number of paths between $s$ and $t$ passing $i$ with length $k$, and $\mu^k(s, t)$ is the number of paths between $s$ and $t$ with a length $k$.
    \item[~] 
        
    \item[--] \textbf{Random-walk betweenness centrality} \cite{newman2005measure}. A random-walk betweenness centrality, also known as a current-flow betweenness centrality, is another popular variant of the betweenness centrality. It models information spreading in a network analogous to an electrical current flow in a circuit. Concretely, the current-flow betweenness centrality of node $i$ is defined as the amount of current flowing through $i$, averaged over all node pairs:
    \begin{equation}
        \Theta_{CF}(i) = \frac{\sum_{s,t \in V} I(s, t \mid i)}{(1/2)n(n - 1)},
    \end{equation}
    where $I(s, t \mid i)$ is the current flowing from $s$ to $t$ that passes $i$. The paper then proves that a message spreading along random walks is equivalent to the above definition.

\end{itemize}


\subsection{Message Passing Based Approaches} \label{sec:MP}
The above mentioned approaches depend solely on the topological information of a network, such as the number of particular subgraphs, the ratio between two subgraphs, the length of shortest paths, or the number of paths. Message passing based approaches further consider the information contained in each node. From a microscopic point of view, in one iteration, only local information is needed at each node. It is worth noticing that the popular graph convolutional network is also based on this idea, i.e, iteratively gathering information from nearby nodes. 
    
    \begin{figure}[h]
    \centerline{\includegraphics[scale = 1]{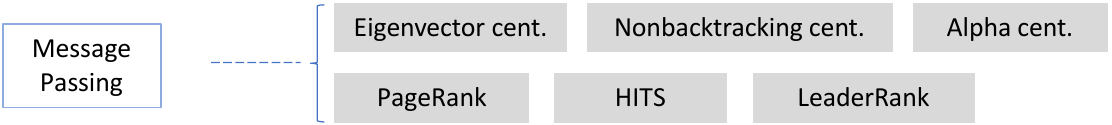}}
    \caption{Message passing based approaches.}
    \label{fig:MP}
    \vspace{0mm}
    \end{figure}
    
    \begin{itemize}[leftmargin=*]
        \item[--] \textbf{Eigenvector centrality} \cite{bonacich1987power}.
        The eigenvector centrality is another classic centrality measure. The idea is that a node's centrality depends on the centralities of its neighbours:
        \begin{equation}
            x(i)=c \sum_{j \in N(i)} x(j),
        \end{equation}
        where $c$ is a normalisation constant. The equation is recursive and computed by starting with a set of initial influence scores and iterating the computation until it converges. In a vectorised form, i.e., $\vec{x} = c\textbf{A}\vec{x}$,  $\vec{x}$ is found to converge to the dominant eigenvector of $\mathbf{A}$ and $c$ converges to the reciprocal of the dominant eigenvalue of $\mathbf{A}$. The eigenvector centrality has some problems in very sparse networks, i.e., the leading eigenvector is localised around nodes of the highest degree and diminishes the effectiveness of quantifying nodes' importance \cite{krivelevich2003largest}.
         \item[~]

        \item[--] \textbf{Nonbacktracking centrality} \cite{martin2014localization}. 
        The nonbacktracking centrality is proposed to address the above mentioned localisation issue. The same as in the eigenvector centrality, a node's centrality is the sum of its neighbours' centralities, but now the neighbours' centralities are calculated without the influence of this node. This is achieved by using the nonbacktracking matrix \cite{hashimoto1989zeta}. The nonbacktracking matrix $\mathbf{B}$, is a $2m \times 2m$ matrix, defined on the directed edges of the graph (undirected edges are converted to bidirectional edges), and elements $\mathbf{B}_{i \rightarrow j, k \rightarrow l}=\delta_{i, l}(1-\delta_{j k})$, where $\delta$ is the Kronecker delta. Then, $e_{j \rightarrow i}$ of the leading eigenvector of $\mathbf{B}$ gives the centrality of node $j$ ignoring the contribution of $i$. Finally, the nonbacktracking centrality of node $i$ is $x(i)= \sum_j \mathbf{A}_{ji} e_{j \rightarrow i}$. The eigenvalues of the nonbacktracking matrix are also found to be useful in a community detection task \cite{krzakala2013spectral}.
        \item[~]
        
        \item[--] \textbf{Alpha centrality}  \cite{bonacich2001eigenvector}. When the eigenvector centrality is applied in directed networks, a node's centrality is determined by those who pointed at it.  Thus, the vector form becomes: $\vec{x} = \frac{1}{\lambda}\textbf{A}^T\vec{x}$. The problem is that nodes with no incoming edges would have zero centrality value. The alpha centrality proposes to solve this problem by taking into account the "external status characteristics". The equation then becomes:  
        \begin{equation}
            \vec{x}=\alpha \textbf{A}^T \vec{x}+\vec{e},
        \end{equation}
         where $\vec{e}$ is a vector of the exogenous sources of characteristics and $\alpha$ is a parameter which reflects the relative importance of a topological structure versus exogenous factors. 
        \item[~]
        
        
        \item[--] \textbf{PageRank} \cite{brin1998anatomy}. PageRank, a popular variation of the eigenvector centrality, is proposed to rank the importance of web pages. Web pages and the links among them are modelled as a directed network, and a page should have a high rank if the sum of the ranks of pages that point to it is high. Specifically, the rank of page $i$ is calculated as: 
        \begin{equation}
            r(i) = c\sum_{j \in N_i^{in}} \frac{r(j)}{d_j^{out}},
        \end{equation}
        where $N_i^{in}$ is the set of pages pointing to $i$ ($i$'s in-neighbours), and $d_j^{out}$ is out-degree of page $j$. In order to deal with the \say{rank sink} problem, where several pages form a loop without other outgoing links, a source of the rank is introduced over all pages (also viewed as a random jumping factor), denoted as a vector $\vec{e}$. Therefore, the rank of page $i$ becomes: $r(i) = c(\sum_{j \in N_i^{in}} \frac{r(j)}{d_j^{out}} + e(i))$, and the corresponding vector form is $\vec{r} = c(\mathbf{A}^T + \vec{e} \times \mathbf{1})\vec{r}$. The PageRank has also been extended to weighted networks \cite{xing2004weighted}, on nonbacktracking matrix \cite{aleja2019non}, and applied to many different areas \cite{gleich2015pagerank}. 
        \item[~]
        
        \item[--] \textbf{HITS} \cite{kleinberg1998authoritative}. Unlike the PageRank which focuses on pages having many incoming links, HITS, abbreviated from a hyperlink induced topic search, proposes to distinguish two roles in the hyperlink structure, i.e., authorities and hubs. Authorities are reliable information sources, and hubs are the websites pointing to them.
        Based on the intuition that an authority should be pointed to by hubs and a hub should point to authorities, an authority weight and a hub weight of page $i$ are thus defined in a mutually dependent manner: 
        \begin{align}
            a(i) = \sum_{j \in N_i^{in}}h(j) && h(i) = \sum_{j \in N_i^{out}}a(j).
        \end{align}
        The corresponding vector forms are: $\vec{a} = \mathbf{A}^T\vec{h}$, and $\vec{h} = \mathbf{A}\vec{a}$. $\vec{a}$ and $\vec{h}$ are updated iteratively, and it is proven that $\vec{a}$ converges to the leading eigenvector of $\mathbf{A}^T\mathbf{A}$, and $\vec{h}$ converges to the leading eigenvector of $\mathbf{A}\mathbf{A}^T$. Based on HITS, ARC (Automatic Resource Compilation) later proposes to incorporate textual information around the link by assigning each link a weight \cite{chakrabarti1998automatic}, and Co-HITS proposes to extend the idea to bipartite networks \cite{deng2009generalized}. 
        \item[~]
        
        \item[--] \textbf{LeaderRank} \cite{lu2011leaders}. In order to solve the above mentioned rank sink problem, the LeaderRank proposes to add a ground node that connects to other nodes via bidirectional links. In the beginning, each node other than the ground node is initialised by one unit of score, and the ground node is initialised to zero. Then, the same as the PageRank, at each iteration, the score of node $i$ is calculated as: $s(i)^{(t)} = c\sum_{j \in N_i^{in}} \frac{s(j)^{(t-1)}}{d_j^{out}}$. After the scores of all nodes reach a steady state, the score of the ground node will be distributed evenly to other nodes, and thus the final score of node $i$ is:
        \begin{equation}
            s(i) = s(i)^c + \frac{s(g)^c}{|V|},
        \end{equation}
        where $s(i)^c$ is the steady score of node $i$, and $s(g)^c$ is the steady score of the ground node. A major advantage of the LeaderRank is that it has no additional parameter that needs to be optimised. Some interesting extensions of the LeaderRank include the weighted LeaderRank that assigns degree-dependent weights onto links associated with the ground node \cite{li2014identifying} and the adaptive LeaderRank that introduces H-index into the weighted mechanism \cite{xu2017identifying}.  
        
        \item[~]
        
    \end{itemize}

\subsection{Hybrid Approaches} 
The methods in the fifth and final category are combinations of previously introduced approaches. 

\begin{figure}[h]
\centerline{\includegraphics[scale = 1]{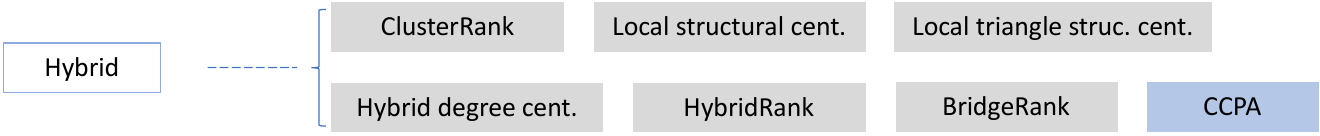}}
\caption{Hybrid Approaches.}
\label{fig:Hybrid}
\vspace{0mm}
\end{figure}

\begin{itemize}[leftmargin=*]
        \item[--] \textbf{ClusterRank} \cite{chen2013identifying}. Previous studies have shown that a large clustering coefficient may slow the spreading process of disease in the entire network \cite{eguiluz2002epidemic, zhou2005maximal}. 
        A ClusterRank thus proposes to consider not only the number of a node's neighbours, but also the negative effect of local clustering when identifying influential nodes. The ClusterRank score of node $i$ is defined as:
        \begin{equation}
            \Theta_{CR}(i) = f(c_i)\sum_{j \in N_i^{out}}(d_j^{out} + 1),
        \end{equation}
        where $c_i = \frac{\sum_{j \in N_i^{out}}|N^{out}(i) \cap N(j)|}{d_i^{out}(d_i^{out} -1)}$ is a modified version of clustering coefficient in directed networks. $f(c_i)$ is a function that is negatively correlated with $c_i$, for example an exponential function $f(c_i) = 10 ^{-c_i}$. Although the ClusterRank is proposed for directed networks, it can be easily extended to undirected networks \cite{chen2013identifying} and weighted networks\cite{wang2017identifying}. Experiments on several real networks demonstrate that the ClusterRank score outperforms the PageRank and the LeaderRank while being more efficient in computation.
        \item[~]
        
        \item[--] \textbf{Local structural Centrality} \cite{gao2014ranking}. Aiming to evaluate the spreading ability of nodes, a local structural centrality essentially extends the local centrality (section~\ref{sec: local paths}) by further considering the connections between higher-order neighbours. The idea is that a node has a better spreading ability when its neighbours are better connected because a neighbour node can be affected directly by the source node or indirectly by another neighbour node. The local structural centrality of node $i$ is defined as:
        \begin{equation}
            \Theta_{LS}(i) = \sum_{j \in N_i}(\alpha |N_j^{1,2}| + (1 - \alpha)\sum_{k \in N_j^{1,2}}c(k)),
        \end{equation}
        where $N_j^{1,2}$ is the node set of 1-hop and 2-hop neighbours of node $j$, and $c(k)$ is the clustering coefficient of node $k$. $\alpha$ is a tunable parameter between $0$ and $1$, balancing a direct and indirect spreading contribution. Notice that the part of the clustering coefficient is multiplied in the ClusterRank when evaluating spreading speed, but added up here  when measuring the spreading ability.   
        \item[~]
        
        \item[--] \textbf{Local triangle structure centrality} \cite{ma2019local}. A local triangle structure centrality (LTSC) proposes to include the triangle proportion of a node, instead of its clustering coefficient when evaluating a node's spreading ability. The triangle proportion is able to indicate the location of a node, whether it is located in a denser or sparser part of a network. LTSC partitions the spreading ability into two parts, i.e., inner spreading ability and outer spreading ability.  Specifically, the local triangle structural centrality of node $i$ is defined as:
        \begin{equation}
            \Theta_{TS}(i) = \sum_{j \in N_i}(d_j(1 + TP(j)) + (\sum_{k \in N_j}d_k - d_j)),
        \end{equation}
        where $TP(j)$ is the triangle proportion of node $j$, calculated by the number of triangles containing $j$ divided by the total number of triangles in the network. For each neighbour $j$ of a given node $i$, the part of $d_j(1 + TP(j)$ is to measure its inner spreading ability, and the part of $\sum_{k \in N_j}d_k - d_j$ is to measure its outer spreading ability. Finally, the local triangle structure centrality of node $i$ is the sum of the spreading abilities of its neighbours.  
        \item[~]
        
        \item[--] \textbf{Hybrid degree centrality} \cite{ma2017identifying}. The spreading probabilities of networks describing diseases, opinions, and rumours should obviously differ. Most existing centrality measures, however, fail to take that into consideration. The performance of centrality measures is sensitive to the spreading probability. The degree centrality, for example, works best with modest spreading probabilities, while the local centrality (section~\ref{sec: local paths}) works better with higher ones \cite{gao2014ranking}. In order to alleviate the sensitivity to different spreading probabilities, a hybrid degree centrality is introduced by integrating the degree centrality and a modified local centrality. The hybrid degree centrality of node $i$ is defined as:
        \begin{equation}
            \Theta_{HD}(i) = (\beta - p) \cdot \alpha \cdot \Theta_D(i) + p \cdot \Theta_{LR}'(i),
        \end{equation}
        where $\Theta_{LR}'(i) = \Theta_{LR}(i) - 2\sum_{j \in N_i}|N_j|$ is the modified local centrality, $p$ is the spreading probability, $\alpha$ and $\beta$ are two tunable parameters. The part contributed by the degree centrality is viewed as a near-source influence, and the part of modified local centrality is a distant influence. 
        \item[~]
        
        \item[--] \textbf{HybridRank} \cite{ahajjam2018identification}. A HybridRank proposes to identify influential spreaders by combining the neighbourhood coreness centrality (section~\ref{sec:1-hop}) and the eigenvector centrality. The reason for integrating these two measures is that they both regard a node as influential if the node is connected to other influential nodes. The hybrid centrality of node $i$ is defined as:
        \begin{equation}
            \Theta_{HR}(i) = \Theta_{NC}(i) \times \Theta_E(i),
        \end{equation}
        where $\Theta_{NC}(i) = \sum_{j \in N_i}ks(j)$ is the neighbourhood coreness of $i$, and $\Theta_E(i)$ is the eigenvector centrality of node $i$. The HybridRank algorithm further suggests that when selecting influential spreaders, the neighbours of selected ones should be neglected in order to maximise the spreading range. The effectiveness of the HybridRank has also been tested in real networks using a SIR model.    
        \item[~]
        
        \item[--] \textbf{BridgeRank} \cite{salavati2018bridgerank}. In order to lower the time complexity of the closeness centrality while keeping comparable performance, a BridgeRank proposes to compute the shortest paths to just a few core nodes in the network. In the BridgeRank algorithm, at first, communities are identified by the Louvain algorithm \cite{blondel2008fast}. Then, core nodes are discovered through calculating the betweenness centralities within each community (one core node per community). Finally, the BridgeRank centrality of each node is defined as a filtered closeness centrality to these core nodes:
        \begin{equation}
            \Theta_{BR}(i) = \frac{1}{\sum_{j \in \mathcal{C}}d(i, j)},
        \end{equation}
        where $\mathcal{C}$ is the set of identified core nodes in each community. The time complexity is therefore reduced from $O(|V|^3)$ to $O(|V|log|V|)$. A modified version that allows multiple core nodes being selected in a community is also introduced \cite{salavati2018bridgerank}. Other community structure based methods include $k$-medoid that uses information transfer probabilities between any node pairs \cite{zhang2013identifying}, and the influence maximization algorithm based on label propagation \cite{zhao2016identification}.  
        \item[~] 
        
        \item[--] \textbf{CCPA} \cite{ahmad2020missing}.
        A common neighbour and centrality based parameterised algorithm, or CCPA, is an approach for a link prediction. It aims to bring together two essential properties of nodes, i.e., the common neighbours and the closeness centrality. The similarity score between a pair of nodes $i$ and $j$ is defined as:
        \begin{equation}
            s(i, j) = \alpha \cdot (|N_i \cap N_j|) + (1 - \alpha) \cdot \frac{|V|}{d(i, j)}.
        \end{equation}
        $|N_i \cap N_j|$ is obviously the part of common neighbours. $\frac{|V|}{d(i, j)}$, reciprocal of the normalised distance between two nodes, is deemed as the closeness centrality of them, since it has a similar form as the classic node closeness centrality. $\alpha \in [0, 1]$ is a user-defined parameter controlling the weight of the two parts. Experiments on real-world datasets suggest that the change in performance (measured in average AUC) caused by the change of $\alpha$ is not significant.   
\end{itemize}

\subsection{Discussion and Outlook}
To end this section, we further discuss graph structural measures in different types of networks and highlight some research avenues for future studies. We then briefly talk about the importance and role of reviewing traditional structural measures in the following part of the survey on GCNs. 

\textit{Dynamic Networks.} Most approaches covered in the survey assume that networks are static or time-independent. Many real-world networks, however, are in fact dynamic, nodes and edges appearing and disappearing over time \cite{holme2012temporal, liao2017ranking}. In telecommunication networks, the connection between agents is often bursty and fluctuates across time; in social networks, relationships among people are typically intermittent and recurrent; in transportation networks, the frequency of public transport service is usually higher in rush hours. This extra dimension of time adds richness and complexity to the graph representation of a system, necessitating the development of more advanced approaches that can leverage temporal information. Many studies have generalised the classic graph structural measures to dynamic networks, including temporal degree centrality\cite{kim2012centrality}, temporal clustering coefficient \cite{nicosia2013graph}, temporal closeness and betweenness centrality \cite{kim2012temporal}, temporal eigenvector centrality \cite{taylor2017eigenvector}, temporal Katz centrality \cite{nicosia2013graph},  temporal motifs \cite{kovanen2011temporal, paranjape2017motifs} and temporal graphlets \cite{hulovatyy2015exploring}. Despite the large number of structural measures proposed for dynamic networks, there are still many open questions to be tackled. For example, what is the impact of the temporal network's structure on the dynamics of processes that occur on it; how to apply temporal measures in inferring spreading chains in incomplete temporal networks, etc.  

\textit{Multilayer Networks.} Sometimes, systems are so complicated that multiple-layered networks are needed to better represent and study them \cite{de2013mathematical, kivela2014multilayer, boccaletti2014structure, bianconi2018multilayer}. For example, a multilayer social network incorporates both friendship and financial relationships among individuals; a multilayer brain network contains both the anatomical brain layer and functional brain layer, and a multilayer transportation network integrates all sorts of transportation. Since interlayer connections cause new structural and dynamic correlations between components, neglecting them or simply aggregating over layers will alter the original topological properties. Therefore, it is desirable to develop structural measures taking interlayer relationships into consideration. Not surprisingly, fundamental single-layer approaches have been largely generalised to multilayer networks, such as multilayer degree, clustering coefficient, closeness and betweenness centralities, \cite{donges2011investigating, de2013mathematical, boccaletti2014structure}, multilayer motifs and graphlets \cite{battiston2017multilayer, dimitrova2020graphlets}, multilayer eigenvector, PageRank and HITS centralities \cite{de2015ranking, halu2013multiplex, de2013centrality}. Some tailor-made approaches for multilayer networks are also recently introduced, for example, the minimal-layers power community index \cite{basaras2017identifying},  and the singular vector of tensor centrality \cite{wang2018new}. The study of multilayer structures, however, is still in an early stage. There is still much room for developing new cross-layer structural approaches that better model inter-layer spreading processes \cite{salehi2015spreading} and captures multiplex dynamics, and controllability \cite{jiang2021controllability}.

\textit{Node/edge attributes.} Network data, besides the pure topological presence, are often accompanied by rich information on node attributes and/or edge attributes, and they are also referred to as labelled networks or attributed networks. Most structural measures, as the name suggests, focus solely on capturing the part of topological properties. Theoretically, message passing approaches are able to include numeric node attributes, such as the initial rank and source of rank in the PageRank \cite{brin1998anatomy}, or the endogenous and exogenous status in the alpha centrality \cite{bonacich2001eigenvector}. In practice though, these features are usually set to identical values for all nodes, for example, all ones for the initial rank and $0.15$ for the source of rank in the PageRank. Multidimensional features are not supported in message passing approaches either. There have also been attempts to integrate node/edge attributes with other graph structural measures. For instance, the degree and betweenness centralities are combined with node attributes in studying criminal networks \cite{bright2015use}; nodes' attributes are used as a threshold in LRIC index \cite{aleskerov2016centrality}; and node/edge attributes are fused into graphlets \cite{rossi2020heterogeneous, jia2021analysing}. We believe there is still great potential for developing novel structural approaches that integrate rich information on nodes and/or edges. It is also worth mentioning  that one reason for the popularity of graph neural networks is that it naturally enables integrating node attributes and some recent works also propose to take edge attributes into account in GNNs \cite{gong2019exploiting, jiang2020co, chen2021edge}.

Finally, we discuss how the traditional structure-based approaches are linked to GCNs. The importance and role of reviewing traditional structural measures in the survey of GCNs are Multifaceted. First, traditional structural approaches, the outcome of decades of Network Science studies, are the precursors and foundations of graph neural networks. For example, the key idea of neighbourhood aggregation and message passing in GCNs can trace back to 1972 when Bonacich proposed the eigenvector centrality \cite{bonacich1972factoring}. Basic network science notions such as the clustering coefficient, motifs and graphlets are utilised in GCNs as well. Second, the taxonomy of traditional approaches from the perspective of structure information inspired us to develop a new taxonomy for GCNs. We will see later how the taxonomy of GCNs from a layer-wise message aggregation scope is similar to that of subgraph count based measures in Section~\ref{sec:subgraph-count}. Third and last, a comprehensive review of traditional structural measures not only helps in revealing their connections to GCN approaches but also benefits the discovery of knowledge gaps. We will see that some GCN approaches are inspired by the traditional message passing based approaches, and that many subgraph count based approaches find their usages in GCNs. However, the connections between GCNs and subgraph formation based or global path based approaches are still largely left undiscovered. 

%% file: 4.local_structure_in_GNN.tex
\section{Structure information in Graph Convolutional Networks} \label{sec: GNN}
After summarising the traditional Network Science structural measures, we are set to review the graph convolutional networks from a novel perspective of graph structural information.

In recent years, graph neural networks, especially graph convolutional networks, have become one of the most prominent research areas in the study of complex networks. It extends the traditional convolutional neural networks to graph data and enables an effective combination of the rich node features information and graph topological structure. Graph convolutional networks have been successfully applied in different types of graph learning tasks, including node classification, link prediction, graph classification and graph clustering. Amongst the large family of graph deep learning approaches \cite{zhang2020deep, ma2021deep}, we particularly focus on graph convolutional networks not only because they have a wider range of applicability, but also because they are the bases of many other graph deep learning approaches, including graph autoencoders, graph reinforcement learning, graph adversarial methods, etc.

There exist several comprehensive surveys on graph neural networks. Bronstein et al. \cite{bronstein2017geometric} provide a thorough review of geometric deep learning, which presents its problems, difficulties, solutions and applications. Hamilton et al. \cite{hamilton2017representation} develop a unified encoder-decoder framework for graph representation learning approaches, bringing together matrix factorisation-based methods, random-walk-based algorithms and graph neural networks. Chami et al. \cite{chami2020machine} later extend the framework by including more recent advancements in the area. Zhang et al. \cite{zhang2019graph} propose a comprehensive review specifically on graph convolutional networks. Zhou et al. \cite{zhou2020graph} introduce a detailed taxonomy after dividing GNNs into several modules, including the propagation module, the sampling module and the pooling module. Wu et al. \cite{wu2020comprehensive} propose to divide GNNs into four categories, i.e., recurrent GNNs, convolutional GNNs, graph autoencoders and spatial-temporal GNNs.

These reviews, when introducing convolutional neural networks, usually focus on the domain of convolutional operations and propose a dichotomy, i.e., the spectral-based methods and the spatial-based methods. However, the line between the two is sometimes blurred. For example, GCN is an approximation of spectral graph convolutions, but it operates directly on graphs --- applying filters acting on the k-hop neighbourhood of the graph in the spatial domain \cite{bronstein2017geometric}. Another recent work also proves that spectral convolutional graph neural networks can be viewed as a particular case of spatial convolutional neural networks \cite{muhammet2020spectral}.

Different from existing reviews, in this survey we primarily, but not exclusively, focus on how local structure plays its role in graph convolutional networks. we propose to categorise GCN approaches from three different perspectives, which are the layer-wise message aggregation scope, the message content, and the overall learning scope. 

\begin{itemize}
    \item Layer-wise message aggregation scope. Analogous to convolutional neural networks, multilayer architecture is one of the key features in graph convolutional networks. Taking the vanilla GCN for example, at each layer, a node gathers information from its 1-hop neighbours. Then from stacking $k$ layers, the node would convolve its $k^{th}$-order neighbourhood. Thereafter, many other approaches propose to apply different scope at each layer, including 2-hop neighbourhood, k-hop neighbourhood, local-random-walk neighbourhood, subgraph neighbourhood, etc. This first structural perspective in GCN design can be summarised into the following question: From where a node aggregates message at each layer? The detailed taxonomy of GCNs from the perspective of layer-wise message aggregation scope and related approaches are given in Subsection 4.1.
    
    \item Message content. Compared to traditional deep learning models such as CNNs and RNNs, the strength of GCNs comes from the ingenious integration of graph structure and node features --- node features are passed through the edges of the graph. In many cases, the feature of nodes is independent of graph structure, such as numerical ratings, word vectors generated from sentences, positional gene sets, immunological signatures, and more. Meanwhile, there are emerging works that include other structural information as part of node features, from the simplest node degree to more complicated distance or subgraph information \cite{hamilton2017inductive,li2020distance, bouritsas2020improving}. This second structural perspective in GCN design can be summarised into the question: What structural information is included in the node feature when initialising or running the message passing scheme? The detailed taxonomy from the message content perspective and the related approaches are given in Subsection 4.2.   
    
    \item Learning scope. Traditional graph representation learning approaches are generally based on matrix factorisation, which thus requires the fixed whole graph. Although the original GCN approach also takes the whole graph's adjacency matrix as input, it has soon been extended to various settings, such as subgraphs, localised subgraphs, and more. To put in a question format, the third structural perspective in a GNN design is: Where GCNs are trained on? or What is/are the input graph/graphs in GCNs? The detailed taxonomy of GCNs from the learning scope perspective and the related approaches are given in Subsection 4.3.     
\end{itemize}

\subsection{Layer-wise message scope}
\begin{figure}[t]
\centerline{\includegraphics[scale = 0.9]{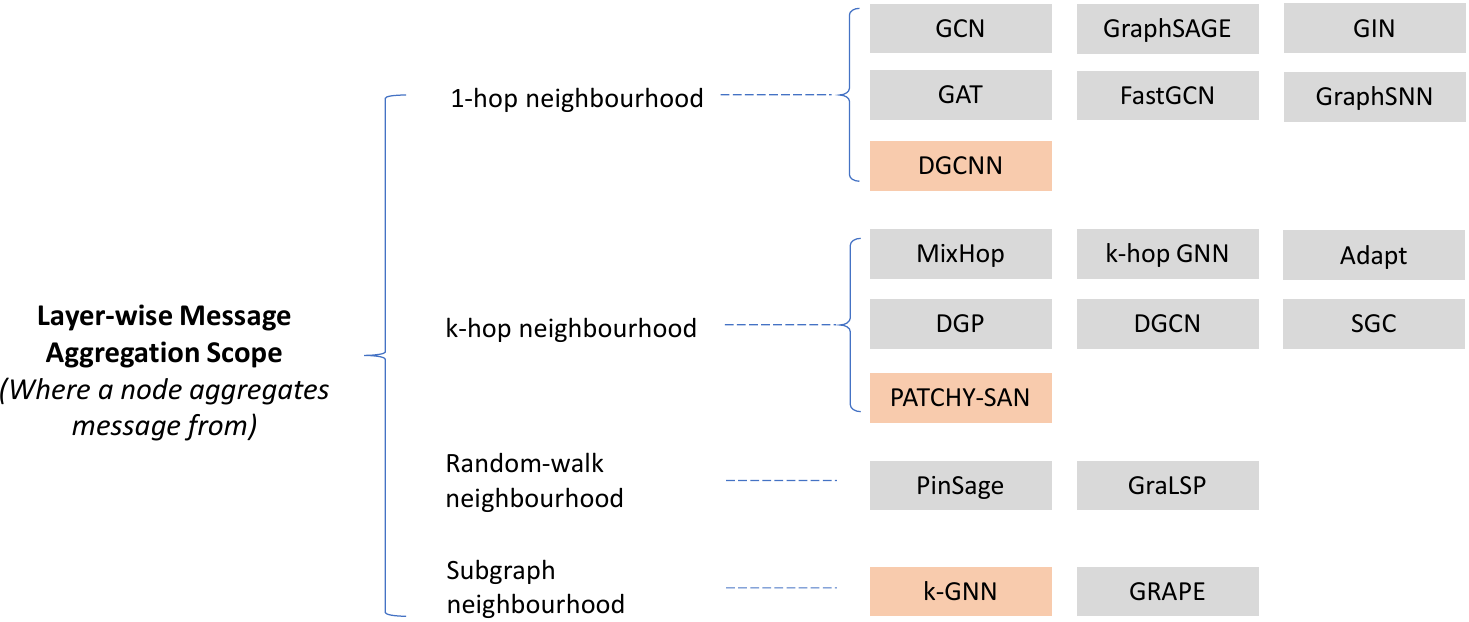}}
\caption{Taxonomy from the Layer-wise Message Aggregation Scope perspective.}
\label{fig:layer-wise-taxonomy}
\vspace{-0mm}
\end{figure}

To begin with, we discuss in detail the first structural perspective in a GCN design, i.e., a layer-wise message scope. By answering the question of where a node aggregates message from at each layer, we divide existing GCN approaches into four categories, which are 1-hop neighbourhood approaches, k-hop neighbourhood approaches, local-random-walk neighbourhood approaches, and subgraph neighbourhood approaches. The taxonomy and representative approaches are given in Figure~\ref{fig:layer-wise-taxonomy}. The colour of the block indicates what task the approach is proposed for: grey is the most common node classification task, orange is a network classification, and blue is the link prediction task which will appear later in Section~\ref{sec:msg-concent}. Notice that a graph representation can be readily obtained via graph pooling, so approaches proposed for a node classification can potentially be applied in a network classification task. Likewise, some approaches proposed for the network classification also generate node representations, making them possible to be used in the node classification.    

\subsubsection{1-hop neighbourhood approaches} \label{sec:1-hop-method}
Many influential GCN approaches adopt the 1-hop neighbourhood aggregation strategy, where a node's representation is iteratively updated through aggregating representations of its neighbours. One iteration happens at one convolutional layer, and after stacking multiple layers, the node's representation is able to capture a wider range of neighbourhoods. 

\textit{GCN}. Motivated by the first-order approximation of localised spectral filters on a graph \cite{defferrard2016convolutional}, GCN proposes the following layer-wise propagation rule operating directly on graphs:  
\begin{equation} \label{eqn:gcn_full}
H^{(l)}=\sigma\left(\hat{A} H^{(l-1)} W^{(l)}\right)
\end{equation},
where $\hat{A} = \tilde{D}^{-\frac{1}{2}} \tilde{A} \tilde{D}^{-\frac{1}{2}}$, $\tilde{A}$ is adjacency matrix with added self-connections, and $\tilde{D}$ is degree matrix of $\tilde{A}$. $H^{(l)}$ is the representations at the $l^{th}$ layer, and $W^{(l)}$ is the learnable weight matrix at $l^{th}$ layer. $\sigma$ denotes a nonlinear activation function such as ReLU. The multiplication of the normalised self-connection added adjacency matrix $\hat{A}$ and the nodes' representation matrix $H$ represents a normalised sum of neighbouring nodes' (and self node's) representation. From a microscopic point of view, the representation of node $v$ at layer $l$ is calculated as:
\begin{equation} \label{eqn: gcn}
h_{v}^{(l)}=\sigma\left(\sum_{u \in \mathcal{N}(v)} \frac{1}{c_{v u}} h_{u}^{(l-1)} W^{(l)}\right),
\end{equation}
where $\mathcal{N}(v)$ is the set of node $v$'s one-hop neighbours (with added self-loops to each node), $c_{v u}=\sqrt{|\mathcal{N}(v)|} \sqrt{|\mathcal{N}(u)|}$ is the normalization constant based on the node degree. The loss is then computed as: $\mathcal{L}=-\sum_{l \in \mathcal{Y}_L} \sum_{f=1}^F Y_{l} \ln Z_{l f}$, where $\mathcal{Y}_L$ is the set of labelled nodes, $Z$ is the output embedding and $F$ are the feature maps. Successfully bringing convolutional operations on graphs, GCN has become one of the most popular graph representation learning approaches. It is worth mentioning that the Iterative Classification Algorithm (ICA) also uses neighbourhood information to train a model \cite{bhagat2011node, neville2000iterative}. The model is then used to iteratively update the labels of nodes in the test set. Obviously, without a multi-layer convolutional network, the scope of ICA in the training stage is strictly limited within the immediate neighbourhood.

\textit{GraphSAGE}. Hamilton et al. later proposed the GraphSAGE framework, which extends the GCN to a more general setting that supports a mini-batch approach and different aggregation functions \cite{hamilton2017inductive}. Specifically, the representation of node $v$ at layer $l$ is given by:
\begin{equation} \label{eqn:sage}
\begin{aligned}
&\mathbf{h}_{\mathcal{N}(v)}^{l} \leftarrow \text { AGGREGATE }_{l}\left(\left\{\mathbf{h}_{u}^{l-1}, \forall u \in \mathcal{N}(v)\right\}\right), 
&\mathbf{h}_{v}^{l} \leftarrow \sigma\left(\mathbf{W}^{l} \cdot \operatorname{CONCAT}\left(\mathbf{h}_{v}^{l-1}, \mathbf{h}_{\mathcal{N}(v)}^{l}\right)\right)
\end{aligned}
\end{equation}

The framework thus gives us the flexibility to choose different aggregator functions, such as mean aggregator (equivalent to GCN), LSTM aggregator and pooling aggregator. Further, unlike GCN which requires full batch gradient descent, GraphSAGE enables mini-batch setting and therefore can also be applied to unseen nodes (also known as inductive learning). In addition, GraphSAGE proposes to sample a fixed-size of neighbours around each node in their aggregation scheme, instead of using all neighbours, which helps to keep the computational cost of each batch fixed.   

\textit{GIN}. 
Although GCN and GraphSAGE have achieved excellent performances in graph learning tasks, especially in node classification tasks, they are unable to distinguish some simple graph structures due to their limits in the neighbourhood aggregation scheme. Graph Isomorphism Network architecture (GIN) \cite{xu2018powerful} is proposed to overcome this shortcoming and is proven to be as powerful as the Weisfeiler-Lehman graph isomorphism test \cite{weisfeiler1968reduction}. Specifically, in order to achieve the same discriminative power as the Weisfeiler-Lehman test, the representation of node $v$ at layer $l$ should be as:
\begin{equation}
h_{v}^{(l)}=\phi^{(l)}\left(h_{v}^{(l-1)},  f^{(l-1)}\left(\left\{h_{u}^{(l-1)}: u \in \mathcal{N}(v)\right\}\right)\right),
\end{equation}
where $f^{(l-1)}$ is a function operating on multisets and $\phi^{(l)}$ is an injective function. The choice of multiset on neighbourhood information aggregation, instead of mean pooling in GCN or max pooling in GraphSAGE, enables it to better preserve neighbourhood structural information. The above representation is then proven to be equivalent to:
\begin{equation}
h_{v}^{(l)}=\operatorname{MLP}^{(l)}\left(\left(1+\epsilon^{(l)}\right) \cdot h_{v}^{(l-1)}+\sum_{u \in \mathcal{N}(v)} h_{u}^{(l-1)}\right),
\end{equation}
where $\epsilon^{(l)}$ is a scalar representing the importance of the focal node, and MLP is used to model the composition of the function $f$ and $\phi$.

\textit{GAT}. Although in the aggregation scheme of the GCN, nodes from the same neighbourhood are assigned different weights by introducing the normalisation term ($c_{vu}$ in Equation~\ref{eqn: gcn}), the approach lacks the flexibility of introducing other weight mechanisms. To overcome this shortcoming, GAT proposes to use a masked self-attentional layer on graphs. \say{Masked} means that only 1-hop neighbours, rather than all other nodes, of a given node, are included in the attention scheme. Specifically, the attention coefficient of an edge $e_{vu}$ at layer $l$ is given by:
\begin{equation}
\alpha_{v u}^{(l)}=\frac{\exp \left(e_{v u}^{(l)}\right)}{\sum_{w \in \mathcal{N}(v)} \exp \left(e_{v w}^{(l)}\right)}, e_{v u}^{(l)}=\operatorname{LeakyReLU}\left(\vec{a}^{(l)}[\mathbf{W}^{(l)}h_{v}^{(l)} \|\mathbf{W}^{(l)} h_{u}^{(l)}]\right),
\end{equation}
where $a^{(l)}$ is a shared feedforward neural network parameterised by a weight vector $\vec{a}$, and $\mathbf{W}^{(l)}$ is a shared linear transformation of input or hidden features. Then the representation of node $v$ at layer $(l+1)$, is obtained through applying the attention coefficients on $v$'s neighbour nodes:
\begin{equation}
h_{v}^{(l+1)}=\sigma\left(\sum_{j \in \mathcal{N}(i)} \alpha_{v u}^{(l)} \mathbf{W}^{(l)} h_{u}^{(l)}\right)
\end{equation} 

\textit{FastGCN.} One issue of the GCN's neighbourhood aggregation scheme is the quick neighbourhood expansion across layers, which largely limits its scalability in large and dense graphs. To address this problem, FastGCN \cite{chen2018fastgcn} proposes to sample a fixed number of nodes at each layer while applying neighbourhood aggregation, so the number of involved nodes is up-bounded by the sample size. Concretely, the representation of nodes at layer $l$ is given by:
\begin{equation}
H^{(l+1)}(v,:)=\sigma\left(\frac{n}{s} \sum_{j=1}^{s} \hat{A}\left(v, u_{j}^{(l)}\right) H^{(l)}\left(u_{j}^{(l)},:\right) W^{(l)}\right),
\end{equation}
where $s$ is the sample size and $n$ is the total number of nodes in a graph. Compared to the node-wise sampling strategy proposed by GraphSAGE, this layer-wise sampling method further improves the computational efficiency of the model. For example, in a 2-layer setup, when 10 nodes are sampled from a node's neighbourhood, there will be a total of $10^2= 100$ nodes involved. In contrast, when 10 nodes are sampled at each layer, the total number of involved nodes is at most $10*2 = 20$. 

\textit{GraphSNN.} A common feature of the above-mentioned approaches is that each node gathers information from its neighbours, that is to say treating the neighbourhood as a 1-hop subtree. A recent work argues that this scheme ignores the rich structure information among the neighbour nodes, and therefore proposes a model named  GraphSNN to treat the neighbourhood as a 1-hop subgraph by including the connections among neighbours \cite{wijesinghe2021new}. Concretely, the work first defines \say{structure coefficients} for each node and its neighbours and generate a weighted adjacency matrix $A_{v u}=w\left(S_{v}, S_{v u}\right)$, where $S_{v}$ is 1-hop neighbourhood subgraph of node $v$, and $ S_{v u}$ is overlap subgraphs of node $v$ and $u$. $w$ is a function on $S_{v}$ and $S_{v u}$ exhibiting properties of local closeness and local denseness, which is designed as $\frac{\left|E_{v u}\right|}{\left|V_{v u}\right| \cdot\left|V_{v u}-1\right|}\left|V_{v u}\right|^{\lambda}$ in the paper. $\lambda$ is a positive value chosen by users. Then, the representation of node $v$ at layer $l$ is generated by:
\begin{equation}
h_{v}^{(l)}=\operatorname{MLP}^{(l)}\left(\gamma^{(l-1)}\left(\sum_{u \in \mathcal{N}(v)} \tilde{A}_{v u}+1\right) h_{v}^{(l-1)}+\sum_{u \in \mathcal{N}(v)}\left(\tilde{A}_{v u}+1\right) h_{u}^{(l-1)}\right),
\end{equation}
where $\gamma^{(l-1)}$ is a learnable scalar parameter, and $ \tilde{A}_{v u}$ is the normalised weighted adjacency matrix. The part before $h_{v}^{(l-1)}$ signifies the focal node's self-importance while the part $\sum_{u \in \mathcal{N}(v)}\left(\tilde{A}_{v u}+1\right)$ before $h_{u}^{(l-1)}$ is to apply different weights on different neighbours based on the overlap subgraph between the focal node and the neighbour node. From this perspective, GraphSNN is also an attention-like scheme that takes the 1-hop subgraph structure into account.

\textit{DGCNN.} In order to apply a GCN on graph-level learning tasks, Deep Graph Convolutional Neural Network (DGCNN) proposes to sort and pool the nodes' representations from multiple graph convolutional layers, then pass them to a traditional CNN architecture, i.e., a one-dimensional convolutional layer followed by dense layers before the final softmax output layer \cite{zhang2018end}. As the GCN can be viewed as \say{a differentiable and parameterised generalisation of the 1-dim Weisfeiler-Lehman algorithm} \cite{kipf2016semi}, each node's representation can be viewed as a \say{continuous colour} at that layer. The order of nodes in DGCNN is thus calculated according to the nodes' representations, i.e., nodes' colours, at the graph convolutional layers (first comparing the representations at the last layer, then the representations at the second-to-last layer when some nodes have the same representation, and so on). Next, in order to fit into the following CNN architecture, the sorted nodes' representation needs to be truncated or extended, which is done by deleting excessive rows or adding zero rows.    
This bridge layer between GCN and CNN is also known as SortPooling. 

\subsubsection{k-hop neighbourhood approaches.}  \label{sec:k-hop}
A natural idea to improve the performance of the GCN is to expand its message aggregation scope at each layer. This leads us to the second subcategory, i.e., k-hop neighbourhood approaches.

\textit{MixHop.} The layer-wise message passing scope of the vanilla GCN is limited to 1-hop neighbours and therefore lacks the ability to mix latent information from neighbours at different distances. MixHop is proposed to address this issue through a higher-order message passing scheme that aggregates information from further neighbours \cite{abu2019mixhop}. Concretely, the convolutional layer is defined as:

\begin{equation}
H^{(i+1)}=\sigma\left(\biggr \|_{j \in K} \widehat{A}^j H^{(i)} W_j^{(i)}\right),
\end{equation}
where $K$ is a set of integers representing the scope, and $\|$ denotes column-wise concatenation. When $K = {1}$, the operation degrades to the vanilla GCN. The paper also proves theoretically that the vanilla GCN cannot recover a 2-hop delta operator and thus cannot represent a general layer-wise neighbourhood mixing. In contrast, MixHop is able to learn a general mixing of information from neighbours at various distances. Their experiments on a synthetic dataset show that MixHop performs significantly better than several baselines on graphs of low levels of homophily. 

\textit{k-hop GNN.} A simple example of the limitation of the 1-hop neighbourhood aggregation is that it cannot distinguish regular graphs of the same size and degree. In order to improve the expressivity of the vanilla GCN, k-hop GCN also proposes to take k-hop neighbours into consideration in the layer-wise aggregation scheme \cite{nikolentzos2020k}. The general model is presented as:

\begin{equation}
\begin{aligned}
&a_v^{(l)}=\operatorname{AGGREGATE}^{(l)}\left(\left\{h_u^{(l-1)} \mid u \in \mathcal{N}_k(v)\right\}\right), 
&h_v^{(l)}=\operatorname{MERGE}^{(l)}\left(h_v^{(l-1)}, a_v^{(l)}\right),
\end{aligned}
\end{equation}
where $ \mathcal{N}_k(v)$ denotes the k-hop neighbourhood of node $v$. Specifically, it adopts an outside-to-inside updating scheme in the aggregation part: gradually updating neighbouring nodes from the furthest to the immediate ones. Each neighbour node $u$ at a distance $d$ from the focal node $v$ goes through two update functions successively:
\begin{equation}
\begin{aligned}
&x_u=\operatorname{UPDATE}_{d, a c r o s s}^{(l)}(u, \mathcal{N}_1(u) \cap R_{d+1}(v)),
&x_u=\operatorname{UPDATE}_{d, within}^{(l)}(u, \mathcal{N}_1(u) \cap R_{d}(v)),
\end{aligned}
\end{equation}
where $R_{d+1}(v)$ or $R_{d}(v)$ denote the set of nodes that are at a distance $d+1$ or $d$ from node $v$. The first function learns representation from node u's neighbours that are $(d+1)$-hop away from node $v$; and the second function learns from node u's neighbours that are $d$-hop away from node $v$. The update functions are defined as:
$\operatorname{UPDATE}(u, S)=\operatorname{MLP}_1\left(\operatorname{MLP}_2\left(x_u\right)+\sum_{w \in S} \operatorname{MLP}_3\left(x_w\right)\right)$. Finally, the representation of a node $v$ is calculated as: $h_v^{(l)}=\operatorname{UPDATE}_{0, a c r o s s}^{(l)}\left(v, \mathcal{N}_1(v)\right)$.
Although this model can capture structural information from the k-hop neighbourhood at a single layer, it requires up to $2k$ update functions and the aggregation scheme is much more complicated and computationally expensive.

\textit{Adapt.} Similar to the FastGCN, Adaptive Sampling GCN (abbreviated as Adapt) adopts the layer-wise sampling strategy in order to accelerate the training of the GCN \cite{huang2018adaptive}. Lower layer sampling is conditioned on the higher layer. Compared to node-wise sampling, layer-wise sampling not only has a fixed number of nodes at each layer but also preserves the connections between lower-layer neighbours and higher-layer parent nodes. Furthermore, the approach proposes to aggregate information from distant nodes via skip connections, i.e., connecting layer $l+1$ with layer $l-1$. Specifically, the skip-connection representation of node $v$ at layer $(l+1)$ is formulated as:
\begin{equation}
h_{v_{\text {skip }}}^{(l+1)}=\sum_{s \in \mathcal{V}^{(l-1)}} \hat{a}_{s k i p}\left(v, s\right) h_s^{(l-1)} W_{\text {skip }}^{(l-1)},
\end{equation}
where $s$ denotes sampled nodes at layer $(l-1)$, $\hat{a}_{s k i p}\left(v, s\right) = \sum_{u \in \mathcal{V}^{(l)}}\hat{a}\left(v, u\right) \hat{a}\left(u, s\right)$, and $W_{\text {skip }}^{(l-1)} = W^{(l-1)} W^{(l)}$. The part of skip-connection is then added to the classic GCN layer before a nonlinear transformation. Therefore, the overall representation of node $v$ is:
\begin{equation}
h_{v}^{(l+1)}=\sigma\left(\sum_{u \in \mathcal{V}^{(l)}} \hat{a}\left(v, u\right) h_u^{(l)} W^{(l)} + h_{v_{\text {skip }}}^{(l+1)}\right).
\end{equation}
That is to say, each node gathers information from both its 1-hop neighbours and 2-hop neighbours. Their experiments on the Cora dataset show that although skip connection does not lead to significant improvement in accuracy, it helps to speed up the convergence. 

\textit{DGP.} Aiming to improve the performance of zero-shot learning tasks on knowledge graphs (directed graphs), Dense Graph Propagation (DGP) proposes to adopt a two-phase propagation scheme on two separate connectivity patterns (one having nodes connected to their ancestors and the other having nodes connected to their descendants) \cite{kampffmeyer2019rethinking}. Furthermore, at each phase, DGP introduces a weighting scheme to include the contributions from distant nodes. Concretely, the overall representation is formulated as: 

\begin{equation} \label{eqn:dgp}
H=\sigma\left(\sum_{k=0}^K \alpha_k^a \hat{A}_k^a \sigma\left(\sum_{k=0}^K \alpha_k^d \hat{A}_k^d X W_d\right) W_a\right),
\end{equation}
where $\hat{A}_k^a$ and $\hat{A}_k^d$ denote the normalised adjacency matrices containing k-hop connections to ancestors and to descendants, respectively. $\alpha_k^a$ and $\alpha_k^d$ are learnable weights denoting contributions from nodes that are k-hop away from a given node. We see from the above equation that DGP can be viewed as consisting of two convolutional layers where the inner layer aggregates information from 1 to k-hop out-neighbours, and the outer layer aggregates information from 1 to k-hop in-neighbours. 

\textit{DGCN.} Directed Graph Convolutional Networks (DGCN) is another attempt to extend the GCN to directed graphs \cite{tong2020directed}. It proposes to expand the receptive field of convolutional operation by considering the first- and second-order proximities. Specifically, they first define the notions of second-order in-degree proximity matrix and second-order out-degree proximity matrix as:
\begin{equation}
\begin{aligned}
&A_{S_{\text {in }}}(u, v)=\sum_w \frac{A_{w, u} A_{w, v}}{\sum_x A_{w, x}}, 
&A_{S_{\text {out }}}(u, v)=\sum_w \frac{A_{u, w} A_{v, w}}{\sum_x A_{x, w}}.
\end{aligned}
\end{equation}
The idea is that if two nodes (a given node and its 2-hop neighbour) share many common in-neighbours (or out-neighbours), they have higher second-order in-degree (or out-degree) proximity. When capturing first-order proximity, they choose to make the adjacency matrix symmetric by ignoring link directions. Then the overall representation at layer $l$ is formulated as:
\begin{equation}
\mathrm{H}^{(l)}=Concact\left(\sigma\left(\hat{\mathrm{A}}_{\mathrm{F}} \mathrm{H}^{(l-1)} \Theta^{(l-1)}\right), \sigma\left(\hat{\mathrm{A}}_{\mathrm{S}_{\mathrm{in}}} \mathrm{H}^{(l-1)} \Theta^{(l-1)}\right), \sigma\left(\hat{\mathrm{A}}_{\mathrm{S}_{\text {out }}} \mathrm{H}^{(l-1)} \Theta^{(l-1)}\right)\right),
\end{equation}
where $\hat{\mathrm{A}}_{\mathrm{F}}$, $\hat{\mathrm{A}}_{\mathrm{S}_{\mathrm{in}}}$ and $\hat{\mathrm{A}}_{\mathrm{S}_{\mathrm{out}}}$ are normalised first-order proximity matrix, normalised second-order in-degree proximity matrix and normalised second-order out-degree proximity matrix, respectively. Notice that although DGP also considers 2-hop neighbours in directed graphs when $k$ equals $2$, DGCN and DGP have different definitions of directed 2-hops.  


\textit{SGC.} In order to improve the efficiency and scalability of GCN, Simple Graph Convolution (SGC) proposes to remove the nonlinear transformation between layers \cite{wu2019simplifying}. They argue that the main advantage of GCN lies in its neighbourhood aggregation scheme, not the nonlinearity between convolutional layers. After removing all nonlinear activations, the final output of the SGC model is represented as follows:
\begin{equation}
\hat{Y}=\operatorname{softmax}\left(\hat{A} \ldots \hat{A} \hat{A} H^{(0)} W^{(1)} W^{(2)} \ldots W^{(L)}\right) = \operatorname{softmax}\left(\hat{A}^LH^{(0)}W\right),
\end{equation}
where $\hat{A}$ is the normalised self-connection added adjacency matrix, $H^{(0)}$ is the input node feature matrix, and $W$ is a single weight matrix. This output representation thus only requires learning a single weight matrix, and the term $\hat{A}^LH^{(0)}$ can be computed directly. Note that the meaning of $\hat{A}^LH^{(0)}$ is the sum of features from k-hop neighbouring nodes. Therefore, the SGC model is actually equivalent to a single convolutional layer where nodes aggregate information from their k-hop neighbours. 

\textit{PATCHY-SAN \cite{niepert2016learning}.} Traditional image-based convolutional networks can be viewed as traversing a node sequence, i.e., a receptive field moving from left to right and from top to bottom. In order to employ the convolutional architecture to graphs where spatial order is missing, PATCHY-SAN proposes to first impose an order on nodes according to a certain ranking algorithm, then construct receptive fields from a fixed number of neighbour nodes for each node in a preselected node sequence. Note here the neighbour nodes are selected by performing a breadth-first search, so it can go beyond 1-hop neighbours. The receptive fields, after being normalised, will then be fed into a one-dimensional convolutional layer and other dense layers. Comparing this CNN-like approach with the GCN, we see that it requires an extra procedure to rank nodes, and there are more hyper parameters to tune, such as the length of node sequence, the stride and the number of neighbour nodes in the receptive field. 

\subsubsection{Random-walk neighbourhood approaches}
Instead of defining neighbourhood based on the distance to the focal node, some GCN approaches adopt a random-walk based neighbourhood, which might enable them to capture random processes on certain types of graphs. 

\textit{PinSage.} In order to apply GCN to web-scale recommender systems, PinSage proposes to construct neighbourhoods via random walks, also referred to as importance-based neighbourhoods \cite{ying2018graph}. The convolutional operation is similar to that of GraphSAGE:

\begin{equation}
\begin{aligned}
&\mathbf{h}_{\mathcal{N}_r(v)}^{l} \leftarrow \gamma\left(\left\{\sigma\left(\mathbf{Q^l}h_u\right) \mid u \in \mathcal{N}_r(v)\right\}, \boldsymbol{\alpha}\right), 
&\mathbf{h}_{v}^{l} \leftarrow \sigma\left(\mathbf{W}^{l} \cdot \operatorname{CONCAT}\left(\mathbf{h}_{v}^{l-1}, \mathbf{h}_{\mathcal{N}_r(v)}^{l}\right)\right),
\end{aligned}
\end{equation}
where ${\mathcal{N}_r(v)}$ is a random-walk neighbourhood, $\gamma$ is an aggregation function, $h_u$ is a set of embeddings of nodes in the neighbourhood, $\boldsymbol{\alpha}$ is a set of weights on nodes in the neighbourhood, $\mathbf{Q^l}$ and $\mathbf{W^l}$ are learnable model parameters. Specifically, ${\mathcal{N}_r(v)}$ comes from simulating a random walk starting from the focal node and calculating the $L_1$-normalised count of visited nodes, then the top $T$ nodes with the highest counts are selected as the neighbourhood in the layer-wise message passing. There are two benefits in this neighbourhood definition: first, the number of nodes involved in the aggregation is fixed, so the cost of the algorithm is predictable; second, the normalised visit counts can be directly used as weights to represent the importance of each node in the neighbourhood. PinSage also introduces some strategies to improve the model's scalability, such as the producer-consumer minibatch construction and a MapReduce pipeline. Notice that PinSage is originally designed for recommender systems which are bipartite networks.

\textit{GraLSP.} Based on the idea that anonymous walks can capture structures through reconstructing local subgraphs \cite{micali2016reconstructing}, GraLSP proposes to adopt random anonymous walks into the neighbourhood aggregation scheme \cite{jin2020gralsp}.  It also combines some other techniques to enhance the model performance, such as adaptive receptive radius, attention and channel-wise amplification. Specifically, the convolutional layer is formulated as:
\begin{equation}
\begin{aligned}
&\mathbf{a}_v^{(k)}=\operatorname{MEAN}_{\mathbf{wk} \in \mathcal{W}^{(i)}, p \in\left[1, r_{\mathbf{wk}}\right]}\left(\lambda_{v, \mathbf{wk}}^{(k)}\left(\mathbf{q}_{v, \mathbf{wk}}^{(k)} \odot \mathbf{h}_{\mathrm{wk}_p}^{(k-1)}\right)\right),
&\mathbf{h}_v^{(k)}=\operatorname{ReLU}\left(\mathbf{W}^{(k)} \mathbf{h}_v^{(k-1)}+\mathbf{U}^{(k)} \mathbf{a}_v^{(k)}\right),
\end{aligned}
\end{equation}
where $wk$ denotes a walk from the set of random walk sequence $\mathcal{W}^{(i)}$, $wk_p$ is the $p$-th node in walk $wk$. $r_{wk}$, $\lambda_{v, \mathbf{wk}}$ and $\mathbf{q}_{v, \mathbf{wk}}$ denote receptive radius, attention coefficient and amplification coefficient, respectively. Adaptive radius is introduced in order to regulate the scope of walks so that nodes that are too far away in the constructed subgraph are excluded while nodes in clustered subgraphs are included. Concretely, it is defined as $r_{wk} =\left\lfloor\frac{2 l}{C_{wk}}\right\rfloor$, where $l$ is walk length and $C_{wk}$ is the number of distinct nodes visited by the walk. Finally, the attention coefficient is introduced to assign different importance to visited nodes, and the channel-wise amplification is used to model the selection of node features.

\subsubsection{Subgraph neighbourhood approaches} \label{sec:subgraph}
In addition to the fixed-hop neighbourhood and random-walk neighbourhood definition in layer-wise message aggregation, some approaches view the neighbourhood as k-node tuples or subgraphs. 

\textit{k-GNN.} As the ability of the GCN to distinguish nonisomorphic graphs is equivalent to that of the 1-dimensional
Weisfeiler-Leman algorithm (1-WL), a k-GNN is proposed to achieve a higher expressivity as that of a  k-WL \cite{morris2019weisfeiler}. Different from the vanilla GCN where each node gathers information from a defined neighbourhood, the k-GNN works on the level of node tuple. Accordingly, the neighbourhood of a k-tuple is defined as other k-tuples containing one node that is not in the focal k-tuple. Specifically, the neighbourhood of k-tuple $s$ is defined as: $N_k(s)=\left\{t \in[V]^k|| s \cap t \mid=k-1\right\}$, where $[V]^k$ is a set of all k-tuples in a given graph. The convolutional operation at layer $l$ is then defined as:
\begin{equation}
h^{(l)}(s)=\sigma\left(h^{(l-1)}(s) \cdot W_1^{(l)}+\sum_{u \in N_k(s)} h^{(l-1)}(u) \cdot W_2^{(l)}\right). 
\end{equation}
At the beginning, $h^{(0)}(s)$ is set as $h^{iso}(s)$, which is a one-hot encoding of the isomorphism type of induced subgraph of $s$. To improve the model's scalability and avoid overfitting, a more restricted k-tuple neighbourhood $LN_k(s)$, named local neighbourhood, is defined as the tuples in $N_k(s)$ also satisfying $(u, v) \in E$ for $u \in s \backslash t$ and $v \in s \backslash t$. In other words, the non-overlapped nodes in a given k-tuple and a neighbouring k-tuple needs to be connected so that the neighbouring k-tuple is a local neighbourhood. Notice that as k-GNN is defined on the k-tuple level, it is unsuitable for node-level tasks.  

\textit{GRAPE.} In order to improve GCN's ability to discriminate graph isomorphism, GRAPE proposes to consider specific subgraph patterns in its layer-wise neighbourhood aggregation \cite{xu2021automorphic}. First, nodes of a given subgraph pattern are grouped into different sets according to their egocentric automorphic equivalences, abbreviated as the Ego-AE set. For example, in a triangle subgraph, the focal node is in one set, and the other two nodes are in another set. Then, different weights are learned for each Ego-AE set. Concretely, node $v$'s Ego-AE sets in a given subgraph $S$ are denoted as: $\left\{\mathcal{AE}_{S, 1}\left(v\right), \ldots, \mathcal{AE}_{S, i}\left(v\right), \ldots, \mathcal{AE}_{S, m}\left(v\right)\right\}$, where $m$ is the total number of AE-set. The convolutional operation for subgraph $S$ is then formulated as:
\begin{equation}
h_S^l(v)=\operatorname{MLP}\left(\sum_i \beta_{S, i} \cdot \sum_{u \in \mathcal{AE}_{S, i}\left(v\right)} h_S^{l-1}\left(u\right)\right),
\end{equation}
where $\beta_{S, i}$ are learnable weights representing the importance of each set $\mathcal{AE}_{S, i}$. Note that the focal node $v$ also belongs to an Ego-AE set. The final embedding of node $v$ at layer $l$ is then achieved through combining embeddings from a set of different subgraph patterns: $h^l(v)=\sum_{S \in \Omega} \alpha^l_S \cdot h_S^l(v)$, where $\Omega$ denotes the set of subgraph patterns, and $\alpha^l_S$ is learnable weight on a given subgraph pattern. This way GRAPE is able to differentiate neighbouring nodes according to their structural roles captured by Ego-AE. Certainly, the approach involves an extra step of choosing subgraph patterns.

\begin{landscape}
\begin{table}[]
\renewcommand{\arraystretch}{1.7}
\begin{tabular}{|l|l|l|l|l|l|}
\hline
\rowcolor[HTML]{b0bec5} 
\large Approach   & \large{Layer-wise aggregation scope}                                                                              & \large Aggregator                & \large Task                & \large{Batch size} & \large{Type of graph} \\ \hline
\rowcolor[HTML]{FFFFFF} 
GCN        & 1-hop neighbourhood                                                                              & Sum/Mean                  & Node \& Graph level & full-batch & General       \\ \hline
\rowcolor[HTML]{f2f2f2} 
GraphSAGE  & 1-hop neighbourhood with sampling                                                                & Flexible choice           & Node \& Graph level & mini-batch & General       \\ \hline
\rowcolor[HTML]{FFFFFF} 
GIN        & 1-hop neighbourhood                                                                              & Multiset                  & Node \& Graph level & mini-batch & General       \\ \hline
\rowcolor[HTML]{f2f2f2} 
GAT        & 1-hop neighbourhood with attention   scheme                                                      & Weighted sum/mean         & Node \& Graph level & mini-batch & General       \\ \hline
\rowcolor[HTML]{FFFFFF} 
FastGCN    & \begin{tabular}[c]{@{}l@{}}1-hop neighbourhood with\\  layer-wise sampling\end{tabular}     & Sum/Mean                  & Node \& Graph level & mini-batch & General       \\ \hline
\rowcolor[HTML]{f2f2f2} 
GraphSNN   & \begin{tabular}[c]{@{}l@{}}1-hop neighbourhood with\\   structural coefficients\end{tabular} & Sum/Mean                  & Node \& Graph level & mini-batch & General       \\ \hline
\rowcolor[HTML]{FFFFFF} 
DGCNN      & 1-hop neighbourhood                                                                              & Sum/Mean                  & Graph level         & full-batch & Directed      \\ \hline
\rowcolor[HTML]{f2f2f2} 
MixHop     & k-hop neighbourhood                                                                              & Sum/Mean                  & Node \& Graph level & full-batch & General       \\ \hline
\rowcolor[HTML]{FFFFFF} 
k-hop GNN  & k-hop neighbourhood                                                                              & Outside-to-insides scheme & Node \& Graph level & mini-batch & General       \\ \hline
\rowcolor[HTML]{f2f2f2} 
Adapt      & 2-hop neighbourhood                                                                              & Sum/Mean                  & Node \& Graph level & mini-batch & General       \\ \hline
\rowcolor[HTML]{FFFFFF} 
DGP        & k-hop neighbourhood                                                                              & Sum/Mean                  & Node \& Graph level & full-batch & Directed      \\ \hline
\rowcolor[HTML]{f2f2f2} 
DGCN       & 2-hop neighbourhood                                                                              & Sum/Mean                  & Node \& Graph level & full-batch & Directed      \\ \hline
\rowcolor[HTML]{FFFFFF} 
SGC        & k-hop neighbourhood                                                                              & Sum/Mean                  & Node \& Graph level & mini-batch & General       \\ \hline
\rowcolor[HTML]{f2f2f2} 
Patchy-SAN & fixed number of nodes                                                                            & CNN-like (weighted sum)   & Graph-level         & mini-batch & General       \\ \hline
\rowcolor[HTML]{FFFFFF} 
PinSage    & random-walk neighbourhood                                                                        & Weighted sum/mean         & Node \& Graph level & mini-batch & Bipartite     \\ \hline
\rowcolor[HTML]{f2f2f2} 
GraLSP     & random-walk neighbourhood                                                                        & Sum/Mean                  & Node \& Graph level & mini-batch & General       \\ \hline
\rowcolor[HTML]{FFFFFF} 
k-GNN      & k-tuple neighbourhood                                                                            & Sum                       & Graph-level         & mini-batch & General       \\ \hline
\rowcolor[HTML]{f2f2f2} 
GRAPE      & subgraph neighbourhood                                                                           & Weighted sum              & Node \& Graph level & mini-batch & General       \\ \hline
\end{tabular}
\caption{Summary of approaches in the first category.}
\label{tab:gcn_cat1}
\end{table}
\end{landscape}

\subsection{Message Content} \label{sec:msg-concent}
\begin{figure}[t]
\centerline{\includegraphics[scale = 0.9]{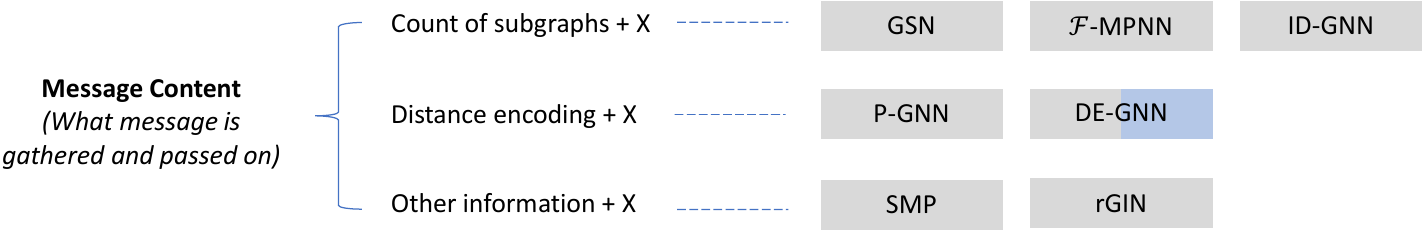}}
\caption{Taxonomy from the message content perspective.}
\label{fig:msg-content}
\vspace{-0mm}
\end{figure}

The superior performance of the GCN lies in its ingenious combination of node attributes and a graph structure with node attributes used as initial representations and then subsequently propagated on the graph through certain convolutional operations. In contrast, some learning approaches only exploit structural information, such as the matrix decomposition based methods \cite{cao2015grarep, ou2016asymmetric} and the random walk based methods \cite{perozzi2014deepwalk, grover2016node2vec}. In situations where no node attributes are provided, a simple structural metric like the node degree is often used as initial representations in the GCNs \cite{hamilton2017inductive}. Another group of approaches further propose to improve the GCN's distinguishability through injecting more complicated structural features 
into the node representations, such as the count of graphlets, distance-based information, etc. The taxonomy and representative approaches
are given in Figure~\ref{fig:msg-content}. 

\subsubsection{Count of subgraphs.} The number of certain subgraphs or substructures is often used as a node feature in traditional network studies \cite{milenkovic2008uncovering}. Some approaches thus propose to include this type of structural information as part of a node representation in the GCN's message passing scheme.  

\textit{GSN.} Graph Substructure Network (GSN) proposes to capture structural features by counting the appearance of particular graphlet orbits and include them as part of node features in the convolutional operation \cite{bouritsas2020improving}. Specifically, node $v$'s representation at layer $l$ is defined as:
\begin{equation}
\mathrm{h}^{l+1}(v)=\mathrm{MLP_1}\left(h^l(v), \sum_{u \in \mathcal{N}(v)} MLP_2\left(\mathbf{h}^t(v), \mathbf{h}^t(u), \mathrm{x}_V(v), \mathrm{x}_V(u), \mathrm{e}(u, v)\right)\right),
\end{equation}
where $\mathrm{x}_V(v)$ and $\mathrm{x}_V(u)$ are structure features of nodes $v$ and $u$, respectively. $\mathrm{e}(u, v)$ is an edge feature if provided. The structural feature is a vector containing the counts of node orbits. For example, if subgraphs 2-path and 3-clique are considered ($G_1$ and $G_2$ in Figure~\ref{fig:motif_graphlet}), the counts of three node orbits will be included in the vector. GSN further introduces a version based on edge orbits, which is formulated as: 
$\mathrm{h}^{l+1}(v)=\mathrm{MLP_1}\left(h^l(v), \sum_{u \in \mathcal{N}(v)} MLP_2\left(\mathbf{h}^t(v), \mathbf{h}^t(u), \mathrm{x}_E(u, v), \mathrm{e}(u, v)\right)\right)$, where $\mathrm{e}(u, v)$ denotes the edge structural feature, i.e., a vector containing the count of edge orbits. The GSN has been proven to be strictly more powerful than the 1-WL test when the chosen subgraphs are not star graphs. Certainly, the choice of subgraphs is the core of this approach, and a larger subgraph will lead to higher computational complexity in a preprocessing step.  

\textit{$\mathcal{F}$-MPNN.} A local graph parameter enabled GNN ($\mathcal{F}$-MPNN) also proposes to include a subgraph count into the GCN \cite{barcelo2021graph}.  $\mathcal{F} = \{P_1^r, ..., P_k^r\}$ is a set of pre-selected subgraph patterns with $r$ referring to a node. The \say{homomorphism count} of each pattern $P_i^r$ for node $v$ in the original graph $G$ is denoted as $hom(P_i^r, G^v)$, which is actually equivalent to the count of a given node orbit (see Section~\ref{sec:motif_vs_graphlet}). Then a structural feature vector $(hom(P_1^r, G^v), ..., hom(P_k^r, G^v))$ is added to the one-hot encoding of node $v$'s label, serving as $v$'s initial feature vector. Concretely, 
the framework is formulated as:
\begin{equation}
\begin{aligned}
&\mathbf{h}_v^{(0)}:=\left(x_v, \operatorname{hom}\left(P_1^r, G^v\right), \ldots, \operatorname{hom}\left(P_{\ell}^k, G^v\right)\right),
&\mathbf{h}_v^{(l)}:=\operatorname{MERGE}\left(\mathrm{x}_v^{(l-1)}, \operatorname{AGGREGATE}\left(\left\{\left\{\mathrm{x}_u^{(l-1)} \mid u \in N(v)\right\}\right\}\right)\right),
\end{aligned}
\end{equation}
where $x_v$ is the one-hot encoding of node $v$'s label, MERGE and AGGREGATE are two MLPs. Since this structural feature is only applied to enhance the initial feature of nodes, it can be used as an add-on to any GCN architecture. Similar to the GSN, the choice of subgraph patterns is the core of $\mathcal{F}$-MPNN. Cycles of length smaller than $10$ and cliques of size smaller than $5$ are used as subgraph patterns in the experiment.   

\textit{ID-GNN.} Identity-aware GNN (ID-GNN) proposes to improve the expressivity of the GCN through distinguishing the root node of the extracted computation graphs from other nodes in its message passing scheme \cite{you2021identity}. It contains two major steps: the first step, named inductive identity colouring, is to uniquely colour the root node in its k-hop ego network; then in the second step, a heterogeneous message passing is applied to all the extracted ego networks. Specifically, the representation of any node $v$ in an extracted computation graph $G_r$ (rooted at node $r$) is formulated as:
\begin{equation}
\begin{aligned}
&\mathbf{m}_u^{(l)}=\operatorname{MSG}_{1[u=r]}^{(l)}\left(\mathbf{h}_u^{(l-1)}\right), 
&\mathbf{h}_v^{(l)}=\operatorname{AGG}^{(l)}\left(\left\{\mathbf{m}_u^{(l)}, u \in \mathcal{N}(v)\right\}, \mathbf{h}_v^{(l-1)}\right).
\end{aligned}
\end{equation}
$\operatorname{MSG}_{1[u=r]}^{(l)}(\cdot)$ means that $\operatorname{MSG}_1^{(l)}(\cdot)$ is applied to the root node while $\operatorname{MSG}_0^{(l)}(\cdot)$ is applied to other nodes. In this way, the representation of the root node is different from that of other nodes and will help distinguish other nodes when propagated to later layers. The approach is inductive since the colouring is based on the extracted computation graphs instead of the original graph. Further, in order to avoid the overhead of extracting ego-networks, ID-GNN-Fast proposes to use the count of cycles as an augmented node feature. Therefore the input node feature is built from concatenating the original node feature and the augmented node feature. 

\subsubsection{Distance information.} Distance measures such as shortest paths between nodes are widely used in traditional network studies \cite{brandes2008variants}. Naturally, some approaches propose to enhance the performance of the GCN through including distance information in their message passing scheme or as an additional initial node feature. 

\textit{P-GNN.} Position-aware graph neural network (P-GNN) proposes to let each node aggregate information from several randomly chosen subsets of nodes, instead of its own 1-hop neighbours \cite{you2019position}. As every node shares the same neighbourhood in P-GNN, distance information is included to indicate the relative position of each node to those subsets. Specifically, given $k$ randomly sampled subsets, $S_i$ denoting the $i$\textsuperscript{th} subset, the representation of node $v$ at layer $l$ is formulated as:
\begin{equation}
\begin{aligned}
&\mathbf{h}_v^{l}=\operatorname{AGG}^{(l)}\left(\mathcal{M}_i^{l-1}, \forall i \in [1,k]\right),
& \mathcal{M}_i^{l-1} = \{ F(d_{uv}, h_u^{l-1}, h_v^{l-1}), \forall u \in S_i\}.
\end{aligned}
\end{equation}
$F$ is a message computation function accounting for both distance information and feature information of a pair of nodes. The output at the last layer is constructed with $\mathcal{M}_i$ being the $i$\textsuperscript{th} embedding dimension, thus making the final representation \say{position aware}. Note that the subsets are resampled at each convolution layer, so that each node can aggregate information from different sets of nodes at each layer. 

\textit{DE-GNN.} Distance-Encoding GNN (DE-GNN) also proposes to improve the GCN's expressivity through adding distance information \cite{li2020distance}. Intuitively, for any given node set $S$ whose representation is to be learnt, other nodes are encoded with their distances to each node of $S$. Formally, DE of node $u$ with regard to the target node set $S$ is defined as:  

\begin{equation}
\begin{aligned}
& \zeta(u \mid S)=\operatorname{AGG}(\{\zeta(u \mid v) \mid v \in S\}), 
&\zeta(u \mid v)=f\left( \left((M)_{u v},\left(M^2\right)_{u v}, \ldots,\left(M^k\right)_{u v}, \ldots\right)\right),
\end{aligned}
\end{equation}
where $M = AD^{-1}$ is a matrix of landing probabilities through random walks,$f$ can be a heuristic function or a learnable neural network. Different distance measures can be captured by the above equation such as the shortest path distance or the generalised PageRank score. DE is denoted as DE-$|S|$ according to the size of set $S$. For example, DE-2 when $|S|=2$. One way of improving the GCN through distance encoding is to use it as an extra node feature: $h_v^{(0)} = CONCAT(x_v, \zeta(v \mid S))$. Another approach is to use DE-1 in the layer-wise aggregation: $h_v^{(l+1)}=f_1\left(h_v^{(l)}, \operatorname{AGG}\left(\left\{(f_2\left(h_u^{(l)}\right), \zeta(u \mid v))\right\}_{u \in V}\right)\right)$. Although DE-GNN adopts minibatch training, the distance information needs to be computed for every node set and for all nodes in its extracted L-hop ego-network, leading to a higher computational cost. Also note that DE-GNN is flexible for tasks on different levels: DE-1 for node-level tasks, DE-2 for link prediction tasks and DE-3 for triangle prediction tasks. To highlight that the DE-GNN is suitable for both node and link-level tasks, we use two colours in its block in Figure~\ref{fig:msg-content}. 

\subsubsection{Other approaches} Apart from the count of certain subgraphs and the distance information, some other information such as the \say{local context matrix} or even random features are also used to enhance the performance of the GCN.

\textit{SMP.} In order to improve the GCN's performances on structure-related tasks, Structural Message Passing (SMP) proposes to maintain a \say{local context matrix} at each node, instead of a feature vector as in the vanilla GCN. Specifically, each node is initialised as a one-hot encoding $\boldsymbol{M}_i^{(0)}=\mathbbm{1}_i \in \mathbbm{R}^{n \times 1}$, and the additional node feature $x_i$ of $v_i$ is appended at the $i$\textsuperscript{th} row: $M_i^{(0)}[i,:]=\left[1, x_i\right] \in \mathbbm{R}^{1+c_X}$. Then the local context matrix of node $v_i$ at layer $l$ is formulated as:
\begin{equation}
M_i^{(l)}=MLP_1^{(l-1)}\left(M_i^{(l-1)}, AGG\left(\left\{MLP_2^{(l-1)}\left(M_i^{(l-1)}, M_j^{(l-1)}, e_{i j}\right)\right\}_{v_j \in N_i}\right)\right).
\end{equation}
$AGG$ is an aggregation function which is by default a normalised sum aggregator: $\left.\sum_{v_j \in N_i} MLP_2^{(l-1)}\left(\boldsymbol{M}_i^{(l)}, \boldsymbol{M}_j^{(l)}, \boldsymbol{e}_{i j}\right) / d_{\mathrm{avg}}\right)$.  In this way, the $j$\textsuperscript{th} row in $M_i$ is the representation node $v_i$ has of node $v_j$. Finally, the vector form representation of node $v_i$ is obtained through applying an equivariant neural network for sets on the rows of its context matrix. Although node ordering is needed when constructing the local context matrix, the learned representation is proven to be order-invariant when $MLP_1$, $MLP_2$ and $AGG$ are permutation equivariant. SMP is shown to excel in various tasks, such as the detection of structural properties including distance, eccentricity connectivity, diameter, etc. 

\textit{rGIN.} Apart from various types of explicit structural features that are being added to the GCN, another work (termed rGIN) proves that the expressive power of the GCN can be enhanced by just adding random features to each node \cite{sato2021random}. Specifically, rGIN first assigns a random value $r_v$ to each node and concatenates it with the original node feature $x_v$, then performs GIN's convolutional operations: 
\begin{equation}
\begin{aligned}
&h_v^{(0)} = \operatorname{MLP}^{(0)}\left(CONCAT(x_v, r_v)\right),
&h_v^{(l)} = \operatorname{MLP}^{(l)}\left(\left(1+\varepsilon^{(l)}\right) h_v^{(l-1)}+\sum_{u \in \mathcal{N}(v)} h_u^{(l-1)}\right).
\end{aligned}
\end{equation}
With this simple modification on the initial node feature, rGIN is proven to be able to distinguish any local structure with high probability. The idea of injecting random features into nodes is that the GCN fails to distinguish graphs with identical node features. For example, a GCN with the node degree as an input feature cannot distinguish a node in a 3-cycle graph from a node in a 6-cycle graph. rGIN is shown to perform well on structure-related tasks such as learning the existence of triangles, learning the local clustering coefficient, and learning the algorithm for the MDS (Minimum Dominating Set) problem. 

\subsection{Learning scope}

\begin{figure}[t]
\centerline{\includegraphics[scale = 0.9]{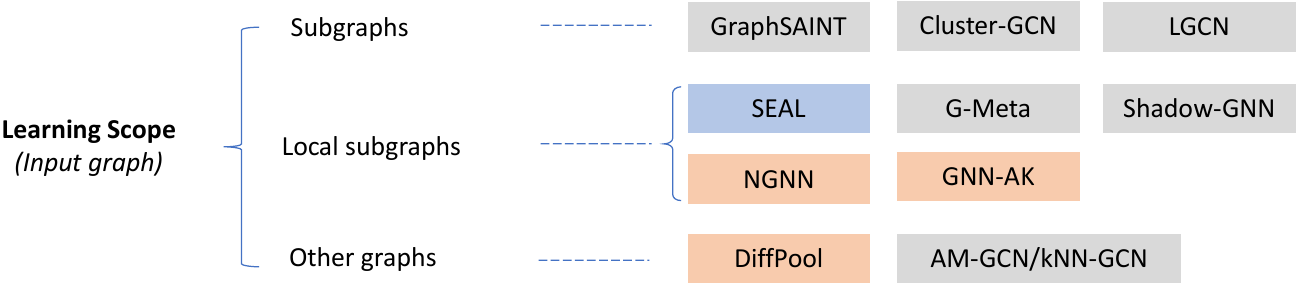}}
\caption{Taxonomy from the learning scope perspective.}
\label{fig:learning-scope}
\vspace{-0mm}
\end{figure}

The third structural perspective on GCNs is regarding the learning scope or the input graph. Whether it is full-batch or mini-batch training, most GCNs still have the whole graph as an input, i.e., in an L-layer GCN, each node has the scope of its L-hop neighbourhood in the original graph. A higher number of layers leads to a neighbourhood explosion and thus higher computational cost. To address this issue, many approaches limit the scope to subgraphs or localised subgraphs while some other methods propose to run the GCN on particular types of generated graphs. The taxonomy and related approaches are given in Figure~\ref{fig:learning-scope}. Again, the block's colour indicates the task the approach is proposed for: grey represents a node classification, blue represents a link prediction, and orange represents a network classification.

\subsubsection{Subgraphs}
An intuitive idea is to limit the training scope to several selected subgraphs instead of the original whole graph, so the neighbourhood is restricted within the sphere of subgraphs no matter how many layers are stacked. 

\textit{GraphSAINT.} In order to enhance the scalability of the GCN, GraphSAINT proposes to train a GCN model iteratively on several sampled subgraphs. Each sampled subgraph $G_s \in \mathbb{G}$ is a mini-batch. The representation of node $v$ in a sampled subgraph $G_s$ is formulated as:
\begin{equation}
h_v^{(l+1)}=\sum_{u \in N_v | G_s} \frac{\tilde{A}_{v, u}}{\alpha_{u, v}}W^{(l)} h_u^{(l)},
\end{equation}
where $u$ is $v$'s neighbour in $G_s$, and $\alpha_{u, v}$ is a coefficient to offset the biases from the sampler. Specifically, $\alpha_{u, v}$ is defined as the probability of edge $(u,v)$ being sampled, divided by the probability of node $v$ being sampled. Given a set of pre-sampled subgraphs $\mathbb{G}$, $\alpha_{u, v} = \frac{C_{u,v}}{C_v}$, where $C_{u,v}$ and $C_v$ are the number of times edge $(u,v)$ and node $v$ appear in $\mathbb{G}$, respectively. Finally, the batch loss is calculated as: $L_{\text {batch }}=\frac{1}{|\mathbb{G}|} \sum_{G_s \in \mathbb{G}} \sum_{v} \frac{L_v}{\lambda_v}$, where $L_v$ is the loss on node $v$ in the GCN's output layer, and $\lambda_v$ is a loss normalisation term computed by the number of node $v$ appearing in $\mathbb{G}$ divided by the total number of nodes in the original graph. Different samplers are integrated within the framework, such as random node sampler, random edge sampler and random walk based sampler. According to the experiment, the random walk based sampler tends to have the best performance.  

\textit{Cluster-GCN.} Also to address the issue of neighbourhood explosion in large graphs, a Cluster-GCN proposes to first partition the whole graph into several clusters according to certain clustering algorithms, then run the GCN on those clusters. Given $c$ clusters, the original adjacency matrix $A$ is approximated as a list of submatrices $A_{11}$, $A_{11}$, ..., $A_{cc}$ at diagonal. The representation of nodes at layer $l$ in the $t$\textsuperscript{th} cluster is thus formulated as: 
\begin{equation}
    H_t^{(l)} = \hat{A}_{tt} H_t^{(l-1)} W^{(l-1)},
\end{equation}
where $\hat{A}_{tt}$ is the normalised version of $A_{tt}$. The loss is then calculated as: $L_t = \frac{1}{\left|V_t\right|} \sum_{i \in V_t} \operatorname{loss}\left(y_i, h_i^{(L)}\right)$. At each iteration, the model weights are updated based on the loss of the cluster. This way, no matter how many convolutional layers are involved, the neighbourhood scope is restricted to one cluster. In order to offset the bias of clustering algorithms, a better version of the Cluster-GCN proposes to randomly form a subgraph with several randomly chosen clusters, then at each iteration, run GCN on one subgraph. Experiments on very large datasets show that the Cluster-GCN is able to train a deeper GCN without time and space overhead and achieves advanced performance.

\textit{LGCN.} A learnable graph convolutional network (LGCN) proposes to transform graph data into a grid-like data structure and apply the traditional convolutional operation on it \cite{gao2018large}. As traditional CNN requires a fixed number of ordered units in the receptive fields, the LGCN proposes to sort features at each dimension and select the k-largest ones to form a grid structure. The transformed data is then fed into a one-dimensional CNN to generate the final representation of the focal node. Specifically, the nodes' representation at layer $l$ is formulated as:
\begin{equation}
    H^{(l+1)} = c(g(H^{(l)}, A, k)), 
\end{equation}
where $k$ is a hyper parameter, $g(\cdot)$ is the function that performs k-largest selection to transfer the original graph data into grid data, and $c(\cdot)$ is a one-dimensional CNN. Furthermore, as Cluster-GCN and GraphSAINT, the LGCN also proposes to train the neural network on subgraphs. Each subgraph is built from randomly selecting a few initial nodes and then expanding adjacent nodes into it using a breadth-first-search algorithm. At each training iteration, multiple subgraphs can be included in a mini-batch. The subgraph training strategy is shown to be more time and space efficient, with only negligible loss in performance.


\subsubsection{Local subgraphs}
Another popular idea to address the computational overhead is training the GCN on local subgraphs. Note that local subgraphs are different from subgraphs in that they are extracted around each node or link. In contrast, subgraphs, as we have discussed earlier, have in general a wider range, without focusing on a node or a link. 

\textit{SEAL.} SEAL is a GCN based framework specially designed for a link prediction task \cite{zhang2018link}. Motivated by the fact that many successful link prediction heuristics, such as the common neighbour, Adamic-Adar and resource allocation, only involve the 1-hop or 2-hop neighbours around a node pair, SEAL proposes to train a GCN on the local subgraphs extracted around each target link. Specifically, the local subgraph is the induced graph from each target node pair and their k-hop neighbours. After having constructed the training data, it further introduces a node labelling procedure to give the target node pair special weights as well as to distinguish the neighbouring nodes in a given local subgraph. Specifically, it labels each node in the target pair as \say{1}, and assigns larger labels to other nodes according to their distances to the target pair. The assigned labels are then concatenated with other features to construct the feature matrix of the local subgraph. In the final step, a GCN is trained on the local subgraphs and their label-enhanced feature matrices. In
the experimental implementation, SEAL chooses to use DGCNN, a GCN model designed for graph classification (see Section~\ref{sec:1-hop-method}), as the default GCN model. Essentially, the link existence problem in the original graph is modelled as a graph classification problem on the extracted local subgraphs.

\textit{G-Meta.} Motivated by the idea that local subgraphs may contain transferable knowledge that can be adapted to unseen tasks, G-Meta proposes to leverage local subgraph information in few-shot graph meta-learning \cite{huang2020graph}. For the node classification task, local subgraphs are constructed as induced graphs from each node and its k-hop neighbours; and when it comes to link prediction, local subgraphs are built as in SEAL. Then a typical GCN is used on these local subgraphs to generate graph embeddings. At last, a prototypical loss and Model-Agnostic Meta-Learning (MAML) algorithm are used to update the GCN's parameters. Specifically, the prototype $t_l$ of label $l$ is calculated through averaging over subgraph embeddings in the support set: $t_l = \frac{1}{N_l}  \sum_{y_j=l} \mathbf{h}_j$. Then for each local subgraph $S_u$ in both support and query set, a class distribution vector $\mathbf{p}$ is calculated as: $\mathbf{p}_l= \frac{\exp \left(-\left\|\mathbf{h}_{S_u}-\mathbf{t}_l\right\|\right)}{\sum_{\hat{l}} \exp \left(-\left\|\mathbf{h}_{S_u}-\mathbf{t}_{\hat{l}}\right\|\right)}$. Finally, the cross-entropy loss is formulated as: $\mathrm{L}(\mathbf{p}, \mathrm{y})=\sum_j \mathbf{y}_j \log \mathbf{p}_j$. Experiments on synthetic and real networks show that local subgraphs are vital for few-shot graph learning. It is worth mentioning that the best performance is yielded when 2-hop neighbours are included. 

\textit{Shadow-GNN.} From the perspective of decoupling the scope (i.e., a receptive field) and the depth (i.e., a number of layers) of the GCN, a Shadow-GNN also proposes to adopt local subgraph as an input \cite{zeng2021decoupling}. Typically on a full graph, the scope of the GCN increases with the number of layers --- an L-layer GCN means an L-hop neighbourhood scope.
As the GCN model is also viewed as a form of Laplacian smoothing that mixes the feature of a node and its neighbours, when the scope becomes too large, node features may be oversmoothed \cite{li2018deeper}. To address this problem, the Shadow-GNN proposes to train the GCN on local subgraphs, so that the scope is bounded by the range of the local subgraphs, regardless of the number of layers. In this setting, the depth can be larger than the scope. It means that nodes in the subgraphs may exchange information multiple times, which could lead to better expressivity. Different subgraph extractors can be selected, such as an L-hop neighbourhood extractor or a random-walk-based extractor. In actual implementation, the scope is set as a 2- or 3-hop neighbourhood while the depth is deeper (3 or 5 layers).  

\textit{NGNN.} A nested graph neural network (NGNN) proposes to apply the local subgraph training strategy on a graph classification task \cite{zhang2021nested}. The extracted local subgraphs, termed rooted subgraphs, are also induced subgraphs from each node and its k-hop neighbours. First, a base GCN is applied on all rooted subgraphs. Taking root node $v$ for example, at layer $l$, any node $u$ in its k-hop rooted subgraph $G_v^k$ is formulated as: 
\begin{equation}
h_{u, G_v^k}^{(l)}=UPDATE^{(l-1)}\left(h_{u, G_v^k}^{(l-1)}, \sum_{w \in N\left(u \mid G_v^k\right)} MSG^{(l-1)}\left(h_{u, G_v^k}^{(l-1)}, h_{w, G_v^k}^{(l-1)}, e_{u w}\right)\right).
\end{equation}
Then, the final representation of root node $v$ at layer $L$ is set to be equal to its rooted subgraph representation obtained from applying a subgraph pooling on all nodes in the subgraph: $h_v=h_{G_v^k}=POOL_1\left(\left\{h_{u, G_v^k}^{(L)} \mid u \in G_v^k\right\}\right).$ With the same base GCN applied on all nodes' rooted subgraphs, the representation of each node can be obtained, and the graph representation can be generated from applying another GCN, termed outer GCN, on those updated node representations. To make it simple, the outer GCN can be just a graph pooling layer: $h_G = POOL_2({h_v \mid v \in G})$. The work theoretically proves that a proper NGNN can discriminate almost all $r$-regular graphs where the vanilla GCN cannot.

\textit{GNN-AK.} GNN-As Kernel (GNN-AK) is another local subgraph based approach for a graph classification problem \cite{zhao2021stars}. Different from the NGNN which directly uses the rooted subgraph embedding to represent each node, the GNN-AK proposes to construct node representation from concatenating three types of embedding, i.e., subgraph embedding, centroid embedding, and context embedding. Centroid embedding is simply the root node representation in its own subgraph, while context embedding is built from the representation of this node in other nodes' rooted subgraphs. It is argued that these two additional embeddings contain information which is not captured in the subgraph embedding. Formally, the representation of node $v$ at layer $l$ is:
\begin{equation}
    h_v^{(l)} = CONCAT(h_{v,\operatorname{centroid}}^{(l)}, h_{v,\operatorname{subgraph}}^{(l)}, h_{v,\operatorname{context}}^{(l)}),
\end{equation}
with $h_{v,\operatorname{centroid}}^{(l)} = h_{v \mid G_v^k}$, $h_{v,\operatorname{subgraph}}^{(l)}=POOL_1(\{h_{i \mid G_v^k} \mid i \in \mathcal{N}_k(v) \})$, and $h_{v,\operatorname{context}}^{(l)}=POOL_2(\{h_{v \mid G_j^k} \forall j \text { s.t. } v \in \mathcal{N}_k(j) \})$. $h_{v \mid G_u^k}$ denotes the representation of node $v$ in node $u$'s rooted subgraph. Then, the final graph representation is obtained from another pooling at the output layer: $h_G = POOL_3(\{h_v^{L} \mid v \in V\})$. It is worth mentioning that a subgraph drop strategy is further introduced to improve the scalability of GNN-AK so that the number of local subgraphs can be much smaller than the number of nodes in the original graph.

\subsubsection{Other types of graphs}
Subgraphs or local subgraphs are still part of the original graphs. In the third subcategory, we see approaches that use differently constructed graphs, such as the coarsened graph and the feature graph. 

\textit{DiffPool.} Analogous to the idea of spatial pooling in a traditional CNN, a DiffPool proposes to learn a graph representation in a hierarchical manner. Nodes at layer $l$ will be collapsed into higher-level cluster nodes at layer $l+1$ via a learned assignment matrix, and after stacking several hierarchical layers, the singular node's embedding at the final layer is viewed as the representation for the whole graph. Concretely, node embedding matrices $Z^{(l)}$ are learned from a GCN (called an embedding GNN), and an assignment matrix $S^{(l)}$ is learned from another GCN, called a pooling GNN:

\begin{equation}
\begin{aligned}
& Z^{(l)} = GNN_{\operatorname{embed}}^{(l)}(A^{(l)}, X^{(l)}),
& S^{(l)} = \operatorname{softmax}(GNN_{\operatorname{pool}}^{(l)}(A^{(l)}, X^{(l)})),
\end{aligned}
\end{equation}
where $A^{(l)}$ and $X^{(l)}$ are the coarsened adjacency matrix and the cluster nodes feature matrix at layer $l$, respectively. The dimension of assignment matrix $S^{(l)}$ is $n_l \times n_{l+1}$, so that each role is one of the $n_l$ nodes at layer $l$ and each column is one of the cluster nodes at layer $l+1$. Then, $A^{(l+1)}$ and $X^{(l+1)}$ which are used as the next layer's inputs are generated as: 
\begin{equation}
\begin{aligned}
& X^{(l+1)} = {S^{(l)}}^T Z^{(l)},
& A^{(l+1)} = {S^{(l)}}^T A^{(l)} S^{(l)}.
\end{aligned}
\end{equation}
The assignment matrix $S^{(L-1)}$ at the penultimate layer is set to be a vector of 1's, so that all nodes will collapse into a single cluster node at the final layer, and the corresponding node embedding is viewed as the representation for the original graph. Note that the number of clusters is a predefined hyperparameter, which is usually set as a percentage of the number of nodes at the previous layer.

\textit{AM-GCN.} An Adaptive Multi-channel Graph Conventional Network (AM-GCN) proposes to not only run the GCN on the original (topological) graph, but also on a feature graph constructed from a feature similarity matrix. Specifically, the similarity matrix is computed using cosine similarity or heat kernel. Then, edges will be added between each node and $k$ other nodes of top similarity scores. The generated feature graph $G_f=(\mathbf{A}_f, \mathbf{X})$ is also called the k-nearest neighbour (kNN) graph. Therefore, the embeddings on the feature graph are formulated as follows: 
\begin{equation}
\mathbf{H}_f^{(l)}=\operatorname{ReLU}\left(\hat{\mathbf{A}}_f \mathbf{H}_f^{(l-1)} \mathbf{W}_f^{(l)}\right),
\end{equation}
where $\hat{\mathbf{A}_f}$ is the normalised feature graph adjacency matrix. Another GCN is used to generate node embeddings on the original graph $G=(\mathbf{A}, \mathbf{X})$: $\mathbf{H}_t^{(l)}=\operatorname{ReLU}\left(\hat{\mathbf{A}} \mathbf{H}_t^{(l-1)} \mathbf{W}_t^{(l)}\right)$. Further, in order to capture the correlation between a topological space and a feature space, a common convolutional module is introduced as: $\mathbf{H}_{ct}^{(l)}=\operatorname{ReLU}\left(\hat{\mathbf{A}} \mathbf{H}_{ct}^{(l-1)} \mathbf{W}_c^{(l)}\right)$;  $\mathbf{H}_{cf}^{(l)}=\operatorname{ReLU}\left(\hat{\mathbf{A}}_f \mathbf{H}_{cf}^{(l-1)} \mathbf{W}_c^{(l)}\right)$. Note that the same weight matrix $\mathbf{W}_c^{(l)}$ is shared in $\mathbf{H}_{ct}^{(l)}$ and $\mathbf{H}_{cf}^{(l)}$. Under this setting, node features are propagated not only in a topological space but also in a feature space.  
The final representation is then obtained through combining the above four embeddings with an attention scheme: $\mathbf{Z}=\alpha_t \cdot \mathbf{H}_t^{(L)}+\alpha_f \cdot \mathbf{H}_f^{(L)}+\alpha_c \cdot (\frac{\mathbf{H}_{ct}^{(L)} + \mathbf{H}_{cf}^{(L)} }{2})$, where $\alpha_t$, $\alpha_f$ and $\alpha_c$ are attention vectors.

\subsection{Discussion}

\subsubsection{Differences between layer-wise scope and overall learning scope}
Here we emphasize the differences between layer-wise scope and overall learning scope. Layer-wise message aggregation scope, or a receptive field, is where a node receives the message from. It can be a 1-hop neighbourhood, k-hop neighbourhood, random-walk neighbourhood, or subgraph neighbourhood according to our taxonomy. Although the receptive field is usually small, distant nodes can exchange messages after stacking multiple GCN layers, causing the well-known neighbourhood explosion issue. Obviously, with a large enough number of layers, a node can exchange information with any other node in the entire graph. 
Overall learning scope, in contrast, is determined by the input graph, which can be the entire original graph, extracted subgraphs or local subgraphs, or coarsened graphs according to our taxonomy. Taking an extracted local subgraph for example, no matter how large the receptive field is or how many layers are stacked, a node can only exchange messages with other nodes in the same subgraph. This naturally solves the neighbourhood explosion issue. A large number of layers on a relatively small subgraph also means that nodes may exchange information multiple times, which is argued to help the GCN \say{better absorb and embed information} \cite{zeng2021decoupling}.      

\subsubsection{Time and space complexity analysis}
In the discussion about complexity, we focus on how the different definitions of the neighbourhood in a convolutional layer influence the cost of computation (corresponding to the layer-wise message scope taxonomy in Figure~\ref{fig:layer-wise-taxonomy}). The time and space complexities of each category are listed in Table~\ref{tab:complexity}.

\begin{table}[]
\renewcommand{\arraystretch}{1.3}
\begin{tabular}{l|l|l}
\textbf{Neighbourhood definition (representative approach)}                                                              & \textbf{Time complexity} & \textbf{ Space complexity} \\ \hline
1-hop neighbourhood (GCN)     &  $O\left(L\left(|V|^2 C+|V| C^2\right)\right)$     & $O\left(\mathrm{L}|V| C+L C^2\right)$   \\ 
1-hop neighbourhood (GraphSAGE)       &  $O(L|V|(sC + C^2))$         & $O\left(\mathrm{L}|V| C+L C^2\right)$   \\
\hline
h-hop neighbourhood (MixHop)                                                                  & $O\left(L\left(|V|^2Ch+|V| C^2\right)\right)$                                    & $O\left(\mathrm{L}|V| C+L C^2\right)$\\ 
h-hop neighbourhood (k-hop GNN)                                                                  & $O(L|V|(k_{max}^hC + C^2))$                                   & $O\left(\mathrm{L}|V| C+L C^2\right)$\\\hline
Random-walk neighbourhood (PinSage)          & $O(L|V|(wl + vlogv + sC + C^2))$    & $O\left(\mathrm{L}|V| C+L C^2\right)$                                     \\ \hline
k-node subgraph neighbourhood (k-GNN)& $O(L \binom{|V|}{k} (k|V|C+C^2))$                                  & $O(L \binom{|V|}{k}C+C^2)$                                    \\ \hline
\end{tabular}
\caption{Time and space complexity from the perspective of a layer-wise message scope. $|V|$ is the number of nodes in the graph, $C$ are node feature channels (assuming the number of features is fixed for all layers), $L$ is the number of convolutional layers, $s$ is the number of sampled nodes, $h$ is the number of hops away from a focal node, $k_{max}$ is maximum node degree, $w$ is the number of random walks, $l$ is the length of a random walk, $v$ is the number of visited nodes, and $k$ is the number of nodes in a subgraph.}
\label{tab:complexity}
\end{table}

First, according to the propagation rule of the vanilla GCN (Equation~\ref{eqn:gcn_full}), which is essentially the multiplication of three matrices $A\in \mathbb{R}^{|V| \times |V|}$, $H\in \mathbb{R}^{|V| \times C}$, and $W\in \mathbb{R}^{C \times C}$, the time complexity at each layer is $O\left(|V|^2 C+|V| C^2\right)$, and thus the overall complexity is $O\left(L\left(|V|^2 C+|V| C^2\right)\right)$. Certainly, when $|V| >> C$, and when the sparsity of adjacency matrix is exploited (for instance through the compressed sparse row format), its time complexity is sometimes expressed as $O(L|E|C)$ \cite{chiang2019cluster, vignac2020building}. As the GCN's space complexity is concerned, we need to store the embeddings of all nodes plus the weight matrix at each layer, which is $O(L|V|C + LC^2)$, or $O(L|V|C)$ when $|V|>>C$.
GraphSAGE illustrates the same propagation procedure from a microscopic view, with a fixed number, denoted $s$, of sampled neighbours involved in the convolutional operation (Equation~\ref{eqn:sage}). The overall time complexity of GraphSAGE is, therefore: $O(L|V|(sC + C^2))$. Notice that when $s$ equals $|V|$, the time complexity of GraphSAGE is the same as the vanilla GCN.  

Then, when each node aggregates messages from its higher-order neighbours, denoted h-hop neighbours here, the propagation rule can be put as: $H^{(l)}=\sigma\left(\hat{A}^h H^{(l-1)} W^{(l)}\right)$. A typical representative is MixHop \cite{abu2019mixhop} (refer to Section~\ref{sec:k-hop}). Thus, the time complexity at each layer is: $O(|V|^2Ch + |V|C^2)$ or $O(|V|^2Ch)$ when $|V|h >> C$, and the space complexity stays unchanged. From a microscopic view, represented by the approach k-hop GNN \cite{nikolentzos2020k}, the time complexity of involving h-hop neighbours in convolutional operation would be $O(L|V|(k_{max}^hC + C^2))$, or $O(L|V|k_{max}^hC)$ when $k_{max}^h >> C$. Clearly, the time complexities of both macroscopic and microscopic algorithms grow with $h$, and when $h$ equals one, they degrade to the versions of 1-hop neighbourhood algorithms, i.e., the vanilla GCN and GraphSAGE.    

Thirdly, approaches with neighbourhood defined on random walks typically include the following steps (represented by PinSage \cite{ying2018graph}): performing $w$ times random walks of length $l$, ranking the visited $v$ nodes based on the visited times, aggregating messages from the top $s$ nodes, and finally applying weight matrix on node representations. Therefore, the overall time complexity is termed as: $O(L|V|(wl + vlogv + sC + C^2))$. Normally, there is no need to record all the random walks, so the space complexity is still $O(L|V|C + LC^2)$. Comparing the time complexity of PinSage with that of GraphSAGE, we see that with the extra step of performing random walks and ranking visited nodes, i.e., the term $wl$ and the term $vlogv$, PinSage is more expensive in computation. 

In the fourth subcategory, we take k-GNN \cite{morris2019weisfeiler} as an example to analyse the complexity of having k-node subgraphs as neighbours. The approach aims to learn embeddings for k-node tuples, and the neighbours of each k-tuple are defined as other k-tuples containing one node that is not in the focal k-tuple(refer to Section~\ref{sec:subgraph}). Each k-tuple aggregates messages from all its k-tuple neighbours, with a time complexity of $O(k|V|C)$. Therefore, on all $\binom{|V|}{k}$ k-tuples and $L$ layers, the overall time complexity is: $O(L \binom{|V|}{k} (k|V|C+C^2))$. To store the embeddings of $\binom{|V|}{k}$ node tuples and the weight matrices at all layers, it requires $O(L \binom{|V|}{k}C+C^2))$ space. This approach is essentially different from the previous ones, in that it is to generate embeddings for k-tuples instead of for each node, resulting in the term $\binom{|V|}{k}$ appearing in both its time and space complexities. Clearly, its complexity grows combinatorially with $k$, and easily surpasses the complexities of all other algorithms when $k$ is relatively large. In practice, however, the value of $k$ generally does not exceed $3$.

\begin{table}[]
\renewcommand{\arraystretch}{1.3}
\begin{tabular}{l|l|l}
\textbf{Message (representative approach)}          & \textbf{Time complexity} & \textbf{ Space complexity} \\ \hline
Node feature X (GCN)     &  $O\left(L\left(|V|^2 C+|V| C^2\right)\right)$     & $O\left(\mathrm{L}|V| C+L C^2\right)$   \\ \hline
Count of graphlets + X (GSN) &  $O(|V|k_{max}^{|S|-1} + L(|V|^2(C+o) + |V|(C+o)^2)$    & $O\left(\mathrm{L}|V|(C+o)+L (C+o)^2\right)$ \\ \hline

Distance information + X (P-GNN)    & $O(|V|^3 + L|V|(Tn + C^2))$      & $O\left(\mathrm{L}|V| C+L C^2\right)$\\  \hline

Random feature + X (rGIN)     &     $O(L(|V|^2(C+r) + |V|(C+r)^2)$    & $O\left(\mathrm{L}|V|(C+r)+L (C+r)^2\right)$     \\ \hline
\end{tabular}
\caption{Time and space complexity from the perspective of a message content. $|V|$ is the number of nodes in the graph, $C$ is node feature channels (assuming the number of features is fixed for all layers), $L$ is the number of convolutional layers, $|S|$ is the maximum size of a set of graphlets, $o$ is the number of orbits in graphlets, $k_{max}$ is maximum node degree, $T$ is the number of anchor sets, $n$ is the maximum number of nodes in an anchor set, and $r$ is the length of the random feature vector.}
\label{tab:complexity2}
\end{table}

In addition, Table~\ref{tab:complexity2} lists the time and space complexities of approaches that include extra node features in the GNNs (corresponding to the message content taxonomy in Figure~\ref{fig:msg-content}). First, when the count of graphlets, or more specifically, the count of node orbits is added to the node features (represented by the approach GSN \cite{bouritsas2020improving}), it requires a preprocessing step to count the number of each node orbit, then performing the general convolutional operation. The cost of counting orbits depends on the size of graphlet $|S|$ and the maximum degree of nodes $k_{max}$. Another difference from the vanilla GCN is that the node feature dimension will increase by the number of orbits, denoted $o$. Therefore, its time complexity is  $O(|V|k_{max}^{|S|-1} + L(|V|^2(C+o) + |V|(C+o)^2)$, and its space complexity is $O\left(\mathrm{L}|V|(C+o)+L (C+o)^2\right)$. Second, when distance information is included, as in the approach P-GNN \cite{you2019position}, it requires first calculating the shortest path distances between all nodes ($O(|V|^3)$ in the typical Floyd-Warshall algorithm), then aggregating message from a number of anchor sets ($T$ anchor sets and each containing at most $n$ nodes). Therefore the time complexity would be $O(|V|^3 + L|V|(Tn + C^2))$. Another less expensive version is to calculate a limited-hop, e.g., h-hop, shortest path distance in the preprocessing step, whose time complexity is $O(|V|k_{max}^h)$. Third and lastly, when random features are included, represented by the approach rGIN \cite{sato2021random}, the impact on time complexity is mainly due to the increase in feature dimension. This is because the cost of generating random features is generally negligible. 

We finally discuss the complexity of approaches that have different learning scopes (corresponding to the taxonomy in Figure~\ref{fig:layer-wise-taxonomy}). For GCNs running on subgraphs, represented by the GraphSAINT \cite{zeng2019graphsaint}, the cost includes two steps, i.e., the subgraph sampling and the training. Given the cost of sampling $T_s$ and a set sampled subgraphs $\mathbb{G}$ (maximum number of nodes in sampled subgraphs denoted $|V_s|$), its complexity is: $O(T_s + |\mathbb{G}|L(|V_s|^2C + |V_s|C^2))$. The cost $T_s$, depending on the choice of the sampler, is normally less expensive than the training. The key term is, therefore, $|\mathbb{G}|L|V_s|^2C$. When $|V_s| << |V|$, subgraph-based approaches significantly reduce the training cost of the GCNs. Similarly, for GCNs running on local subgraphs, exemplified by the Shadow-GNN \cite{zeng2021decoupling}, the two steps are
extracting local subgraphs (extraction cost is denoted as $T_e$, the maximum number of nodes in extracted local subgraphs is denoted as $|V_l|$), and training the GCN on them. Therefore, the time complexity is $O(T_e + |V|L(|V_l|^2C + |V_l|C^2))$. Note that local subgraphs are usually extracted at each node, so the number of extracted subgraphs equals the number of nodes $|V|$. Given that $|V_l| << |V|$ (we should also have $|V_l| << |V_s|$), local subgraph based GCNs are generally much faster in training than full graph or subgraph based GCNs. 


%% file: 5.conclusion.tex
\section{Discussion and Outlook} \label{sec:discussion}
After reviewing the traditional structural measures and the graph convolutional networks, we are set to answer the following research question: How are these two classes of methods related, especially how traditional structural measures of Network Science can inform GCN methods? In this section, we first briefly discuss the performance of GCNs in major learning tasks, then move on to drawing connections between GCNs and traditional structure based approaches, and finally introduce three future directions.

\subsection{GCN's performance in learning tasks}
Convolutional Neural Networks have been shown to be state-of-the-art in various tasks in the area of image processing, including image classification, object detection, and semantic segmentation \cite{li2021survey, gu2018recent}. GCNs have also achieved promising performances in various graph-related tasks. As an extension of CNNs in graph data, GCNs, since their appearance, have received a lot of attention and are viewed as state-of-the-art by default. However, there are works showing that simple heuristics from traditional network science achieve a comparative performance of GCNs \cite{zhang2018link, yun2021neo}, or even beat them in link prediction and network reconstruction tasks \cite{mara2022empirical}. A recent paper shows that simply feeding heuristics derived from nodes similarity scores in a logistic regression model can achieve the best performance in link prediction among many deep learning approaches, including GCNs \cite{mara2022empirical}. In addition, Katz index is the top performer in the network reconstruction task, followed by VGAE which uses GCN as the graph encoder \cite{kipf2016variational}. 

Although the majority of GCN approaches focus on node classification and graph classification tasks, they rarely include structural heuristic-based methods as baselines in the experiment. This overlook could hinder a comprehensive evaluation of the performance of graph convolutional networks. Additionally, comparing GCN approaches with traditional heuristic-based methods could help to better understand the strengths and limitations of GCNs. We believe a closer integration of graph deep learning approaches and traditional network science approaches would immensely benefit both communities, and revealing the connections between the two classes of methods lays the foundation of this integration. 


\subsection{Connections between traditional network science approaches and GCNs} 
Based on the current literature, the connections between GCNs and traditional structure based approaches are observed via the following four aspects. The first aspect covers the foundations of GCNs in traditional Network Science; the second aspect focuses on their similarities in dealing with directed networks; the third and final aspect cover two typical applications of traditional structural information in GCNs: (i) number of graphlets and (ii) distance information. 
\subsubsection{Message passing based approaches and GCN}
As we have seen in message passing based approaches (Section~\ref{sec:MP}), a node's influential score or centrality is calculated through iteratively aggregating the scores of its neighbours until it converges. Taking the eigenvector centrality, for example, the centrality of node $i$, denoted $x(i)$, is formulated as:
\begin{equation*}
            x(i)=c \sum_{j \in N(i)} x(j).
\end{equation*}
$x$, a vector of all nodes' centralities, is found to converge to the dominant eigenvector of the adjacency matrix $A$, and $c$ converges to the reciprocal of the dominant eigenvalue of $A$. 

Interestingly, graph convolutional networks adopt the same idea of neighbourhood aggregation, and the iteration process is implemented through the usage of multiple layers. Taking the vanilla GCN for example, we have the following convolutional operation: 
\begin{equation*}
h_{v}^{(l)}=\sigma\left(\sum_{u \in \mathcal{N}(v)} \frac{1}{c_{v u}} h_{u}^{(l-1)} W^{(l)}\right).
\end{equation*}
Comparing the above two expressions, one major difference is obviously the appearance of weight matrices: in eigenvector centrality, the influential score is directly calculated from forward propagation (e.g., a power iteration), while in the GCN, weight matrices are updated in the backward propagation with the help of labelled samples. Another subtle yet significant difference is that GCNs allow rich node features (n-dimensional vector for each node), while traditional message passing approaches, such as the eigenvector centrality, alpha centrality or PageRank, only support using a numeric value that represents the node's importance or influence. These two points are also the main reasons why GCNs have quickly gained popularity --- the learnable setting makes GCNs suitable for various types of tasks, and the support of rich node features makes them appropriate for different types of real-world data. Despite the advancements and popularity of graph convolutional networks, traditional network science approaches remain important in the field. They have a strong theoretical foundation, which can provide insights into the underlying mechanisms of networked systems. Furthermore, traditional approaches are often more computationally efficient than deep learning approaches, making them more practical for certain types of tasks or data. Overall, the continued use and development of traditional network science approaches alongside newer methods, such as GCNs, can help to deepen our understanding of complex networked systems and advance the field as a whole.

\subsubsection{Dealing with link direction}
When directions of links are considered, we observe interesting connections between the traditional message passing approach HITS \cite{kleinberg1998authoritative} and the recent graph convolutional approach DGP \cite{kampffmeyer2019rethinking}. HITS proposes to distinguish two roles in webpages, i.e., authorities and hubs. Authorities, being reliable information sources, are pointed by hubs (based on incoming edges to the node), while hubs, acting as a home page or library, point to authorities (based on outgoing edges from the node). An authority score and a hub score are defined in a mutually dependent way: 
\begin{equation*}
\begin{aligned}
& a(i) = \sum_{j \in N_i^{in}}h(j),
& h(i) = \sum_{j \in N_i^{out}}a(j).
\end{aligned}
\end{equation*}
Interestingly, DGP, as a graph convolutional approach, proposes to distinguish link direction through a two-phase propagation scheme, i.e., one phase capturing outgoing connections and the other capturing incoming connections (find more in Section~\ref{sec:k-hop}): 
\begin{equation*}
H=\sigma\left(\sum_{k=0}^K \alpha_k^a \hat{A}_k^a \sigma\left(\sum_{k=0}^K \alpha_k^d \hat{A}_k^d X W_d\right) W_a\right).
\end{equation*}
Clearly, the major difference here is that in DGP one type of connection is stacked on top of another, and therefore only one representation is learnt, instead of two scores as in HITS. Besides, k-hop outgoing/incoming connections are included at once in one convolutional layer. Another GCN approach that applies exactly the same idea of distinguishing outgoing edges and incoming edges is Asymmetric GNN, or AGNN \cite{tan2021asymmetric}. It proposes a one-way message passing that only operates on the outgoing or incoming edges of a graph. Two embeddings are then generated for each node to model their roles of sending and receiving information.
It is also possible to design a one-way GCN at particular layers, while still considering both types of edges in other layers, which could allow the model to focus on different aspects of the graph structure at different stages of processing.

\subsubsection{Number of graphlets}
The number of graphlets, or more specifically, node orbits or edge orbits are important topological features around individual nodes or edges (find more in Section~\ref{sec:motif_vs_graphlet}). In a traditional non-learning setting, a vector composed of the counts of a chosen set of node orbits is used to distinguish the roles of nodes \cite{milenkovic2008uncovering, jia2022encoding}. Weights of the orbits, when introduced, are calculated from hand-coded function. In graph convolutional networks, the count of graphlets is added as additional features in the message passing scheme, as we have seen in GSN \cite{bouritsas2020improving}, $\mathcal{F}$-MPNN \cite{barcelo2021graph}, and ID-GNN \cite{you2021identity}. Taken GSN for example, node orbits $x_V(u)$, $x_V(v)$ or edge orbits $x_E(u,v)$ are introduced as follow:
\begin{equation*}
\mathrm{h}^{l+1}(v)=\mathrm{MLP_1}\left(h^l(v), \sum_{u \in \mathcal{N}(v)} MLP_2\left(\mathbf{h}^t(v), \mathbf{h}^t(u), \mathrm{x}_V(v), \mathrm{x}_V(u), \mathrm{e}(u, v)\right)\right),
\end{equation*}
\begin{equation*}
\mathrm{h}^{l+1}(v)=\mathrm{MLP_1}\left(h^l(v), \sum_{u \in \mathcal{N}(v)} MLP_2\left(\mathbf{h}^t(v), \mathbf{h}^t(u), \mathrm{x}_E(u, v), \mathrm{e}(u, v)\right)\right).
\end{equation*}
Obviously, in a learning setting, the weights on all types of features, including the count of graphlets, are learned in the training stage. Another interesting difference between non-learning approaches and GCN approaches is that the former chooses to include all node or edge orbits within a given size, while the latter tends to focus on specific substructures like cycles or cliques. One open problem in using graphlets or orbits in GCNs is determining which ones to choose. Existing approaches have focused on using cliques and/or cycles within a specific range \cite{bouritsas2020improving, barcelo2021graph, you2021identity}, without providing much rationale for this choice. While these types of graphlets and orbits are crucial in some contexts, it is likely that other types could also be exploited to improve the performance of GCN models. There is still much to be explored in terms of the utility of different graphlets and orbits in GCN models, and further research in this area could lead to advances in the field.


\subsubsection{Distance information}
The path related information is largely used in traditional structural measures, such as in closeness centrality, betweenness centrality, $\kappa$-path centrality, etc. Taking the closeness centrality, for example, it is defined as the reciprocal of the average shortest path from the focal node $i$ to all other nodes:\begin{equation*}
        \Theta_C(i) = \frac{|V| - 1}{\sum_{j \in V, j \neq i}d(i, j)}.
    \end{equation*}
The value of a node's closeness centrality is directly used to describe the node's capacity of spreading information on the graph. Unsurprisingly, the distance information is also made of use in graph convolutional networks, as we have seen in P-GNN \cite{you2019position}  and DE-GNN \cite{li2020distance}. In P-GNN, for example, the distance between a node and several anchor sets is included in the convolutional operation: 
\begin{equation*}
\begin{aligned}
&\mathbf{h}_v^{l}=\operatorname{AGG}^{(l)}\left(\mathcal{M}_i^{l-1}, \forall i \in [1,k]\right),
& \mathcal{M}_i^{l-1} = \{ F(d_{uv}, h_u^{l-1}, h_v^{l-1}), \forall u \in S_i\}.
\end{aligned}
\end{equation*}
In DE-GNN, the distance information between node $v$ and a target node set $S$ is used as an extra initial node feature:  $h_v^{(0)} = CONCAT(x_v, \zeta(v \mid S))$. Recall that this idea of including extra structural features as additional initial node features is also found in $\mathcal{F}$-MPNN \cite{barcelo2021graph}, ID-GNN \cite{you2021identity}, and rGIN \cite{sato2021random}.

\subsection{Future directions}
Although recent years have witnessed the great success of graph convolutional networks in various domains, there are still many open problems to be solved and a lot of room for further exploration \cite{bronstein2017geometric, zhou2020graph, zhang2020deep}. Except for the frequently mentioned directions, such as proposing GCNs for more complicated types of networks or to further increase the expressivity or scalability of GCNs, we would like to point out three potential directions which combine the traditional graph analysis approaches and GCN approaches.

\textit{Exploring the applicability of more structural measures in GCNs}. We have seen appearances of various structural measures in GCNs, from the simplest node degree \cite{hamilton2017inductive} to the much more complicated distance information \cite{you2019position} and graphlet orbits \cite{bouritsas2020improving}. However, there are many other traditional structural measures that have yet to be fully explored in the context of GCNs. For example, subgraph formation based measures, such as the clustering coefficient \cite{watts1998collective} and the closure coefficient \cite{yin2019local,jia2020closure}, could be incorporated as node-level features or used to weight the edges of the graph \cite{velivckovic2017graph}. Global path based measures, such as the closeness or betweenness centrality measures, can be used to guide the sampling of nodes, edges or subgraphs when constructing the training set for a GCN \cite{zhou2020graph}. For example, we could use closeness centrality to select the nodes that are most influential in the graph and build subgraphs based on these nodes as the input to the GCN. 
It would be interesting to see how these and other structural measures could be utilised in GCNs to improve performance on certain tasks or in particular types of networks.

\textit{Improving the explainability of GCNs / Guiding the choice of GCNs via traditional structural measures}.
When it comes to the explainability of GCN models, existing methods, represented by perturbation-based methods, mostly focus on generating explanations for a trained GCN \cite{ying2019gnnexplainer, yuan2022explainability}. There are, however, still many questions to be answered, such as how different GCNs perform differently on different types of networks, and what are the reasons for these differences. An analysis from a structural information perspective can provide more insights into how different GCN models extract and utilise graph structural information, and how the information may differ across different GCN models and graph types. This can help to better understand the strengths and limitations of different GCN models and how to effectively apply them in different scenarios. Moreover, in view of the large collection of GCN models and their composition modules, it is difficult to decide which one to choose and how to set it up for the targeted dataset and task \cite{hu2020open}. Traditional structural measures could be used as indices for selecting the appropriate GCN model and the related modules. For example, for graphs that are rich in triangles, a particular GCN would be a better choice, while for graphs where quadrangles are overrepresented, another GCN model should be selected.

\textit{Integrating edge features in GCNs.}  While the vanilla GCN primarily focuses on aggregating and passing information from neighbouring nodes, it is important to consider the role of edge attributes in many real-world networks. For example, in consumer review networks, the ratings of products are often labelled on the edges, and in social networks, the type and frequency of interactions are labelled on the edges. Integrating edge features into GCNs could not only enhance the applicability of the model but also increase the accuracy and relevance of its predictions.
There are works that naively include edge features in GCN or propose a tailor-made model to encompass them \cite{bouritsas2020improving, gong2019exploiting}. However, there is still much to be learned about the utility of traditional edge-level structural measures in GCN models, such as the edge orbits \cite{hovcevar2016computation}, the edge clustering coefficient \cite{wang2011identification}, the local path index \cite{lu2009similarity}, etc. Further research in this area is likely to yield valuable insights and improvements to the performance of GCN models.


\section{Conclusion} \label{sec:conclusion}


The complexity of graph data mainly comes from its intricate topological structures. Mining and exploiting graph structural information have always been one of the focal points in the study of graphs. A large amount of work in traditional network science proposes various types of structural measures, especially local structural measures, to characterise and study complex networks. When more nodes or edges are involved, such approaches, however, become infeasibly complicated. Graph convolutional networks, on the other hand, are proposed to automatically extract relevant features from nodes' neighbourhoods, and in this manner, avoid choosing and  manually calculating structural metrics.   

In order to reveal the connections between the two classes of methods, especially how traditional structural measures can inform GCNs, in this paper, we first reviewed the traditional structure-based approaches in Network Science and proposed a new taxonomy encompassing many seemingly unrelated concepts from a structural perspective. With this prerequisite knowledge, we then extend the scope to the prominent and powerful graph convolutional networks, and provide a Network Science perspective on them --- review and classify GCNs from three structural angles, which are the layer-wise message aggregation scope, the message content, and the overall learning scope. Furthermore, we extensively discuss the connections between the traditional structural approaches and the graph convolutional networks and suggest three future research directions in the joint research area. We believe that the well-established foundations of traditional structure-based approaches in Network Science not only form the basis for GCNs but also could, and probably should, serve as a driving force for their future advances.
